\documentclass[aps,prx,reprint,nofootinbib,superscriptaddress,longbibliography,floatfix]{revtex4-2}
\usepackage{amssymb,bm,amsthm,mathtools,mathrsfs,dsfont}
\usepackage{graphicx,array,booktabs}
\usepackage{needspace}
\usepackage[colorlinks=true,allcolors=blue,breaklinks=true]{hyperref}
\newtheorem{theorem}{Theorem}
\newtheorem{proposition}{Proposition}
\newtheorem{corollary}{Corollary}
\newtheorem{lemma}{Lemma}
\newcommand{\lsp}{\mkern 2mu}

\DeclarePairedDelimiter{\ket}{\lvert}{\rangle}
\DeclarePairedDelimiter{\bra}{\langle}{\rvert}
\DeclarePairedDelimiterX{\braket}[2]{\langle}{\rangle}{#1\,\delimsize\vert\,#2}
\newcommand{\ketbra}[2]{\ket{#1}\mkern-2mu\bra{#2}}
\newcommand{\proj}[1]{\ketbra{#1}{#1}}
\DeclarePairedDelimiter{\norm}{\lVert}{\rVert}
\DeclarePairedDelimiterX{\inner}[2]{\langle}{\rangle}{#1,#2}

\DeclareMathOperator{\tr}{tr}
\DeclareMathOperator{\rk}{rank}
\DeclareMathOperator{\Rea}{Re}
\DeclareMathOperator{\Ima}{Im}
\DeclareMathOperator{\Var}{Var}

\DeclareMathOperator{\supp}{supp}
\DeclareMathOperator{\Beta}{Beta}
\DeclareMathOperator{\Sym}{Sym}

\DeclareMathOperator{\spanop}{span}

\newcommand{\bbC}{\mathbb{C}}
\newcommand{\bbE}{\mathbb{E}}
\newcommand{\bbF}{\mathbb{F}}
\newcommand{\bbP}{\mathbb{P}}

\newcommand{\bbR}{\mathbb{R}}

\newcommand{\bbone}{\mathds{1}}

\newcommand{\caA}{\mathcal{A}}
\newcommand{\caB}{\mathcal{B}}
\newcommand{\caD}{\mathcal{D}}
\newcommand{\caE}{\mathcal{E}}

\newcommand{\caH}{\mathcal{H}}
\newcommand{\caJ}{\mathcal{J}}
\newcommand{\caK}{\mathcal{K}}
\newcommand{\caL}{\mathcal{L}}
\newcommand{\caM}{\mathcal{M}}
\newcommand{\caN}{\mathcal{N}}
\newcommand{\caO}{\mathcal{O}}
\newcommand{\caP}{\mathcal{P}}

\newcommand{\caS}{\mathcal{S}}

\newcommand{\rmd}{\mathrm{d}}
\newcommand{\rme}{\mathrm{e}}
\newcommand{\rmi}{\mathrm{i}}
\newcommand{\reg}{\mathrm{reg}}
\newcommand{\rmB}{\mathrm{B}}
\newcommand{\rmbas}{\mathrm{bas}}

\newcommand{\rmH}{\mathrm{H}}
\newcommand{\Ha}{\mathrm{Haar}}
\newcommand{\SIC}{\mathrm{SIC}}
\newcommand{\Stab}{\mathrm{Stab}}
\newcommand{\rmstab}{\mathrm{stab}}
\newcommand{\rmcl}{\mathrm{cl}}
\newcommand{\rmdod}{\mathrm{dod}}
\newcommand{\rmeq}{\mathrm{eq}}
\newcommand{\rmerr}{\mathrm{err}}
\newcommand{\rmopt}{\mathrm{opt}}
\newcommand{\rmph}{\mathrm{ph}}
\newcommand{\rmphase}{\mathrm{phase}}

\newcommand{\rmres}{\mathrm{res}}

\newcommand{\rmsucc}{\mathrm{succ}}
\newcommand{\rmtar}{\mathrm{tar}}
\newcommand{\designthree}{\text{3-design}}

\newcommand{\scrP}{\mathscr{P}}
\newcommand{\fkE}{\mathfrak{E}}
\newcommand{\fkR}{\mathfrak{R}}
\newcommand{\fkr}{\mathfrak{r}}

\newcommand{\bmr}{\boldsymbol{r}}
\newcommand{\bmu}{\boldsymbol{u}}
\newcommand{\bmv}{\boldsymbol{v}}

\newcommand{\bmsigma}{\boldsymbol{\sigma}}
\newcommand{\bfone}{\mathbf{1}}

\newcommand{\sfT}{\mathsf{T}}
\newcommand{\sfP}{\mathsf{P}}

\newcommand{\hrho}{\widehat{\rho}}
\newcommand{\hp}{\widehat{p}}
\newcommand{\heta}{\widehat{\eta}}
\newcommand{\ba}{\bar{a}}
\newcommand{\oa}{\overline{a}}
\newcommand{\IdTan}[1]{\operatorname{id}_{\caH_{#1}}}
\newcommand{\IdMat}[1]{\mathrm{Id}_{#1}}

\newcommand{\thref}[1]{Theorem~\ref{#1}}
\newcommand{\pref}[1]{Proposition~\ref{#1}}
\newcommand{\coref}[1]{Corollary~\ref{#1}}
\newcommand{\lref}[1]{Lemma~\ref{#1}}
\newcommand{\eref}[1]{Eq.~\eqref{#1}}
\newcommand{\sref}[1]{Sec.~\ref{#1}}
\newcommand{\aref}[1]{Appendix~\ref{#1}}
\newcommand{\rcite}[1]{Ref.~\cite{#1}}

\providecommand{\bmv}{\bm{v}}
\providecommand{\bmu}{\bm{u}}

\makeatletter
\AtBeginDocument{\immediate\write\@auxout{\string\citation{apsrev42Control}}}
\makeatother

\begin{document}
\raggedbottom
\title{Sharp Uniform Fisher Information Bounds from Projective Designs}
\author{Zhiyao Yang}
\thanks{These authors contributed equally to this work.}
\author{Bowen Xiao}
\thanks{These authors contributed equally to this work.}
\affiliation{State Key Laboratory of Surface Physics, Department of Physics, and Center for Field Theory and Particle Physics, Fudan University, Shanghai 200433, China}
\affiliation{Institute for Nanoelectronic Devices and Quantum Computing, Fudan University, Shanghai 200433, China}
\affiliation{Shanghai Research Center for Quantum Sciences, Shanghai 201315, China}
\author{Huangjun Zhu}
\email{zhuhuangjun@fudan.edu.cn}
\affiliation{State Key Laboratory of Surface Physics, Department of Physics, and Center for Field Theory and Particle Physics, Fudan University, Shanghai 200433, China}
\affiliation{Institute for Nanoelectronic Devices and Quantum Computing, Fudan University, Shanghai 200433, China}
\affiliation{Shanghai Research Center for Quantum Sciences, Shanghai 201315, China}
\affiliation{Hefei National Laboratory, Hefei 230088, China}
\begin{abstract}
Fixed measurements for multiparameter pure-state metrology must remain informative without knowing the input state. We quantify this requirement by the Fisher floor, the smallest fraction of the quantum Fisher information that a measurement retains over all pure states and tangent directions. We determine the Fisher guarantees, namely lower bounds on Fisher floors, implied by projective design order alone in every dimension $d\ge2$. These guarantees are sharp at order $t=2$ and every odd order $t\ge3$, and finite weighted designs attain them; at even orders, our bounds are nearly tight. As the order increases, the guarantees approach one half, the single-copy quantum limit. Sampling and spectral completion of any $3$-design yield, with probability $1-\caO(1/d)$, a measurement with linearly many outcomes and a dimension-independent Fisher floor. Such a measurement separates all pure states in Hellinger distance, so pure-state maximum-likelihood estimation from $N$ copies attains mean infidelity $\caO(d/N)$ uniformly over pure states, optimal up to a universal constant factor.
\end{abstract}
\maketitle

\begingroup
\small
\makeatletter
\let\l@subsubsection\@gobbletwo
\makeatother
\tableofcontents
\endgroup

\section{Introduction}
\label{sec:intro}

Quantum metrology concerns the estimation of unknown parameters encoded in quantum states, with applications ranging from precision sensing to quantum state characterization~\cite{GiovannettiLloydMaccone2011,SzczykulskaBaumgratzDatta2016}. Quantum Fisher information (QFI) provides a fundamental benchmark for estimation precision~\cite{BraunsteinCaves1994}. In multiparameter estimation, measurements optimal for different parameters can be incompatible, leading to trade-offs in the accessible information~\cite{GillMassar2000,Matsumoto2002,WangChenYuan2026}. Moreover, an optimal measurement may depend on the unknown state or the parameters of interest. In tomography of an unknown state, a measurement can be informative for typical states yet nearly blind to some changes of particular states. This motivates fixed single-copy measurements, described by positive-operator-valued measures (POVMs), that retain a uniform fraction of the QFI across states and parameter directions. We quantify this property by the Fisher floor: the largest constant $c$ such that the classical Fisher information of a POVM is at least $c$ times the QFI at every pure state and in every tangent direction. A positive Fisher floor protects against poorly resolved directions before the state is known.

For pure-state estimation, Fisher-symmetric measurements attain the optimal single-copy trade-off at a prescribed state~\cite{LiEtAl2016}. They retain one half of the QFI in every tangent direction, the largest fraction that the Gill--Massar bound allows simultaneously in all directions~\cite{GillMassar2000}. The continuous covariant Haar measurement achieves this fraction uniformly over all pure states~\cite{Hayashi1998,LuSha2025}. Projective $2$-designs also yield two-copy measurements that are Fisher symmetric at every pure state~\cite{ZhuHayashi2018}. Finite randomized measurements offer a practical alternative and have proved useful in quantum state estimation and classical shadow tomography~\cite{HuangKuengPreskill2020,ElbenEtAl2023}.

How does the structure of a fixed measurement determine its Fisher floor? We address this question through complex projective designs, ensembles of pure states that reproduce the Haar moments up to a given order $t$~\cite{Scott2006,RoyScott2007}. Designs are natural candidates: symmetric informationally complete (SIC) ensembles and complete sets of mutually unbiased bases are $2$-designs, multiqubit stabilizer states form a $3$-design that can be measured with Clifford circuits, and designs underlie many randomized-measurement protocols. For projective $3$-designs, a dimension-independent Fisher guarantee, that is, a lower bound on the Fisher floor, is already known~\cite{ZhouChen2026}, with related results on approximate unitary designs~\cite{DuEtAl2026}. We seek the largest Fisher guarantee implied by design order alone and ask whether finite designs attain it.

We determine the sharp Fisher guarantees implied by design order alone at order $t=2$ and every odd order $t\ge3$, and we construct finite weighted designs that attain them in every dimension; in particular, the known $3$-design bound is sharp. The $2$-design guarantee decays as $1/d$. In contrast, at every odd order $t\ge3$, the guarantee falls short of one half by at most $1/(t+1)$, uniformly in $d$, so the guarantees approach the single-copy quantum limit as the order increases. At even orders, we obtain lower bounds strictly above the preceding odd-order values and upper bounds from finite constructions; the gap between them vanishes as the order increases, uniformly in $d$. Particular designs can exceed these guarantees. SIC measurements attain the $2$-design guarantee, whereas complete multiqubit stabilizer measurements have Fisher floor $1/3$, strictly above the $3$-design guarantee for $d>2$. We also determine the Fisher floors of the five Platonic-solid qubit measurements; the dodecahedral measurement likewise exceeds its design guarantee.

We next address measurement resources. A positive Fisher floor requires linearly many outcomes, and linearly many outcomes can suffice to identify every pure state~\cite{ConcaEtAl2015}. Identifiability alone, however, does not provide a dimension-independent Fisher guarantee. We show that sampling $\caO(d)$ projectors from any $3$-design, followed by spectral completion, produces, with probability at least $1-1/(2d)-4^{-d}$, a POVM with $\caO(d)$ outcomes and a dimension-independent Fisher floor. Sampling $\caO(d)$ orthonormal bases whose vectors form a $3$-design achieves the same without completion, using $\caO\left(d^2\right)$ outcomes; for stabilizer bases, we show that at least $\log_2d+1$ bases are necessary.

Finally, we connect Fisher floors with global statistical guarantees. For finite rank-one POVMs, a positive Fisher floor implies global Hellinger separation between the output distributions of any two pure states, and we obtain a sharp separation bound for $3$-designs. This separation yields guarantees for phase retrieval and pure-state discrimination. It also yields uniform $\caO(d/N)$ mean infidelity for pure-state maximum-likelihood estimation from $N$ copies, for every measurement produced by the sampling construction on its success event; this rate is optimal up to a constant factor. Locally, the pairing of Fisher eigenvalues shows that, for $3$-designs, efficient estimators at regular states have asymptotic mean infidelity at most $4/3$ times the optimal value, which coincides with the optimum attainable by collective measurements on many copies~\cite{Matsumoto2002}.

These results provide a design-based criterion, available before any data are collected, for how much information a fixed readout retains about an unknown pure state. They apply directly to stabilizer and other randomized measurements used in tomography and shadow estimation, and they quantify how far such measurements can be reduced in size without losing uniform sensitivity.

The paper is organized as follows. Section~\ref{sec:setting} introduces the measurement model, projective designs, and the Fisher floor, and establishes a trace identity and the pairing of Fisher eigenvalues. Section~\ref{sec:hier} presents the central result of the paper: after explaining why the canonical shadow-variance argument uses only the first three design moments, it establishes the sharp design-order Fisher guarantees in \thref{thm:hier}, determines the Fisher floors of SIC, stabilizer, and Platonic measurements, and outlines the proofs. Section~\ref{sec:samp} derives an outcome constraint and constructs sampled measurements with $\caO(d)$ outcomes. Section~\ref{sec:est} turns Fisher floors into global separation and applies it to phase retrieval, discrimination, and reconstruction. Section~\ref{sec:discuss} discusses implications and open problems. Proofs are collected in the appendices.

\section{Framework: measurements, designs, and Fisher floors}
\label{sec:setting}

A fixed readout must resolve changes in different directions of an unknown state. For example, a computational-basis measurement on a qubit detects population changes but is insensitive to the relative phase. This section introduces the measurement model, projective designs, and the Fisher floor, and records geometric properties used throughout. Proofs are given in \aref{app:geometry}.

\subsection{Measurement model and notation}

Let $\caH$ be a $d$-dimensional complex Hilbert space ($d\ge2$), identified with $\bbC^d$ when coordinates are needed. Denote by $\caL(\caH)$ and $\caL^{\rmH}_0(\caH)$ the spaces of linear operators and traceless Hermitian operators, respectively; denote by $\caD(\caH)$ the set of density operators and by $\caP(\caH)\subset\caD(\caH)$ the set of pure states. Here $\tr$ denotes trace and $\bbone$ is the identity operator on $\caH$. The symbols $\sfT$ and $\dagger$ denote transpose and conjugate transpose, respectively. We use $\rho,\sigma$ for state operators and write a pure state as $\rho=\proj\psi$ with a unit ket $\ket\psi$; extrema over state kets range over unit vectors in $\caH$. For rank-one measurement projectors, write $\phi=\proj\phi$. We use $\norm{\cdot}_\ell$ for the Schatten $\ell$-norm on operators, $\norm{\cdot}_2$ for the Euclidean norm on vectors, and $\norm{\cdot}_1$ for the usual $\ell_1$-norm on probability vectors. For vectors $x,y\in\caH$, write $\inner{x}{y}=x^\dagger y$ and $\inner{x}{y}_{\bbR}=\Rea\inner{x}{y}$. The symbols $\bbE_\rho$ and $\Var_\rho$ denote expectation and variance over measurement outcomes at state $\rho$.

A finite POVM $\caM=\{E_i\}_{i=1}^M$ has nonzero elements $E_i\succeq0$ satisfying $\sum_iE_i=\bbone$. At a state $\rho$, its Born probabilities are $p_i(\rho)=\tr(E_i\rho)$, with $p_\rho=(p_i(\rho))_{i=1}^M$. A rank-one POVM has elements
\begin{equation}
E_i=w_i\phi_i,\quad \phi_i=\proj{\phi_i},\quad
\sum_{i=1}^{M}w_i\phi_i=\bbone,
\label{eq:povm}
\end{equation}
where $\ket{\phi_i}\in\caH$ are unit kets and $w_i>0$. Thus $\sum_iw_i=d$, and the associated ensemble weights are $\pi_i=w_i/d$. A state $\rho$ is \emph{regular with respect to $\caM$} if $p_i(\rho)>0$ for every $i$, and \emph{singular} otherwise. When the state is fixed, we abbreviate $p_i=p_i(\rho)$. We use the same POVM independently on each copy. Throughout, $M$ counts labeled measurement outcomes and $N$ counts measured copies.

\subsection{Projective designs and structured ensembles}

A complex projective $t$-design (or state $t$-design) reproduces Haar tensor moments through order $t$~\cite{Hoggar1982,Scott2006,RoyScott2007}. Throughout, a $t$-design means a finite weighted ensemble $\caE=\{(\pi_i,\phi_i)\}_i$, with $\pi_i>0$ and $\sum_i\pi_i=1$, that satisfies
\begin{equation}
\sum_i\pi_i\phi_i^{\otimes s}
=\frac{P_{[s]}}{D_{[s]}},\qquad
D_{[s]}=\binom{d+s-1}{s},
\label{eq:des}
\end{equation}
exactly for $1\le s\le t$. Here $P_{[s]}$ projects onto the symmetric subspace of $\caH^{\otimes s}$, and $P_{[s]}/D_{[s]}=\bbE_{\Ha}\phi^{\otimes s}$ is the Haar tensor moment; $\bbE_{\Ha}$ denotes expectation over the normalized unitarily invariant measure on pure states. We set $D_{[0]}=1$. Tracing out tensor factors shows that the order-$t$ identity implies the lower-order identities. In particular, for fixed operators $A_1,\ldots,A_s$ and $s\le t$, the design and Haar averages of $\prod_{j=1}^s\tr(A_j\phi)$ coincide, by contracting \eref{eq:des} with $A_1\otimes\cdots\otimes A_s$. Denote by $\fkE_t(d)$ the set of all $t$-designs in dimension $d$.

The $s=1$ identity gives $\sum_i d\pi_i\phi_i=\bbone$. Hence every $t$-design $\caE$ defines a rank-one POVM $\caM_\caE=\{d\pi_i\phi_i\}_i$, which we call the POVM associated with $\caE$.

Standard examples are symmetric informationally complete (SIC) ensembles~\cite{Zauner2011,RenesEtAl2004,FuchsHoangStacey2017} and complete sets of mutually unbiased bases~\cite{Ivanovic1981,WoottersFields1989,DurtEtAl2010}, which are $2$-designs~\cite{Scott2006,RoyScott2007}, and multiqubit stabilizer states, which form a $3$-design~\cite{KuengGross2015,Webb2016,Zhu2017}. A SIC ensemble consists of $d^2$ equally weighted rank-one projectors satisfying
\begin{equation}
\tr(\phi_i\phi_j)=\frac1{d+1}\qquad(i\ne j).
\label{eq:sic-overlap}
\end{equation}
For $n$ qubits, $d=2^n$, the stabilizer states form the Clifford orbit of $\ket0^{\otimes n}$, where the Clifford group is the unitary normalizer of the Pauli group~\cite{Gottesman1997}. Equivalently, each stabilizer state spans a joint eigenspace of a maximal commuting Pauli subgroup. With global phases identified, the distinct stabilizer projectors form an equal-weight $3$-design~\cite{KuengGross2015,Webb2016,Zhu2017}.

\subsection{Fisher information and Fisher floors}\label{sec:fisher-defs}

\begin{figure*}[!t]
\centering
\raisebox{-0.00000pt}[153.89244pt][0.00000pt]{\includegraphics[width=469.20215pt,height=153.89244pt]{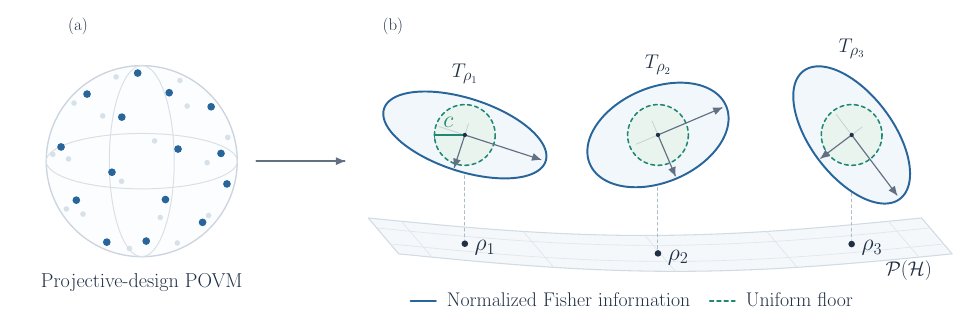}}
\caption{Fisher floor of a fixed design measurement. (a)~Rank-one elements of the POVM, shown as points on a sphere. (b)~At each pure state $\rho=\proj\psi$, the blue ellipsoid represents $\{G_{\psi,\caM}\nu\colon\nu\in\caH_\psi,\ \norm{\nu}_2\le1\}$, with principal semiaxes equal to the eigenvalues of $G_{\psi,\caM}$. The dashed green circles represent balls of radius $c$. Every ellipsoid contains such a ball exactly when $G_{\psi,\caM}\succeq c\IdTan\psi$ at every pure state, equivalently $I_{\rho,\caM}\succeq cJ_\rho$.}
\label{fig:unif-fi-idea}
\end{figure*}

Fix a pure state $\rho=\proj\psi\in\caP(\caH)$. The horizontal space $\caH_\psi=(\spanop_{\bbC}\{\ket\psi\})^\perp$ is a real Hilbert space of dimension $2d-2$ with the inner product $\inner{\cdot}{\cdot}_{\bbR}$. Each horizontal ket $\ket\nu\in\caH_\psi$ defines the tangent operator $\Delta_{\psi,\nu}$, and these operators span the real tangent space $T_\rho\subset\caL^{\rmH}_0(\caH)$ of the pure-state manifold at $\rho$:
\begin{align}
\Delta_{\psi,\nu}&:=\ketbra\psi\nu+\ketbra\nu\psi,
\label{eq:tan}\\
T_\rho&:=\spanop_{\bbR}\{\Delta_{\psi,\nu}\colon\ket\nu\in\caH_\psi\}.
\label{eq:tangent-space}
\end{align}
The map $\ket\nu\mapsto\Delta_{\psi,\nu}$ is a real-linear isomorphism from $\caH_\psi$ onto $T_\rho$. Every nonzero $\Delta\in T_\rho$ can be written as $\Delta=r\Delta_{\psi,\nu}$ with $r>0$ and $\ket\nu\in\caH_\psi$ a unit ket.

We use the symmetric logarithmic derivative (SLD) quantum Fisher information~\cite{BraunsteinCaves1994}. For $\Delta\in T_\rho$, one SLD along $\Delta$ is $L_\Delta=2\Delta$, which gives \eref{eq:qfi} below. For a finite POVM $\caM=\{E_i\}_{i=1}^M$, the directional derivative of the Born probabilities is $\partial_{\Delta}p_i(\rho)=\tr(E_i\Delta)$, and the classical Fisher information is given by \eref{eq:score-fi}~\cite{GillMassar2000}:
\begin{align}
J_\rho(\Delta,\Delta)&=2\tr\left(\Delta^2\right),
\label{eq:qfi}\\
I_{\rho,\caM}(\Delta,\Delta)&=
\sum_{i\colon p_i(\rho)>0}\frac{[\tr(E_i\Delta)]^2}{p_i(\rho)}.
\label{eq:score-fi}
\end{align}
Both bilinear forms follow by real polarization. In particular, $J_\rho(\Delta_{\psi,\nu},\Delta_{\psi,\nu})=4$ for every unit ket $\ket\nu\in\caH_\psi$. Outcomes with $p_i(\rho)=0$ are omitted from \eref{eq:score-fi}; as a result, the Fisher information can be discontinuous at singular states.

We define the normalized Fisher operator $G_{\psi,\caM}$ as the real self-adjoint operator on $\caH_\psi$ with the quadratic form in \eref{eq:normalized-fi-form} on unit kets, extended by homogeneity and real polarization. Thus $G_{\psi,\caM}$ compares the classical Fisher information with the QFI direction by direction; we call its eigenvalues the normalized Fisher eigenvalues at $\rho$. The Fisher floor of $\caM$ is the infimum of the least normalized Fisher eigenvalue over all pure states:
\begin{align}
\inner{\nu}{G_{\psi,\caM}\nu}_{\bbR}&=\frac14 I_{\rho,\caM}(\Delta_{\psi,\nu},\Delta_{\psi,\nu}),
\label{eq:normalized-fi-form}\\
\eta(\caM)&=\inf_{\ket\psi}\lambda_{\min}(G_{\psi,\caM}).
\label{eq:eta}
\end{align}
Equivalently, $\eta(\caM)$ is the largest $c$ for which $I_{\rho,\caM}\succeq cJ_\rho$ at every pure state. It thus measures the sensitivity of $\caM$ at the least favorable state and tangent direction, as illustrated in Fig.~\ref{fig:unif-fi-idea}. A \emph{Fisher guarantee} is a lower bound on the Fisher floor, valid for a single measurement or for every measurement in a specified class. A Fisher guarantee $c$ for a class is \emph{sharp} if no larger constant is a Fisher guarantee for that class, and it is attained if some measurement in the class has $\eta(\caM)=c$.

The trace $\tr_{\bbR}$ and identity $\IdTan{\psi}$ refer to $\caH_\psi$ regarded as a real space. For a rank-one POVM as in \eref{eq:povm}, write $a_i=\braket{\phi_i}{\psi}$, so that $p_i(\rho)=w_i|a_i|^2$. Then \eref{eq:score-fi} gives, for each unit ket $\ket\nu\in\caH_\psi$,
\begin{equation}
\inner{\nu}{G_{\psi,\caM}\nu}_{\bbR}
=\sum_{i\colon p_i(\rho)>0}w_i
\frac{[\Rea(\ba_i\braket{\phi_i}{\nu})]^2}{|a_i|^2}.
\label{eq:hor-fi}
\end{equation}
The Fisher operator depends on the representative $\ket\psi$ of $\rho$ only up to conjugation. Replacing $\ket\psi$ by $\rme^{\rmi\alpha}\ket\psi$ leaves $\caH_\psi$ unchanged, and the real orthogonal map $\ket\nu\mapsto\rme^{\rmi\alpha}\ket\nu$ conjugates the two Fisher operators. Hence the spectrum of $G_{\psi,\caM}$ depends only on $\rho$.

\subsection{Trace identity and Fisher spectral pairing}\label{sec:fisher-geom}

For a rank-one POVM and a pure state $\rho=\proj\psi$, define
\begin{equation}
Z_\psi=\sum_{i\colon p_i(\rho)=0}w_i\phi_i.
\label{eq:zero-op}
\end{equation}
Each summand annihilates $\ket\psi$, so $Z_\psi$ acts on $\caH_\psi$. We write $(Z_\psi)_{\bbR}$ for the same operator on $\caH_\psi$ regarded as a real Hilbert space.

\begin{proposition}[Trace identity and spectral pairing]\label{prop:geometry}
For every finite rank-one POVM $\caM=\{E_i=w_i\phi_i\}_{i=1}^M$ and every pure state $\rho=\proj\psi$,
\begin{equation}
\tr_{\bbR}G_{\psi,\caM}=d-1-\sum_{i\colon p_i(\rho)=0}w_i,
\label{eq:trace-defect}
\end{equation}
and the complex structure $\caJ$, given by multiplication by $\rmi$ on $\caH_\psi$, satisfies
\begin{equation}
G_{\psi,\caM}+\caJ G_{\psi,\caM}\caJ^{-1}
=\IdTan{\psi}-(Z_\psi)_{\bbR}.
\label{eq:pairing}
\end{equation}
At a regular state, $Z_\psi=0$ and the eigenvalues occur in pairs $1/2\pm s_j$, with $0\le s_j\le1/2$.
\end{proposition}

Equation~\eqref{eq:hor-fi} explains the pairing: replacing $\ket\nu$ by $\rmi\ket\nu$ exchanges, up to sign, the real and imaginary parts of $\ba_i\braket{\phi_i}{\nu}$, so the quadratic forms in the two directions add up to $\sum_{i\colon p_i(\rho)>0}w_i|\braket{\phi_i}{\nu}|^2$. At a regular state, the $2d-2$ normalized Fisher eigenvalues therefore sum to $d-1$ and average to $1/2$; improving sensitivity in one member of a spectral pair reduces it in the other. The continuous covariant Haar POVM attains the value $1/2$ in every direction at every pure state~\cite{Hayashi1998,LuSha2025}. This benchmark concerns the full pure-state model. A restricted parameter submodel can have a larger ratio of classical to quantum Fisher information, while a bound $I_{\rho,\caM}\succeq cJ_\rho$ transfers to any differentiable pure-state parameter model by evaluating both forms on its coordinate tangent directions. A positive Fisher floor controls the imbalance within spectral pairs throughout the state manifold, while zero-probability outcomes additionally reduce the trace of $G_{\psi,\caM}$.

Because the Fisher information can jump at singular states, we also define the regular-state Fisher floor
\begin{equation}
\eta_{\reg}(\caM)
:=\inf_{\substack{\ket\psi\\p_i(\rho)>0\ \forall i}}
\lambda_{\min}(G_{\psi,\caM}).
\label{eq:reg-eta}
\end{equation}
The next proposition shows that the two floors coincide and that the infimum in \eref{eq:eta} is attained.

\begin{proposition}[All-state and regular-state Fisher floors]\label{prop:reg-eq}
Let $\caM=\{E_i\}_{i=1}^M$ be a finite POVM on $\caH$ with nonzero elements. Then
\begin{equation}
0\le\eta_{\reg}(\caM)=\eta(\caM)
=\min_{\ket\psi}\lambda_{\min}(G_{\psi,\caM})\le\frac12.
\label{eq:reg-eq}
\end{equation}
\end{proposition}

The upper bound $\eta(\caM)\le1/2$ is consistent with the Gill--Massar bound~\cite{GillMassar2000} for simultaneous estimation of all pure-state parameters.

For finite POVMs, a positive Fisher floor is equivalent to pure-state identifiability.

\begin{proposition}[Pure-state identifiability and Fisher positivity]\label{prop:pure-ic}
For a finite POVM $\caM$ on $\caH$, the following conditions are equivalent:
\begin{enumerate}
\renewcommand{\labelenumi}{(\arabic{enumi})}
\item\label{cond:pure-identifiability} $\caM$ distinguishes all pure states.
\item\label{cond:pointwise-fisher} $I_{\rho,\caM}(\Delta,\Delta)>0$ for every pure state $\rho$ and nonzero $\Delta\in T_\rho$.
\item\label{cond:uniform-fisher} $\eta(\caM)>0$.
\end{enumerate}
\end{proposition}

\Needspace{6\baselineskip}
\section{Fisher guarantees from projective designs}
\label{sec:hier}

Design moments constrain the Fisher information uniformly over unknown states. We distinguish the guarantee shared by all $t$-designs from improvements due to the structure of particular ensembles. The design Fisher guarantee at order $t$ is
\begin{equation}
\eta_{\mathrm{des}}(d,t):=\inf_{\caE\in\fkE_t(d)}\eta(\caM_\caE),
\label{eq:des-cls-eta}
\end{equation}
where $\fkE_t(d)$ is the set of $t$-designs in dimension $d$ defined in \sref{sec:setting}. By \pref{prop:reg-eq}, replacing $\eta$ by $\eta_{\reg}$ in \eref{eq:des-cls-eta} does not change its value. Individual designs can have larger Fisher floors. Since every $(t+1)$-design is also a $t$-design, $\eta_{\mathrm{des}}(d,t)$ is nondecreasing in $t$.

We begin with shadow reconstruction, which links the accuracy of observable estimation to Fisher floors. For designs, the canonical reconstruction recovers the known $3$-design guarantee, but its variance involves only the first three design moments. This limitation motivates the direct moment bounds in \thref{thm:hier}, which improve with design order and are attained by finite designs at second and odd orders.

\subsection{From shadow variance to Fisher guarantees}
\label{sec:shadows}

The results of this subsection are proved in \aref{app:shadow-fi}. Classical shadows provide a classical representation of an unknown state $\sigma\in\caD(\caH)$ from randomized measurements~\cite{HuangKuengPreskill2020,ElbenEtAl2023,InnocentiEtAl2023}. In each shot, a randomly chosen measurement is applied to a fresh copy of the state, and the setting and outcome are recorded. In the standard protocol, a random unitary precedes a computational-basis measurement; the observed basis state is mapped back by the inverse unitary and then processed by the inverse of the average measurement channel. This produces a Hermitian reconstruction matrix, or classical shadow, for each shot. A collection of such matrices can be used to estimate many observable expectations by classical postprocessing.

Averaged over the random settings, the whole protocol is equivalent to a single POVM $\caM=\{E_i\}$, where each outcome $i$ records both the setting and the measurement result. Classical postprocessing assigns to each outcome $i$ a fixed Hermitian operator $\hrho_i$, which need not be a density operator. We require the reconstruction to be unbiased:
\begin{equation}
\sum_i p_i(\sigma)\hrho_i=\sigma
\quad\text{for every }\sigma\in\caD(\caH).
\label{eq:shadow-mean}
\end{equation}
For a Hermitian observable $O$, $\tr(O\hrho_i)$ is then an unbiased estimate of $\tr(O\sigma)$, with single-shot variance
\begin{equation}
\Var_\sigma[\tr(O\hrho_i)]
=\sum_i p_i(\sigma)
\bigl[\tr(O\hrho_i)-\tr(O\sigma)\bigr]^2.
\label{eq:shadow-var-def}
\end{equation}
Given $N$ independent records $i_1,\ldots,i_N$, the sample mean $N^{-1}\sum_{j=1}^N\tr(O\hrho_{i_j})$ estimates $\tr(O\sigma)$ with variance $\Var_\sigma[\tr(O\hrho_i)]/N$. The reconstruction variance therefore determines the number of copies needed for a prescribed precision. Extensions include shallow circuits~\cite{BertoniEtAl2024,HuChoiYou2023}, robust reconstruction~\cite{ChenEtAl2021Robust,HuEtAl2025RobustShallow}, and common randomized measurements that use classical state approximations~\cite{VermerschEtAl2024CRM,YangEtAl2025CRM}. Inverse Fisher information also characterizes the sample complexity of shadow tomography in the high-precision regime~\cite{ChenGongZhou2026}.

Uniform variance bounds for traceless observables also yield Fisher guarantees on pure states. Such bounds have been derived for projective $3$-designs and several structured ensembles~\cite{HuangKuengPreskill2020,YangChenZhu2026,MaoYiZhu2025,ParkTeoJeong2025}.

\begin{figure}[!t]
\centering
\raisebox{-0.00000pt}[169.26228pt][0.00000pt]{\includegraphics[width=241.30907pt]{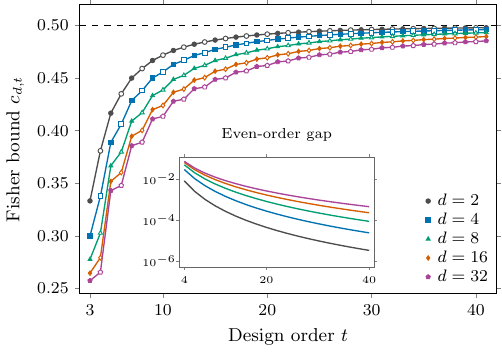}}
\caption{Fisher guarantees from \thref{thm:hier} for $t\ge3$. Filled markers give the sharp odd-order values; open markers give the even-order lower bounds $c_{d,t}$. The inset shows the gap between the even-order upper and lower bounds in \eref{eq:even-bracket}. The dashed line marks the benchmark $1/2$.}
\label{fig:des-hier}
\end{figure}

\begin{proposition}[Reconstruction variance and pure-state Fisher information]
\label{prop:shadow-fi}
Suppose a finite POVM $\caM$ admits an unbiased reconstruction as in \eref{eq:shadow-mean}, with
\begin{equation}
\Var_\sigma[\tr(O\hrho_i)]\le C\norm{O}_2^2
\label{eq:shadow-var}
\end{equation}
for every state $\sigma$ and traceless Hermitian $O$, where $C>0$. Then
\begin{equation}
\eta(\caM)\ge\frac1{2C}.
\label{eq:shadow-fi-bnd}
\end{equation}
\end{proposition}

The proof uses \eref{eq:shadow-var} only for pure states $\rho$ and observables $O\in T_\rho$. The variance-based bounds in Table~\ref{tab:shadow-fi-ens} follow from this criterion; the exact qubit stabilizer Fisher floor is established in \pref{prop:stab-fi}. The criterion is not limited to designs of order three. For instance, odd-prime qudit stabilizer states form $2$-designs but not $3$-designs~\cite{Zhu2017}. For such ensembles, design moments alone give only the guarantee $c_{d,2}$ in \thref{thm:hier}, which decays as $1/d$, whereas shadow-variance bounds still give Fisher guarantees independent of the number of qudits.

\begin{table}[!ht]
\caption{Lower bounds on Fisher floors of some typical POVMs. $\caM_{\designthree}$ denotes a POVM from a projective $3$-design; $\caM_{\rmstab(2,n)}$, the complete $n$-qubit stabilizer POVM in dimension $d=2^n$; $\caM_{\rmstab(q,n)}$, the complete $n$-qudit stabilizer POVM in dimension $d=q^n$, where $q$ is an odd prime; $\caM_{\rmphase(L)}$, the POVM combining phase $3$-designs over $L$ mutually unbiased reference bases, with $2\le L\le d+1$~\cite{YangChenZhu2026}; and $\caM_{\rmeq+Z}$, the equal-probability mixture of equatorial-stabilizer and computational-basis measurements on $n$ qubits, with $d=2^n$~\cite{ParkTeoJeong2025}. The star marks an exact Fisher floor; the other entries are lower bounds.}
\label{tab:shadow-fi-ens}
\centering
\begingroup
\small
\setlength{\tabcolsep}{5pt}
\renewcommand{\arraystretch}{1.2}
\begin{tabular}{@{}lc@{}}
\toprule
Measurement ensemble & Fisher guarantee\\
\midrule
$\caM_{\designthree}$~\cite{ZhouChen2026,HuangKuengPreskill2020} & $(d+2)/[4(d+1)]$\\
$\caM_{\rmstab(2,n)}$~(\pref{prop:stab-fi}) & $(1/3)^{\star}$\\
$\caM_{\rmstab(q,n)}$~\cite{MaoYiZhu2025,ZhuMaoYi2024}~(\pref{prop:stab-fi}) & $1/[4(q-1)]$\\
$\caM_{\rmphase(L)}$~\cite{YangChenZhu2026} & $(L-1)^2/[2L(3L+1)]$\\
$\caM_{\rmeq+Z}$~\cite{ParkTeoJeong2025} & $1/60$\\
\bottomrule
\end{tabular}
\endgroup
\end{table}

For a $3$-design, the canonical reconstruction is $\hrho_i=(d+1)\phi_i-\bbone$~\cite{HuangKuengPreskill2020,InnocentiEtAl2023}. For pure states and tangent observables, its variance satisfies \eref{eq:shadow-var} with $C=2(d+1)/(d+2)$, and \pref{prop:shadow-fi} then gives the Fisher guarantee $(d+2)/[4(d+1)]$ of \rcite{ZhouChen2026}. The canonical reconstruction has the same form for every design of order $t\ge3$, and its variance involves only the first three moments. Consequently, this approach cannot improve on the $3$-design guarantee for designs of higher order.

\subsection{Design-order guarantees}

To exploit higher design moments, we work directly with the normalized Fisher quadratic form in \eref{eq:normalized-fi-form}. For $d\ge2$, define the design-order coefficients by
\begin{equation}
\begin{aligned}
c_{d,2}&:=\frac1{d+1+\sqrt{d+1}\lsp},\\
c_{d,2k-1}&:=\frac12\left[1-\frac d{k(d+k-1)}\right],
\end{aligned}
\label{eq:des-coef}
\end{equation}
where $k\ge2$. At even orders $t=2k\ge4$, set
\begin{equation}
c_{d,2k}:=\frac12\left[1-\frac d{k(d+k)}\right]+\frac{d+1}{8k^2(d+k)^2}.
\label{eq:even-coef}
\end{equation}
For fixed $d$, the coefficients $c_{d,t}$ increase strictly with the order $t$: $c_{d,2}<c_{d,3}$ and $c_{d,2k-1}<c_{d,2k}<c_{d,2k+1}$ for $k\ge2$.

\begin{theorem}[Fisher guarantees from projective designs]\label{thm:hier}
In every dimension $d\ge2$, the design Fisher guarantee satisfies
\begin{equation}
\eta_{\mathrm{des}}(d,t)=c_{d,t}
\label{eq:hier}
\end{equation}
at $t=2$ and every odd order $t\ge3$, and this value is attained by a finite $t$-design. At even order $t=2k\ge4$,
\begin{equation}
c_{d,2k}\le\eta_{\mathrm{des}}(d,2k)\le\frac12\left[1-\frac d{(k+1)(d+k-1)}\right].
\label{eq:even-bracket}
\end{equation}
\end{theorem}

At odd order $t=2k-1$, an attaining design $\caE_*$ on $\caH$ admits a unit ket $\ket{\psi_*}$ satisfying
\begin{equation}
G_{\psi_*,\caM_{\caE_*}}=c_{d,2k-1}\IdTan{\psi_*}.
\label{eq:sharp-iso}
\end{equation}

At even order, $c_{d,2k}>c_{d,2k-1}$, so the additional design moment strictly improves the guarantee. The optimal even-order guarantees remain undetermined, but the gap between the bounds in \eref{eq:even-bracket} is less than $1/[2k(k+1)]$ uniformly in $d$ and is $\caO\left(k^{-4}\right)$ at fixed $d$. The $t=3$ lower bound was established in \rcite{ZhouChen2026}; our finite attaining designs show that it is sharp. A corresponding bound for multiplicative-error approximate unitary $3$-design measurements was obtained in \rcite{DuEtAl2026}.

For odd $t=2k-1$, the deficit from the benchmark is
\begin{equation}
\frac12-c_{d,2k-1}=
\frac{2d}{(t+1)(2d+t-1)}.
\label{eq:ord-gap}
\end{equation}
It is of order $t^{-2}$ at fixed dimension, and of order $t^{-1}$ when $d$ is at least a fixed positive multiple of $t$. At fixed $d$, both $c_{d,2k-1}$ and $c_{d,2k}$ tend to $1/2$ as $k\to\infty$. At fixed $k\ge2$, both tend to $(k-1)/(2k)$ as $d\to\infty$, while $dc_{d,2}\to1$. In particular, the $3$-design guarantee tends to $1/4$ as $d\to\infty$. Figure~\ref{fig:des-hier} compares the sharp odd-order guarantees with the even-order lower bounds.

For a design $\caE=\{(\pi_i,\phi_i)\}_i$ on $\caH$ and a pure state $\rho$ on $\caH$, let $\alpha_0(\rho):=\sum_{i\colon p_i(\rho)=0}\pi_i$ denote the total weight of projectors orthogonal to $\rho$.

\begin{corollary}[Sharp zero-probability weight]\label{cor:zero-wt}
Let $\caE$ be a $t$-design on $\caH$, with $t\ge2$ and $k=\lceil t/2\rceil$. For every pure state $\rho$ on $\caH$,
\begin{equation}
\alpha_0(\rho)\le\frac{d-1}{d+k-1}\begin{cases}
\dfrac1k,&t=2k-1,\\[5pt]
\dfrac1{k+1},&t=2k.
\end{cases}
\label{eq:zero-wt}
\end{equation}
For every $t\ge2$ and $d\ge2$, equality holds for some finite $t$-design and pure state.
\end{corollary}

The corollary links zero-probability outcomes to the Fisher trace. By \eref{eq:trace-defect}, $\tr_{\bbR}G_{\psi,\caM_\caE}=d-1-d\alpha_0(\rho)$, so a large zero-probability weight forces a small Fisher trace. At odd order, the bound in \eref{eq:zero-wt} is equivalent to the trace inequality implied by \thref{thm:hier}, with equality if and only if $G_{\psi,\caM_\caE}=c_{d,2k-1}\IdTan\psi$. At even order, a design attaining \eref{eq:zero-wt} yields the upper bound in \eref{eq:even-bracket}.

\subsection{Structured measurements: SIC, stabilizer, and Platonic}
\label{sec:examples}

\begin{table}[!tb]
\caption{Fisher floors of the equal-weight Platonic qubit POVMs. The vertices define the outcomes, $t$ is the design strength, and the design guarantee follows from \thref{thm:hier}; $c_{2,2}=1/(3+\sqrt{3}\lsp)$. All five exact floors are stated in \pref{prop:dod-fi}.}
\label{tab:plat-fi}
\centering
\begingroup
\small
\setlength{\tabcolsep}{3pt}
\renewcommand{\arraystretch}{1.15}
\begin{tabular}{@{}lcccc@{}}
\toprule
Solid & $M$ & $t$ & Design guarantee & Fisher floor $\eta$\\
\midrule
Tetrahedron & $4$ & $2$ & $c_{2,2}$ & $c_{2,2}$\\
Octahedron & $6$ & $3$ & $1/3$ & $1/3$\\
Cube & $8$ & $3$ & $1/3$ & $1/3$\\
Icosahedron & $12$ & $5$ & $5/12$ & $5/12$\\
Dodecahedron & $20$ & $5$ & $5/12$ & $(15-\sqrt{5}\lsp)/30$\\
\bottomrule
\end{tabular}
\endgroup
\end{table}

\subsubsection{SIC and stabilizer measurements}

Let $\caM_{\SIC}$ denote the POVM associated with a SIC ensemble, defined in \sref{sec:setting}.

\begin{proposition}[Fisher floor of SIC measurements]\label{prop:sic-fi}
If a SIC exists in dimension $d\ge2$, then every SIC POVM satisfies
\begin{equation}
\eta(\caM_{\SIC})=c_{d,2}.
\label{eq:sic-fi-bnds}
\end{equation}
\end{proposition}

Similarly, let $\caM_{\rmstab(2,n)}=\{d\phi/M_n\colon \phi\in\Stab_n\}$ be the POVM associated with the equal-weight ensemble of all $n$-qubit stabilizer states, where $\Stab_n$ is the set of distinct $n$-qubit stabilizer projectors and $M_n=|\Stab_n|$. For an odd prime $q$, let $\caM_{\rmstab(q,n)}$ denote the analogous $n$-qudit stabilizer POVM in dimension $d=q^n$.

\Needspace{14\baselineskip}
\begin{proposition}[Fisher floor of stabilizer measurements]\label{prop:stab-fi}
Let $n\ge1$. For qubits, $d=2^n$, the complete stabilizer POVM satisfies
\begin{equation}
\eta(\caM_{\rmstab(2,n)})=\frac13,
\label{eq:stab-fi-bnd}
\end{equation}
with normalized Fisher information $1/3$ in every tangent direction at every stabilizer state. For an odd prime $q$ and $d=q^n$, the complete qudit stabilizer POVM satisfies
\begin{equation}
\eta(\caM_{\rmstab(q,n)})\ge\frac1{4(q-1)}.
\label{eq:stab-qd-bnd}
\end{equation}
\end{proposition}

For $d>2$, the qubit Fisher floor $1/3$ strictly exceeds the $3$-design guarantee $c_{d,3}=(d+2)/[4(d+1)]$; the proof exploits the Pauli structure of the full stabilizer ensemble. The qudit bound follows from \pref{prop:shadow-fi} and the reconstruction variance~\cite{MaoYiZhu2025,ZhuMaoYi2024}; it is independent of $n$ at fixed $q$.

\subsubsection{Platonic-solid qubit measurements}

\begin{figure*}[!t]
\centering
\raisebox{-0.00000pt}[137.72934pt][0.00000pt]{\includegraphics[width=499.29845pt,height=137.72934pt]{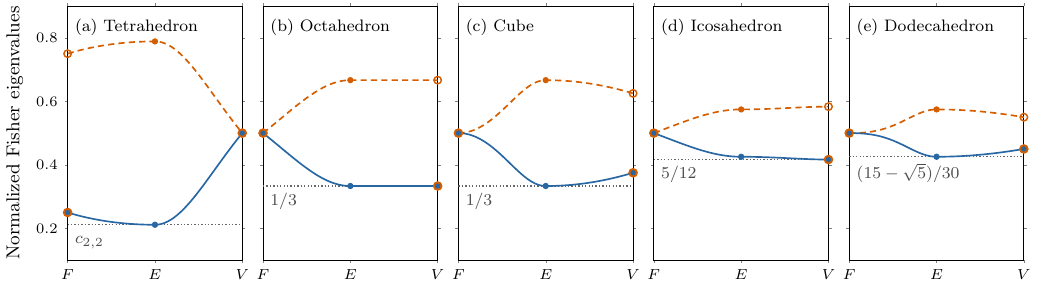}}
\caption{Normalized Fisher eigenvalues for the five Platonic qubit POVMs along a symmetry path. The path follows the shorter great-circle arcs from a face-center direction ($F$) to the midpoint of an incident edge ($E$) and then to a vertex of that edge ($V$); each arc occupies half the horizontal axis. In each panel, the blue solid curve shows the least eigenvalue, and the orange dashed curve shows the largest. Filled markers show the eigenvalues at $F$, $E$, and $V$, with concentric blue and orange markers indicating coincident eigenvalues. At singular endpoints, both eigenvalues coincide at the lower marker, and open orange circles show the limit of the largest eigenvalue from nearby regular states. The horizontal gray dotted line marks the Fisher floor listed in Table~\ref{tab:plat-fi}.}
\label{fig:plat-qb}
\end{figure*}

The five Platonic solids give natural qubit examples~\cite{DeckerEtAl2003,DeBurghEtAl2008}. For a solid with $M$ vertices, let $\bmv_i$ be the unit Bloch vectors pointing to its vertices. The equal-weight POVM has elements $E_i=(\bbone+\bmv_i\cdot\bmsigma)/M$, where $\bmsigma=(\sigma_x,\sigma_y,\sigma_z)$ is the Pauli vector. For a pure state $\rho(\bmr)=(\bbone+\bmr\cdot\bmsigma)/2$ with $\norm{\bmr}_2=1$, put $\bmv_{i,\perp}=\bmv_i-(\bmv_i\cdot\bmr)\bmr$. On the Bloch tangent plane $\bmr^\perp$, the normalized Fisher operator is represented by
\begin{equation}
F_{\bmr}=\frac1M
\sum_{i\colon 1+\bmv_i\cdot\bmr>0}
\frac{\bmv_{i,\perp}\bmv_{i,\perp}^{\sfT}}
{1+\bmv_i\cdot\bmr},
\label{eq:qb-fi-mat}
\end{equation}
where zero-probability terms are omitted. Its two eigenvalues coincide with those of $G_{\psi,\caM}$ for $\rho(\bmr)=\proj\psi$, so
\begin{equation}
\eta(\caM)=\inf_{\norm{\bmr}_2=1}
\lambda_{\min}(F_{\bmr}).
\label{eq:qb-eta-bloch}
\end{equation}

Figure~\ref{fig:plat-qb} plots both eigenvalues along the path $F\to E\to V$. Along this path, the least eigenvalue is continuous, whereas the largest can drop at singular states.

For the tetrahedron, octahedron, cube, and icosahedron, the Fisher floor equals the design guarantee of \thref{thm:hier}, because some state attains it. The tetrahedron is a SIC, so \pref{prop:sic-fi} gives $c_{2,2}$; analytic qubit SIC Fisher spectra along special Bloch directions are given in \rcite{Zhu2023Projective}. The octahedral POVM corresponds to measuring a uniformly chosen Pauli observable. Its projectors form a $3$-design, and both normalized Fisher eigenvalues at a Pauli eigenstate equal $c_{2,3}=1/3$, so this six-outcome POVM attains the $3$-design guarantee. The cube is also a $3$-design, and its least normalized Fisher eigenvalue $1/3$ occurs at an edge-midpoint direction. The icosahedron and dodecahedron are $5$-designs~\cite{DeBurghEtAl2008}. At an icosahedral vertex, both normalized Fisher eigenvalues equal $c_{2,5}=5/12$. The dodecahedron behaves differently.

\begin{proposition}[Fisher floors of Platonic qubit measurements]\label{prop:dod-fi}
The equal-weight qubit POVMs defined by the vertices of the five Platonic solids have the Fisher floors listed in Table~\ref{tab:plat-fi}. In particular, the dodecahedral POVM $\caM_{\rmdod}$ has Fisher floor
\begin{equation}
\eta(\caM_{\rmdod})=\frac{15-\sqrt{5}\lsp}{30},
\label{eq:dod-floor}
\end{equation}
attained precisely at Bloch directions toward edge midpoints of the dodecahedron.
\end{proposition}
Thus the dodecahedral Fisher floor strictly exceeds the $5$-design guarantee $5/12$. Proposition~\ref{prop:dod-fi} is proved in \aref{app:dod-fi}.

\begin{figure}[!t]
\centering
\raisebox{-0.00000pt}[170.38736pt][0.00000pt]{\includegraphics[width=240.70622pt,height=170.38736pt]{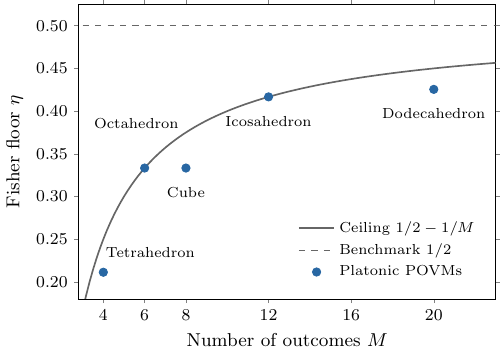}}
\caption{Fisher floor versus outcome count for the Platonic qubit POVMs in Table~\ref{tab:plat-fi}. All five values are exact. The universal ceiling follows from \pref{prop:res} at $d=2$ and is attained by the octahedron and icosahedron.}
\label{fig:qb-res}
\end{figure}

\subsection{Proof overview}
\label{sec:hier-strat}

This subsection outlines the proofs of \thref{thm:hier} at second and odd orders and of \pref{prop:stab-fi}. A polynomial method extends the known $3$-design bound~\cite{ZhouChen2026} to all odd orders, finite attaining designs establish sharpness, and an induction over qubits gives the stabilizer floor. The complete proofs are given in Appendices~\ref{app:hier} and~\ref{app:structured-fi}.

\subsubsection{Polynomial moment bounds}

Let $\caE=\{(\pi_i,\phi_i)\}_i$ be a $t$-design on $\caH$ with associated POVM $\caM_\caE$. Fix orthonormal kets $\ket\psi,\ket\nu$, put $\rho=\proj\psi$ and $\Delta=\Delta_{\psi,\nu}$ as in \eref{eq:tan}, and write $G=G_{\psi,\caM_\caE}$ for the normalized Fisher operator in \eref{eq:normalized-fi-form}. For each design projector $\phi_i$, let $u_i=\tr(\rho\phi_i)$ and $x_i=\tr(\Delta\phi_i)$. Since $E_i=d\pi_i\phi_i$, \eref{eq:score-fi} gives
\begin{equation}
\inner{\nu}{G\nu}_{\bbR}
=\frac d4\sum_{i\colon u_i>0}\pi_i\frac{x_i^2}{u_i}.
\label{eq:m-ov-fi}
\end{equation}
At order two, the denominator $u_i$ prevents direct use of the design identity. A pointwise polynomial certificate bounds the Fisher contribution using quantities whose averages are fixed by the second moment. Averaging then gives $c_{d,2}$; the certificate and its proof are in \lref{lem:2des-cert}.

For higher orders, use the probability measure
\begin{equation}
\mu_d(\rmd u)=d(1-u)^{d-1}\rmd u
\label{eq:m-rad-mu}
\end{equation}
on $[0,1]$, which arises from the Haar average of $x^2h(u)$ for polynomials $h$; see \eref{eq:design-radial-moment}. We separate $1/u$ into a polynomial part controlled by design moments and a nonnegative remainder. Choose a real polynomial $r$ with $r(0)=1$, $r(u)\ge0$ on $[0,1]$, and $\deg r\le t-1$. Then $h(u)=[1-r(u)]/u$ is a polynomial of degree at most $t-2$, so $x^2h(u)$ has degree at most $t$ in $\phi$. Substituting $1/u=h(u)+r(u)/u$ and replacing the average of $x^2h(u)$ with its Haar average gives
\begin{align}
\inner{\nu}{G\nu}_{\bbR}&=\frac12[1-\bbE_{\mu_d}r]+R_r,
\label{eq:m-minorant}\\
R_r&:=\frac d4\sum_{i\colon u_i>0}\pi_i\frac{x_i^2r(u_i)}{u_i}\ge0.
\label{eq:m-minorant-rem}
\end{align}
At odd order $2k-1$, let $K_{k-1}$ be the reproducing kernel for polynomials of degree at most $k-1$ in $L^2(\mu_d)$. Choosing $r=q_{k-1}^2$, where
\begin{equation}
q_{k-1}(u)=\frac{K_{k-1}(u,0)}{K_{k-1}(0,0)},\quad
\bbE_{\mu_d}q_{k-1}^2=\frac d{k(d+k-1)},
\label{eq:m-poly-opt}
\end{equation}
gives $c_{d,2k-1}$ after dropping $R_r$. By \lref{lem:end}, $q_{k-1}$ minimizes $\bbE_{\mu_d}q^2$ among polynomials of degree at most $k-1$ with $q(0)=1$, thereby optimizing the lower bound for $r=q^2$.

\subsubsection{Finite attaining designs}

At odd order, the roots of $q_{k-1}$ and the endpoints $0,1$ support a positive radial quadrature. At each node $u$, combine an angular design ket $\ket\xi\in\caH_\psi$ with finitely many relative phases $\theta$:
\begin{equation}
\ket\phi=\sqrt{u}\lsp\ket\psi+\sqrt{1-u}\lsp\rme^{\rmi\theta}\ket\xi.
\label{eq:m-lift-ket}
\end{equation}
Phase averaging removes unequal powers on the ket and bra sides; the angular design and radial quadrature reproduce the surviving tensor moments. The quadrature weight $\alpha_0=(d-1)/[k(d+k-1)]$ at $u=0$ and the trace identity \eref{eq:trace-defect} give
\begin{equation*}
\tr_{\bbR}G=d-1-d\alpha_0=(2d-2)c_{d,2k-1}.
\end{equation*}
Together with the lower bound $G\succeq c_{d,2k-1}\IdTan\psi$, this forces $G=c_{d,2k-1}\IdTan\psi$. At order two, a separate finite construction attains $c_{d,2}$.

\subsubsection{Stabilizer Fisher floor: induction}

By \pref{prop:reg-eq}, it suffices to prove the lower bound at regular states. The complete multiqubit stabilizer measurement can be realized in two stages: first measure a uniformly chosen nonidentity Pauli operator $P$, then apply the complete $(n-1)$-qubit stabilizer measurement in the observed eigenspace of $P$. Merging outcomes with the same projector recovers $\caM_{\rmstab(2,n)}$, so this realization has the same Fisher information.

For fixed $P$, the first stage yields the sign $\pm1$ with probability $p_\pm$. Let $\ket{\psi_\pm}$ be the normalized post-measurement states, and let $\ket{\nu_\pm}$ be the corresponding horizontal tangent vectors induced by $\ket\nu$. Write $\mathcal F_P$ for one quarter of the classical Fisher information of the two-stage measurement with $P$ fixed, along $\Delta_{\psi,\nu}$. The chain rule for Fisher information gives
\begin{equation}
\mathcal F_P=\frac14\sum_{\pm}\frac{\dot p_\pm^{\,2}}{p_\pm}
+\sum_{\pm}p_\pm\mathcal F^{(n-1)}_{\psi_\pm}(\nu_\pm),
\label{eq:m-stab-chain}
\end{equation}
where $\dot p_\pm$ is the derivative of $p_\pm$ along $\Delta_{\psi,\nu}$, and $\mathcal F^{(n-1)}$ is the normalized Fisher quadratic form of the $(n-1)$-qubit measurement. The induction hypothesis bounds the second sum below by $\sum_\pm p_\pm\norm{\nu_\pm}_2^2/3$. Averaging over $P$ and using bounds on Pauli expectation values gives the floor $1/3$. At stabilizer states, the zero-probability weight and the trace identity force $G_{\psi,\caM_{\rmstab(2,n)}}=\IdTan\psi/3$.

\section{Measurements from sampled projective designs}
\label{sec:samp}

Projective $t$-designs have at least $d^2$ elements for $t\ge2$ and, at fixed $t$, at least of order $d^t$~\cite{RoyScott2007}; the multiqubit stabilizer $3$-design even has superpolynomially many elements in $d$. A full design measurement can therefore be costly. In this section, we ask how many outcomes a dimension-independent Fisher floor requires and show that $\caO(d)$ outcomes suffice. We first derive a universal outcome constraint. We then show that sampling $\caO(d)$ projectors from any $3$-design, followed by spectral completion, gives such a floor with probability at least $1-1/(2d)-4^{-d}$. Sampling $\caO(d)$ complete bases whose vectors form a $3$-design also gives such a floor, using $\caO\left(d^2\right)$ labeled outcomes without completion. Proofs are given in \aref{app:samp}.

\Needspace{6\baselineskip}
\begin{proposition}[Outcome constraint for Fisher floors]\label{prop:res}
Every finite rank-one POVM $\caM$ with $M$ outcomes satisfies
\begin{equation}
\eta(\caM)\le\frac12\left(1-\frac dM\right).
\label{eq:out-ub}
\end{equation}
\end{proposition}

Equation~\eqref{eq:out-ub} shows that a dimension-independent Fisher floor requires linearly many outcomes. More precisely, a Fisher floor of at least $1/2-\varepsilon$, with $0<\varepsilon<1/2$, requires $M\ge d/(2\varepsilon)$. The octahedral and icosahedral qubit POVMs attain \eref{eq:out-ub} at $M=6$ and $M=12$, respectively, and hence maximize $\eta$ among rank-one qubit POVMs with these outcome counts; see Fig.~\ref{fig:qb-res}. By \pref{prop:pure-ic}, a positive Fisher floor also entails the known lower bound $M\ge4d-\caO(\log d)$ for pure-state identifiability~\cite{HeinosaariEtAl2013}. Conversely, quantum measurements with linearly many outcomes can identify almost all pure states~\cite{FlammiaEtAl2005} or every pure state~\cite{Finkelstein2004,HeinosaariEtAl2013}; for example, generic phase-retrieval frames of $4d-4$ vectors identify all pure states~\cite{ConcaEtAl2015}. Identifiability alone, however, does not provide a Fisher floor bounded away from zero independently of $d$.

\subsection{Projector sampling with spectral completion}

The shadow criterion in \sref{sec:shadows} requires unbiased reconstruction of arbitrary density operators and hence at least $d^2$ outcomes. For pure states, far fewer outcomes suffice: we sample $\caO(d)$ elements of a $3$-design and add at most $d$ residual elements.

Draw projectors $\phi_1,\ldots,\phi_m$ independently according to the ensemble weights, allowing repetitions, and put
\begin{equation}
S_m=\frac dm\sum_{i=1}^m\phi_i,\qquad
E_i=\frac{d}{2m}\phi_i.
\label{eq:samp-povm}
\end{equation}
The sampled elements sum to $S_m/2$. We call a realization \emph{feasible} if $S_m\preceq2\bbone$. Then the residual $\bbone-S_m/2$ is positive semidefinite and has the spectral decomposition
\begin{equation}
\bbone-\frac{S_m}{2}
=\sum_{j=1}^{r_{\rmres}}\lambda_j\proj{e_j},
\label{eq:spec-comp}
\end{equation}
where $r_{\rmres}\le d$, $\lambda_j>0$, and the $\ket{e_j}$ are orthonormal eigenkets. Adding the rank-one residual elements $\lambda_j\proj{e_j}$ defines the completed POVM $\caM_m$, with at most $m+d$ labeled outcomes. The completed POVM is fixed and applied independently to each copy.

Let $\Xi_d$ denote the Hermitian matrices with spectrum $(1,-1,0,\ldots,0)$. For $\beta>0$, define the absolute signal and the construction success event by
\begin{gather}
F_m(\Delta):=\frac dm\sum_{i=1}^m|\tr(\phi_i\Delta)|,
\label{eq:samp-frm}\\
\Omega_m(\beta):=\left\{S_m\preceq2\bbone,\
\inf_{\Delta\in\Xi_d}F_m(\Delta)\ge\beta\right\}.
\label{eq:samp-event}
\end{gather}
The first condition ensures feasible completion; the second states that $F_m(\Delta)\ge\beta$ for every $\Delta\in\Xi_d$.

\Needspace{9\baselineskip}

\begin{theorem}[Linear sampling with a dimension-independent Fisher floor]\label{thm:samp}
There is a universal constant $C>0$ with the following property. Let $\caE$ be a $3$-design in dimension $d\ge2$, and let $m\ge Cd$. Then the event $\Omega_m:=\Omega_m(1/4)$ in \eref{eq:samp-event} satisfies
\begin{equation}
\bbP(\Omega_m)\ge1-\frac1{2d}-4^{-d}.
\label{eq:samp-prob}
\end{equation}
On this event, the completed POVM $\caM_m$ satisfies
\begin{equation}
I_{\rho,\caM_m}\succeq c_*J_\rho
\label{eq:samp-fi}
\end{equation}
at every pure state, where $c_*=1/256$. For $m=\lceil Cd\rceil$, the total outcome count is $M\le m+d=\caO(d)$.
\end{theorem}

The constant $C$ depends on the universal covariance constant $C_6$ through \eref{eq:samp-const-val}; we do not specify a numerical value. Neither $C$ nor $c_*$ is optimized. The proof combines a uniform rank-two signal bound with sample-covariance control~\cite{Tikhomirov2018}, using the projective moments through order three. These tools also appear in design-based matrix recovery~\cite{KuengEtAl2017,Gilles2026,HuangLi2026Recovery}; here they yield a dimension-independent Fisher floor for the completed POVM. In \sref{sec:ph-ret}, we show that the same event also yields stable phase retrieval.

\begin{figure}[!t]
\centering
\raisebox{-0.00000pt}[167.83368pt][0.00000pt]{\includegraphics[width=244.92094pt,height=167.83368pt]{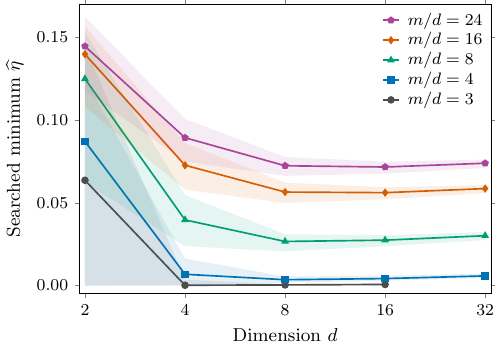}}
\caption{Numerical upper estimates of the Fisher floor after sampling and spectral completion of the stabilizer $3$-design. Each parameter point is based on $1000$ measurement realizations. Curves and bands show the median and the $10$th--$90$th percentiles over realizations whose completion succeeds; at $d=32$ and $m/d=3$, no completion succeeded. Completion fractions and search details are given in \aref{app:numerics}.}
\label{fig:samp-fi}
\end{figure}

Figure~\ref{fig:samp-fi} shows numerical upper estimates of the Fisher floor for POVMs sampled from the stabilizer $3$-design. They illustrate typical behavior but do not certify a floor, and their dependence on the search budget is discussed in \aref{app:numerics}. The parent stabilizer POVM has Fisher floor $1/3$ and superpolynomially many outcomes in $d$. Sampling and completion reduce the outcome count to $\caO(d)$ while retaining a positive Fisher floor on $\Omega_m$.

\subsection{Orthonormal-basis sampling}

Fisher-information concentration has previously been used to analyze low-rank tomography with random-basis measurements~\cite{AcharyaKypraiosGuta2016,AcharyaGuta2017}. We sample complete orthonormal bases from a finite mixture whose associated projector ensemble is a projective $3$-design. With high probability, $\caO(d)$ sampled bases give a dimension-independent Fisher floor.

\begin{proposition}[Sampling orthonormal bases whose vectors form a $3$-design]\label{prop:basis-samp}
Consider a finite mixture of orthonormal-basis measurements whose associated projector ensemble is a projective $3$-design. Let $\caM_L^{\rmbas}$ be the equal mixture of $L$ bases drawn independently according to the mixture weights. For $L\ge2700d$,
\begin{equation}
\bbP\left\{\eta\left(\caM_L^{\rmbas}\right)\ge\frac1{64}\right\}\ge1-\rme^{-d}.
\label{eq:basis-fisher-guarantee}
\end{equation}
\end{proposition}

Stabilizer bases provide a natural instance. Each $n$-qubit stabilizer basis can be measured by a Clifford circuit followed by computational-basis readout, and the uniform mixture of all stabilizer bases has the stabilizer $3$-design~\cite{KuengGross2015,Webb2016,Zhu2017} as its associated projector ensemble. Hence \thref{thm:samp} and \pref{prop:basis-samp} both apply. How many stabilizer bases are needed? The following proposition gives a lower bound.

\Needspace{7\baselineskip}
\begin{proposition}[A lower bound on the number of stabilizer bases]\label{prop:stab-basis-obstruction}
Let $d=2^n$, with $n\ge1$. Any $L$ stabilizer bases with $1\le L\le n$ fail to distinguish some pair of orthogonal pure states. Every mixture of these bases therefore has $\eta=0$.
\end{proposition}

Thus at least $n+1=\log_2d+1$ stabilizer bases are necessary even for pure-state identifiability, whereas \pref{prop:basis-samp} gives a dimension-independent Fisher floor with high probability using $\caO(d)$ bases. The optimal number of bases for such a floor remains open.

\section{Statistical guarantees from Fisher floors}
\label{sec:est}

\begin{figure*}[!t]
\centering
\includegraphics[width=494pt]{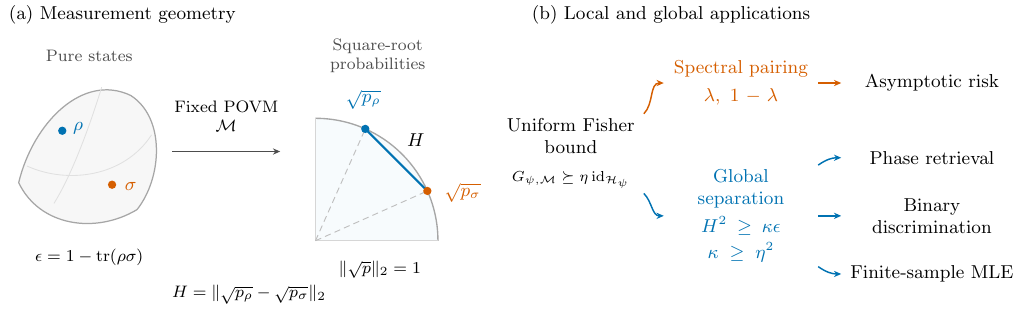}
\caption{Local and global implications of a positive Fisher floor for a fixed rank-one POVM on pure states. (a)~Square-root Born probabilities lie in the nonnegative orthant of the unit sphere; the Hellinger distance $H$ is their Euclidean chord length. (b)~Fisher spectra and pairing control the asymptotic risk of efficient estimators at regular states. Global separation, quantified by $\kappa$, yields phase-retrieval stability, binary discrimination bounds, and finite-sample reconstruction guarantees.}
\label{fig:applications-overview}
\end{figure*}

We convert Fisher floors into global separation of pure-state output distributions and stability of weighted amplitude maps. We then apply these bounds to discrimination between given candidate states and to local asymptotic and finite-sample state reconstruction (Fig.~\ref{fig:applications-overview}). Proofs are given in \aref{app:est}.

\subsection{Global Hellinger separation and phase retrieval}
\label{sec:glob-stab}

\subsubsection{Hellinger geometry and Fisher information}
\label{sec:hel-fi}

Fix a finite rank-one POVM $\caM$ and write $p_\rho$ for its output distribution at a pure state $\rho$. For probability vectors $p,q$ on the same outcome set, define the Bhattacharyya coefficient and Hellinger distance by
\begin{gather}
B(p,q):=\sum_i\sqrt{p_iq_i},
\label{eq:bhatt}\\
H^2(p,q):=\sum_i\bigl(\sqrt{p_i}\lsp-\sqrt{q_i}\lsp\bigr)^2=2[1-B(p,q)].
\label{eq:hel-def}
\end{gather}
Thus $0\le B\le1$ and $0\le H^2\le2$. With the squared convention for classical fidelity, put
\begin{equation}
F_{\rmcl}(p,q)=B(p,q)^2,\qquad
\epsilon_{\rmcl}(p,q)=1-F_{\rmcl}(p,q).
\label{eq:cl-inf}
\end{equation}
These quantities satisfy
\begin{equation}
\epsilon_{\rmcl}=H^2\left(1-\frac{H^2}{4}\right),\quad
\frac{H^2}{2}\le\epsilon_{\rmcl}\le H^2.
\label{eq:hel-cl-inf}
\end{equation}
For pure states $\rho=\proj\psi$ and $\sigma=\proj\varphi$, write $\epsilon=\epsilon(\rho,\sigma)=1-\tr(\rho\sigma)$ for the quantum infidelity. Monotonicity of fidelity under measurement~\cite{FuchsVDG1999} gives $B(p_\rho,p_\sigma)\ge|\braket{\psi}{\varphi}|=\sqrt{1-\epsilon}$. Substituting this into $H^2=2(1-B)$ yields
\begin{equation}
H^2(p_\rho,p_\sigma)\le2\left(1-\sqrt{1-\epsilon}\lsp\right)\le2\epsilon.
\label{eq:hel-in-ub}
\end{equation}

The square-root representation of $H$ gives the local Fisher geometry. For a smooth pure-state model $\rho(\theta)$, let $\theta$ be the real parameter vector, $\delta\theta$ a small parameter displacement, and $I_\theta$ the classical Fisher information matrix of $\caM$ in these coordinates. At a regular point with $p_i(\rho(\theta))>0$ for all $i$, the expansion as $\delta\theta\to0$ is~\cite{Wootters1981,BraunsteinCaves1994}
\begin{equation}
H^2(p_{\rho(\theta)},p_{\rho(\theta+\delta\theta)})
=\frac14\delta\theta^{\sfT}I_\theta\delta\theta
+o\left(\norm{\delta\theta}_2^2\right).
\label{eq:loc-hel-fi}
\end{equation}
At states with vanishing Born probabilities, a one-sided Hellinger expansion may contain terms absent from the Fisher information in \eref{eq:score-fi}. The centered pairs in \pref{prop:sym-fi} cancel these contributions and recover the Fisher coefficient in \eref{eq:sym-fi}.

\begin{proposition}[Symmetric amplitudes and Fisher information]\label{prop:sym-fi}
Let $\ket\psi,\ket \nu\in\caH$ be orthonormal. For real $\theta$, define $\ket{\psi_\pm(\theta)}=\cos\theta\ket\psi\pm\sin\theta\ket \nu$ and $\rho_\pm(\theta)=\proj{\psi_\pm(\theta)}$. For every finite rank-one POVM,
\begin{equation}
\lim_{\theta\to0}\frac{H^2(p_{\rho_+(\theta)},p_{\rho_-(\theta)})}
{\epsilon(\rho_+(\theta),\rho_-(\theta))}
=\inner{\nu}{G_{\psi,\caM}\nu}_{\bbR}.
\label{eq:sym-fi}
\end{equation}
\end{proposition}

To quantify how much pure-state distinguishability survives the measurement at finite separation, define the global Hellinger separation coefficient of $\caM$ by
\begin{equation}
\kappa(\caM)=
\inf_{\substack{\rho,\sigma\in\caP(\caH)\\\rho\ne\sigma}}
\frac{H^2(p_\rho,p_\sigma)}{\epsilon(\rho,\sigma)}.
\label{eq:kappa}
\end{equation}

\begin{theorem}[Fisher floors and global statistical stability]\label{thm:fi-glob}
Let $\caM$ be a finite rank-one POVM and write $\eta=\eta(\caM)$ for its Fisher floor. For all pure states $\rho,\sigma$, with $\epsilon=\epsilon(\rho,\sigma)$,
\begin{gather}
\norm{p_\rho-p_\sigma}_1\ge\eta\norm{\rho-\sigma}_1,
\label{eq:fi-glob-l1}\\
\eta^2\epsilon\le H^2(p_\rho,p_\sigma)\le2(1-\eta)\epsilon,
\label{eq:fi-glob-hel}\\
\eta^2\epsilon\le\epsilon_{\rmcl}(p_\rho,p_\sigma)\le\epsilon.
\label{eq:fi-glob-cl-inf}
\end{gather}
Consequently,
\begin{equation}
\eta^2\le\kappa(\caM)\le\eta.
\label{eq:fi-glob-sandwich}
\end{equation}
\end{theorem}

A Fisher floor controls infinitesimal distinguishability; \thref{thm:fi-glob} turns this local control into a quantitative separation of every pair of pure states. The key step is to regard the normalized difference of two pure states as a common tangent direction at every state on a suitable phase circle: averaging the Fisher information over that circle bounds the distance between the two outcome distributions. Thus a dimension-independent Fisher floor also gives dimension-independent global stability, which underlies the reconstruction guarantees below.

For $3$-designs, \thref{thm:fi-glob} and $\eta\ge c_{d,3}$ give the Hellinger coefficient $c_{d,3}^2$. The third design moment yields the stronger coefficient $c_{d,3}$, as stated next. These coefficients approach $1/16$ and $1/4$, respectively, as $d$ increases.

\begin{proposition}[Sharp global separation for $3$-designs]\label{prop:glob-3}
For every $\caE\in\fkE_3(d)$ and all pure states $\rho,\sigma$,
\begin{equation}
H^2(p_\rho,p_\sigma)\ge c_{d,3}\epsilon(\rho,\sigma).
\label{eq:3-des-glob-sep}
\end{equation}
Moreover,
\begin{equation}
\min_{\caE\in\fkE_3(d)}\kappa(\caM_\caE)=c_{d,3}.
\label{eq:sharp-glob-3}
\end{equation}
The finite extremal design in \thref{thm:hier} at $k=2$ attains $\kappa=\eta=c_{d,3}$.
\end{proposition}

The Hellinger lower bound extends to higher odd design orders.

\begin{proposition}[Odd-order Hellinger guarantee]\label{prop:hel}
Let $d\ge2$ and $k\ge2$. For every $\caE\in\fkE_t(d)$ with $t\ge2k-1$ and all pure states $\rho,\sigma$,
\begin{equation}
H^2(p_\rho,p_\sigma)\ge b_{d,k}\epsilon(\rho,\sigma),
\label{eq:des-hel}
\end{equation}
where
\begin{equation}
b_{d,k}=\frac13\left[1-
\frac{(d-2)_{k-1}}{\binom{k+2}{3}(d+1)_{k-1}}\right]
\label{eq:hel-hier}
\end{equation}
and $(g)_j=g(g+1)\cdots(g+j-1)$, with $(g)_0=1$, denotes the rising factorial. In particular, $b_{d,2}=c_{d,3}=(d+2)/[4(d+1)]$.
\end{proposition}

The proof is similar to that of the odd-order Fisher bound in \thref{thm:hier}. The coefficients satisfy
\begin{equation}
\lim_{k\to\infty}b_{d,k}=\frac13,\qquad
\lim_{d\to\infty}b_{d,k}=\frac13\left[1-\binom{k+2}{3}^{-1}\right].
\label{eq:hel-lim}
\end{equation}
For $k>2$, the optimality of this bound remains open.

\begin{figure*}[!t]
\centering
\raisebox{-0.00000pt}[154.04399pt][0.00000pt]{\includegraphics[width=500.36331pt,height=154.04399pt]{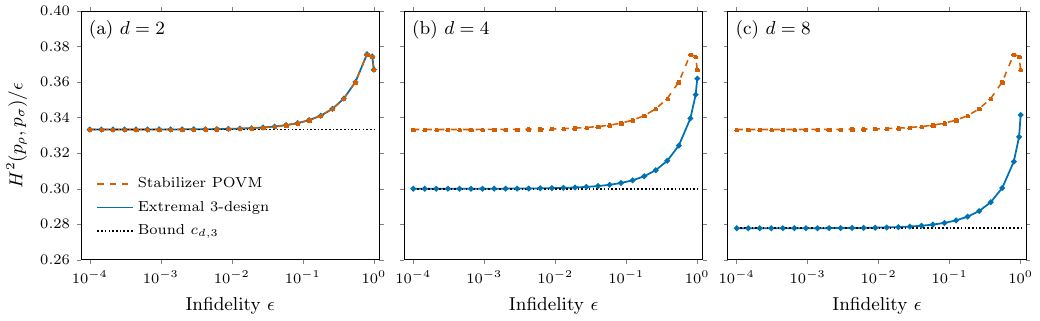}}
\caption{Global separation for fixed $3$-design measurements. Panels (a)--(c) correspond to $d=2,4,8$, respectively. In each panel, the blue solid curve shows the smallest $H^2(p_\rho,p_\sigma)/\epsilon(\rho,\sigma)$ found for the extremal $3$-design in \thref{thm:hier}, and the orange dashed curve shows the corresponding result for the complete stabilizer POVM. The black dotted line marks the rigorous bound $c_{d,3}$. The extremal POVMs have $6$, $156$, and $4696$ outcomes, respectively, and satisfy $\eta=\kappa=c_{d,3}$. The searched minima are numerical upper estimates of the true infima. At $d=2$, the two measurements coincide. Construction and search details are given in \aref{app:numerics}.}
\label{fig:glob-sep}
\end{figure*}

Figure~\ref{fig:glob-sep} compares the extremal $3$-design with the complete stabilizer measurement. The searched separation ratios approach $c_{d,3}$ for the extremal design, illustrating the sharpness of \eref{eq:sharp-glob-3}. For the complete stabilizer measurement, \pref{prop:stab-fi} gives $\eta=1/3$, and \eref{eq:fi-glob-sandwich} implies $\kappa\le1/3$. The searched ratios at $d=4,8$ remain above $1/3$ and approach it at small separations, suggesting $\kappa=1/3$. A matching uniform Hellinger lower bound for general multiqubit dimensions has not been established.

\begin{proposition}[Global separation for the sampled POVM]\label{prop:samp-sep}
On the event $\Omega_m=\Omega_m(1/4)$ defined in \eref{eq:samp-event}, the completed POVM of \thref{thm:samp} satisfies
\begin{equation}
\kappa(\caM_m)\ge c_*.
\label{eq:samp-kappa}
\end{equation}
For the complete-basis measurement in \pref{prop:basis-samp}, $L\ge2700d$ also gives $\kappa\left(\caM_L^{\rmbas}\right)\ge1/64$ with probability at least $1-\rme^{-d}$.
\end{proposition}

For distinct pure states, $(\rho-\sigma)/\sqrt{\epsilon(\rho,\sigma)}\lsp\in\Xi_d$, so the bound on $F_m$ improves the generic coefficient $c_*^2$ obtained from \thref{thm:fi-glob} to $c_*$.

\subsubsection{Application to phase retrieval}
\label{sec:ph-ret}

Phase retrieval reconstructs complex signals, modulo a global phase, from the magnitudes of their measurement coefficients~\cite{BalanCasazzaEdidin2006,BandeiraEtAl2014}. For pure states $\rho=\proj\psi$ and $\sigma=\proj\varphi$, define the weighted amplitude map on the corresponding unit vectors by
\begin{gather}
\caA_{\caM}(\ket\psi):=\bigl(\sqrt{w_i}\lsp|\braket{\phi_i}{\psi}|\bigr)_{i=1}^M,
\label{eq:ph-ret-amp-map}\\
\norm{\caA_{\caM}(\ket\psi)-\caA_{\caM}(\ket\varphi)}_2
=H(p_\rho,p_\sigma).
\label{eq:hel-amp-id}
\end{gather}
The Euclidean distance modulo global phase is
\begin{equation}
\begin{aligned}
\Delta_{\rmph}(\rho,\sigma)
&:=\min_{\theta\in[0,2\pi]}
\norm*{\ket\psi-\rme^{\rmi\theta}\ket\varphi}_2\\
&\,=\sqrt{2\bigl(1-|\braket{\psi}{\varphi}|\bigr)}.
\end{aligned}
\label{eq:ph-quot-dist}
\end{equation}
Since $\Delta_{\rmph}^2/2\le\epsilon\le\Delta_{\rmph}^2$, the definition of $\kappa$ and POVM completeness give the two-sided bound~\cite{BalanZou2016,XiaXuXu2025}
\begin{equation}
\sqrt{\frac{\kappa(\caM)}2}\lsp\Delta_{\rmph}(\rho,\sigma)
\le H(p_\rho,p_\sigma)\le\Delta_{\rmph}(\rho,\sigma).
\label{eq:ph-ret-stab}
\end{equation}
For $\eta(\caM)>0$, \thref{thm:fi-glob} gives the lower Lipschitz constant $\eta(\caM)/\sqrt{2}$. Propositions~\ref{prop:glob-3} and~\ref{prop:samp-sep} give $\sqrt{c_{d,3}/2}\lsp$ and $\sqrt{c_*/2}$, respectively, both bounded away from zero independently of $d$. The sampled construction uses $\caO(d)$ outcomes at $m=\lceil Cd\rceil$.

\subsection{Pure-state discrimination with a fixed measurement}
\label{sec:disc}

Consider two given candidate pure states $\rho$ and $\sigma$. A fixed measurement can be reused across candidate pairs, with the classical decision rule using their known outcome distributions. Fixed-measurement discrimination has been studied through POVM-induced distinguishability norms~\cite{MatthewsEtAl2009,KuengZhuGross2016}. Our bound in \eref{eq:fi-glob-l1} expresses it in terms of the Fisher floor. For equal priors, let $P_{\rmsucc}^{\caM}$ be the maximal success probability using $\caM$ and classical postprocessing, and $P_{\rmsucc}^{\rmopt}$ the optimum over single-copy measurements. The classical decision rule and Helstrom formula~\cite{Helstrom1976} give $P_{\rmsucc}^{\caM}=1/2+\norm{p_\rho-p_\sigma}_1/4$ and $P_{\rmsucc}^{\rmopt}=1/2+\norm{\rho-\sigma}_1/4$. Hence
\begin{equation}
P_{\rmsucc}^{\caM}-\frac12
\ge\eta(\caM)\left(P_{\rmsucc}^{\rmopt}-\frac12\right).
\label{eq:qsd-bias}
\end{equation}
The fixed measurement retains at least a fraction $\eta(\caM)$ of the optimal discrimination advantage for every pure-state pair. For the complete stabilizer POVM, \pref{prop:stab-fi} together with \eref{eq:fi-glob-l1} gives the dimension-independent fraction $1/3$ in \eref{eq:qsd-bias}, improving the coefficient $1/6$ in Corollary~1 of \rcite{KuengZhuGross2016}. As a continuous benchmark, the Haar POVM retains at least one half of the optimal discrimination advantage for every pure-state pair~\cite{KuengZhuGross2016}.

Hellinger separation also controls the error after repeated measurements. Apply $\caM$ independently to $N$ copies and use the optimal classical test between $p_\rho^{\otimes N}$ and $p_\sigma^{\otimes N}$. Independence makes the sum over outcome strings factor into $N$ identical overlaps~\cite{FuchsVDG1999}:
\begin{equation}
B\left(p_\rho^{\otimes N},p_\sigma^{\otimes N}\right)=B(p_\rho,p_\sigma)^N.
\label{eq:ov-tensor}
\end{equation}
For equal priors, the optimal classical test has error probability
\begin{equation}
P_{\rmerr}^{\caM,N}(\rho,\sigma)
=\frac12-\frac14\norm*{p_\rho^{\otimes N}-p_\sigma^{\otimes N}}_1,
\label{eq:rep-disc-ex}
\end{equation}
which is one half of the sum, over records, of the smaller of the two record probabilities. Bounding each minimum by the geometric mean of the two probabilities and using \eref{eq:ov-tensor} gives $P_{\rmerr}^{\caM,N}\le B(p_\rho,p_\sigma)^N/2$. Finally,
\begin{equation}
B=1-\frac{H^2}{2}
\le1-\frac{\kappa(\caM)\epsilon}{2}
\le\rme^{-\kappa(\caM)\epsilon/2},
\label{eq:disc-affinity}
\end{equation}
so
\begin{equation}
P_{\rmerr}^{\caM,N}(\rho,\sigma)
\le\frac12 B(p_\rho,p_\sigma)^N
\le\frac12\rme^{-N\kappa(\caM)\epsilon(\rho,\sigma)/2}.
\label{eq:rep-disc}
\end{equation}
For $\kappa(\caM)>0$ and target error $0<\delta<1/2$, two pure states with infidelity $\epsilon=\epsilon(\rho,\sigma)>0$ satisfy $P_{\rmerr}^{\caM,N}\le\delta$ if
\begin{equation}
N\ge\left\lceil\frac{2}{\kappa(\caM)\epsilon}
\log\frac1{2\delta}\right\rceil,
\label{eq:disc-N-cnt}
\end{equation}
where $\log$ is natural. Designs of order $t\ge3$ have $\kappa\ge c_{d,3}\ge1/4$ by \pref{prop:glob-3}, so $\lceil8\epsilon^{-1}\log[1/(2\delta)]\rceil$ copies suffice.

The same conclusion holds for the sampled measurements of \sref{sec:samp}. On the event $\Omega_m$ in \eref{eq:samp-event}, \pref{prop:samp-sep} gives $\kappa(\caM_m)\ge c_*>0$. The same fixed POVM therefore tests any given pair with $\caO\left[\epsilon^{-1}\log(1/\delta)\right]$ copies, independently of $d$. For $0<\epsilon\le1/2$ and $0<\delta\le1/4$, this matches optimal unrestricted pure-state discrimination~\cite{ChengEtAl2025} up to constant factors.

\subsection{Pure-state reconstruction}
\label{sec:rec}

\subsubsection{Local asymptotic reconstruction risk}
\label{sec:loc-rec}

Inverse Fisher information has also been used to characterize quantum-learning sample complexity in the small-error regime~\cite{KwonLieJiang2026}. At a fixed regular state $\rho=\proj\psi$ with $G_{\psi,\caM}\succ0$, consider an asymptotically efficient pure-state estimator $\hrho_N$ from $N$ copies, with convergent second moments under the assumptions in \aref{app:loc-risk}~\cite{VanDerVaart1998,Hradil1997}. In a regular identifiable parameterization, write $I_\theta$ and $J_\theta\succ0$ for the classical and quantum Fisher information matrices. A Fisher bound $I_\theta\succeq cJ_\theta$, $c>0$, gives the leading covariance bound $I_\theta^{-1}/N\preceq J_\theta^{-1}/(cN)$.

The mean infidelity of such an estimator has the expansion~\cite{Zhu2014Overcomplete,GillMassar2000,ZhuHayashi2018}
\begin{align}
\bbE_\rho[1-\tr(\rho\hrho_N)]&=\frac{R_{\rho,\caM}}N+o\left(N^{-1}\right),
\label{eq:loc-risk-coef}\\
R_{\rho,\caM}&:=\frac14\tr_{\bbR}G_{\psi,\caM}^{-1}.
\label{eq:loc-risk-trace}
\end{align}
For pure-state estimates, Hilbert--Schmidt mean squared error is twice the infidelity~\cite{Zhu2014Overcomplete,Scott2006}. Spectral pairing in \pref{prop:geometry} gives the following bound on the leading coefficient.

\begin{proposition}[Asymptotic infidelity of efficient estimators]\label{prop:loc-risk}
Let $\caE\in\fkE_t(d)$ with $t\ge2$, and let $\rho\in\caP(\caH)$ be regular with respect to $\caM_\caE$. Set $c=c_{d,t}$. Then
\begin{equation}
d-1\le R_{\rho,\caM_\caE}
\le\frac{d-1}{4c(1-c)}.
\label{eq:loc-risk-bnd}
\end{equation}
\end{proposition}

Fisher-symmetric measurements with $G=\IdTan{\psi}/2$ attain the optimal local single-copy value $R=d-1$~\cite{GillMassar2000,LiEtAl2016}. For pure states, this value coincides with the optimal asymptotic infidelity coefficient attainable by collective measurements on many copies~\cite{Matsumoto2002}, and with the leading coefficient of the collective Haar-average optimum $(d-1)/(N+d)$ discussed in \sref{sec:fin-rec}. For design orders $3$, $5$, and $7$, \eref{eq:loc-risk-bnd} gives
\begin{equation}
\frac{R_{\rho,\caM_\caE}}{d-1}\le
\begin{cases}
4/3,&t=3,\\
9/8,&t=5,\\
16/15,&t=7.
\end{cases}
\label{eq:loc-risk-rat}
\end{equation}
Although the $3$-design Fisher guarantee approaches $1/4$ as $d\to\infty$, the leading infidelity coefficient exceeds the optimal value $d-1$ by at most a factor $4/3$. The even-order coefficient $c_{d,2k}$ gives a smaller factor than the preceding odd-order guarantee. In general, the upper bound on $R_{\rho,\caM_\caE}/(d-1)$ is $1/[1-(1-2c)^2]$, whose excess over one is $\caO\left[(1-2c)^2\right]$ as $c\to1/2$. For the complete multiqubit stabilizer measurement, Propositions~\ref{prop:geometry} and~\ref{prop:stab-fi} give $R_{\rho,\caM_{\rmstab(2,n)}}/(d-1)\le9/8$ at every regular state.

\subsubsection{Finite-sample maximum-likelihood reconstruction}
\label{sec:fin-rec}

The preceding analysis describes the asymptotic risk at a fixed regular state. We now derive a finite-sample guarantee uniform over pure states by combining likelihood estimation with global Hellinger separation. Given $X_1,\ldots,X_N$ drawn independently from $p_\rho$, define the pure-state maximum-likelihood estimator (MLE) by
\begin{equation}
\hrho_N\in\arg\max_{\sigma\in\caP(\caH)}
\prod_{j=1}^N p_\sigma(X_j).
\label{eq:mle-def}
\end{equation}
The continuous likelihood attains its maximum on the compact pure-state space. If several states maximize it, we select one by a fixed rule.

\Needspace{10\baselineskip}
\begin{proposition}[Finite-sample likelihood reconstruction]\label{prop:mle-risk}
Let $\caM$ be a fixed finite rank-one POVM with $M$ outcomes and $\kappa(\caM)\ge\kappa_0>0$. For every $N\ge1$, the pure-state MLE in \eref{eq:mle-def} satisfies
\begin{equation}
\begin{aligned}
\sup_{\rho\in\caP(\caH)}\bbE_\rho\epsilon(\rho,\hrho_N)
&\le\frac{2(M-1)}{\kappa_0N},\\
\sup_{\rho\in\caP(\caH)}\bbE_\rho\epsilon(\rho,\hrho_N)
&\le\frac{4d\log(1+8N\sqrt{d}\lsp)+3}{\kappa_0N}.
\end{aligned}
\label{eq:mle-risk}
\end{equation}
\end{proposition}

The first bound follows from likelihood optimality and the expected relative entropy of the empirical distribution with respect to the true outcome distribution. The second uses a finite family of pointwise upper bounds on the square-root Born probabilities. Both bounds hold uniformly over pure states, including states with zero Born probabilities. By \thref{thm:fi-glob}, a Fisher floor $\eta(\caM)\ge\eta_0>0$ permits $\kappa_0=\eta_0^2$.

For designs of order $t\ge3$, we may take $\kappa_0=c_{d,3}\ge1/4$ by \pref{prop:glob-3}. For any target mean infidelity $\epsilon_{\rmtar}\in(0,1)$, the second bound in \eref{eq:mle-risk} then gives the sufficient copy count
\begin{equation}
N=\caO\!\left(\frac d{\epsilon_{\rmtar}}\log\frac{2d}{\epsilon_{\rmtar}}\right).
\label{eq:mle-copy-design}
\end{equation}

The logarithmic factor disappears for the sampled measurements of \sref{sec:samp}. On the event $\Omega_m$ in \eref{eq:samp-event}, \pref{prop:samp-sep} gives $\kappa(\caM_m)\ge c_*$, and the completed POVM has $M\le m+d$ outcomes. The first bound in \eref{eq:mle-risk} therefore gives
\begin{equation}
\sup_{\rho\in\caP(\caH)}\bbE_\rho\epsilon(\rho,\hrho_N)
\le\frac{2(m+d-1)}{c_*N}.
\label{eq:mle-sampled-risk}
\end{equation}
At $m=\lceil Cd\rceil$, the same target mean infidelity is achieved with
\begin{equation}
N=\caO\!\left(\frac d{\epsilon_{\rmtar}}\right),\qquad M=\caO(d).
\label{eq:mle-copy}
\end{equation}
The collective Haar-average optimum $(d-1)/(N+d)$ lower-bounds the minimax infidelity risk~\cite{BrussMacchiavello1999,Zhu2022Estimation}. Thus, for $m=\lceil Cd\rceil$ and any fixed sampled POVM in $\Omega_m$, pure-state MLE attains the minimax rate $\min\{1,d/N\}$ up to universal constants. This statement concerns statistical efficiency; computing the MLE efficiently is a separate problem, discussed in \sref{sec:discuss}.

\section{Discussion and outlook}
\label{sec:discuss}

We have shown that projective design order alone determines a hierarchy of Fisher guarantees for fixed measurements of unknown pure states. At second and odd orders, these guarantees are sharp, and finite designs attain them, so design moments alone cannot certify more. At even orders, our lower and upper bounds are close but do not yet coincide. Particular ensembles can nevertheless exceed their design guarantees: the dodecahedral qubit measurement and, for $d>2$, the complete multiqubit stabilizer measurement have Fisher floors strictly above the corresponding design guarantees. For experiments, these results provide a criterion, available before any data are collected, for how much information a fixed readout retains about an unknown pure state.

Our analysis also separates two figures of merit that are often conflated. The Fisher floor controls the least sensitive direction, whereas the asymptotic risk of an efficient estimator depends on all Fisher eigenvalues. Spectral pairing links each weak direction to a strong complementary direction. Consequently, for $3$-designs, the Fisher guarantee approaches $1/4$ in large dimension, while the leading infidelity coefficient remains within a factor $4/3$ of the optimal value. Zero-probability outcomes offer a complementary view of sharpness: design order limits their total weight, and the trace identity converts this weight into a Fisher-trace deficit. At odd order, saturating the weight bound together with the spectral lower bound forces isotropic attainment, although the weight bound alone does not control the least eigenvalue.

We have further shown that dimension-independent Fisher floors survive a drastic reduction in measurement size. Sampling and spectral completion of any $3$-design give, with high probability, a measurement with only $\caO(d)$ outcomes. Through global Hellinger separation, every such measurement discriminates given pure states with a dimension-independent number of copies and supports pure-state maximum-likelihood estimation with uniform mean infidelity $\caO(d/N)$. The local and global results serve distinct purposes: the former describe precision near a regular state, and the latter guarantee stable inference over all pure states. Complete-basis sampling avoids completion and, for stabilizer bases, requires only Clifford circuits and computational-basis readout, but our sufficient bound uses $\caO(d)$ bases and $\caO\left(d^2\right)$ labeled outcomes. These constructions indicate that randomized tomography and shadow protocols for pure states can be compressed to linearly many outcomes without losing uniform sensitivity.

Several questions remain open. On the mathematical side, we would like to determine the optimal even-order Fisher guarantees, to decide whether the odd-order Hellinger bounds can be improved beyond third order, and to find the largest Fisher floor achievable with a prescribed number of outcomes or bases. For stabilizer bases, the gap between the necessary $\log_2d+1$ bases and the sufficient $\caO(d)$ bases is particularly wide. On the practical side, reducing the constants in our sampling bounds would lower the measurement cost. It is equally important to certify a positive Fisher floor for a given measurement, to quantify its stability under errors in the design moments and under measurement noise, and to develop computationally efficient reconstruction methods that achieve uniform mean infidelity $\caO(d/N)$. Extending these guarantees to low-rank mixed states is another natural direction.

\emph{Note added---}After obtaining our main results, we learned of the preprint by Wanqing Li and Jiaxin Liu (arXiv:2611.00599). They also derive the exact Fisher floors of SIC and complete multiqubit stabilizer measurements, spectral pairing, and Fisher bounds based on shadow variances; our results on these topics were obtained independently. Our work establishes sharp Fisher guarantees for $2$-designs and designs of every odd order $t\ge3$, and derives global Hellinger separation and reconstruction bounds. We also construct POVMs with $\caO(d)$ outcomes and a dimension-independent Fisher floor by sampling and completion from projective $3$-designs.

\emph{Acknowledgments}---We thank Luyu Lu, Datong Chen, Jianxiang Ma, and Xinyang Shu for inspiring discussions. This work is supported by National Natural Science Foundation of China (Grant No.\ 92576101),  Shanghai Science and Technology Innovation Action Plan (Grant No.\ 24LZ1400200), Quantum Science and Technology-National Science and Technology Major Project (Grant No.\ 2024ZD0300101), and National Key Research and Development Program of China (Grant No.\ 2022YFA1404204).

\bibliographystyle{apsrev4-2}
\bibliography{references}

@article{ConcaEtAl2015,
  author = {Conca, A. and Edidin, D. and Hering, M. and Vinzant, C.},
  title = {An algebraic characterization of injectivity in phase retrieval},
  journal = {Appl. Comput. Harmon. Anal.},
  volume = {38},
  pages = {346--356},
  year = {2015},
  doi = {10.1016/j.acha.2014.06.005}
}

@article{FuchsVDG1999,
  author = {Fuchs, C. A. and van de Graaf, J.},
  title = {Cryptographic distinguishability measures for quantum-mechanical states},
  journal = {IEEE Trans. Inf. Theory},
  volume = {45},
  pages = {1216--1227},
  year = {1999},
  doi = {10.1109/18.761271}
}

@article{Zhu2017,
  author = {Zhu, H.},
  title = {Multiqubit {Clifford} groups are unitary $3$-designs},
  journal = {Phys. Rev. A},
  volume = {96},
  pages = {062336},
  year = {2017},
  doi = {10.1103/PhysRevA.96.062336}
}

@article{BraunsteinCaves1994,
  author = {Braunstein, S. L. and Caves, C. M.},
  title = {Statistical distance and the geometry of quantum states},
  journal = {Phys. Rev. Lett.},
  volume = {72},
  pages = {3439--3443},
  year = {1994},
  doi = {10.1103/PhysRevLett.72.3439}
}

@article{GillMassar2000,
  author = {Gill, R. D. and Massar, S.},
  title = {State estimation for large ensembles},
  journal = {Phys. Rev. A},
  volume = {61},
  pages = {042312},
  year = {2000},
  doi = {10.1103/PhysRevA.61.042312}
}

@article{LiEtAl2016,
  author = {Li, N. and Ferrie, C. and Gross, J. A. and Kalev, A. and Caves, C. M.},
  title = {{Fisher}-symmetric informationally complete measurements for pure states},
  journal = {Phys. Rev. Lett.},
  volume = {116},
  pages = {180402},
  year = {2016},
  doi = {10.1103/PhysRevLett.116.180402}
}

@article{ZhuHayashi2018,
  author = {Zhu, H. and Hayashi, M.},
  title = {Universally {Fisher}-symmetric informationally complete measurements},
  journal = {Phys. Rev. Lett.},
  volume = {120},
  pages = {030404},
  year = {2018},
  doi = {10.1103/PhysRevLett.120.030404}
}

@article{Zhu2022Estimation,
  author = {Zhu, H.},
  title = {Quantum measurements in the light of quantum state estimation},
  journal = {PRX Quantum},
  volume = {3},
  pages = {030306},
  year = {2022},
  doi = {10.1103/PRXQuantum.3.030306}
}

@article{ZhouChen2026,
  author = {Zhou, S. and Chen, S.},
  title = {Randomized measurements for multiparameter quantum metrology},
  journal = {PRX Quantum},
  volume = {7},
  pages = {010314},
  year = {2026},
  doi = {10.1103/s27y-gbrp}
}

@article{HeinosaariEtAl2013,
  author = {Heinosaari, T. and Mazzarella, L. and Wolf, M. M.},
  title = {Quantum tomography under prior information},
  journal = {Commun. Math. Phys.},
  volume = {318},
  pages = {355--374},
  year = {2013},
  doi = {10.1007/s00220-013-1671-8}
}

@article{XiaXuXu2025,
  author = {Xia, Y. and Xu, Zhiqiang and Xu, Zili},
  title = {Stability in phase retrieval: Characterizing condition numbers and the optimal vector set},
  journal = {Math. Comp.},
  volume = {94},
  pages = {2931--2960},
  year = {2025},
  doi = {10.1090/mcom/4042}
}

@article{BalanZou2016,
  author = {Balan, R. and Zou, D.},
  title = {On {Lipschitz} analysis and {Lipschitz} synthesis for the phase retrieval problem},
  journal = {Linear Algebra Appl.},
  volume = {496},
  pages = {152--181},
  year = {2016},
  doi = {10.1016/j.laa.2015.12.029}
}

@article{KuengEtAl2017,
  author = {Kueng, R. and Rauhut, H. and Terstiege, U.},
  title = {Low rank matrix recovery from rank one measurements},
  journal = {Appl. Comput. Harmon. Anal.},
  volume = {42},
  pages = {88--116},
  year = {2017},
  doi = {10.1016/j.acha.2015.07.007}
}

@article{Gilles2026,
  author = {Gilles, T.},
  title = {Stable low-rank matrix recovery from $3$-designs},
  journal = {Appl. Comput. Harmon. Anal.},
  volume = {84},
  pages = {101887},
  year = {2026},
  doi = {10.1016/j.acha.2026.101887}
}

@article{MatthewsEtAl2009,
  author = {Matthews, W. and Wehner, S. and Winter, A.},
  title = {Distinguishability of quantum states under restricted families of measurements with an application to quantum data hiding},
  journal = {Commun. Math. Phys.},
  volume = {291},
  pages = {813--843},
  year = {2009},
  doi = {10.1007/s00220-009-0890-5}
}

@misc{DuEtAl2026,
  author = {Du, Z. and Cheng, S. and Ye, H. and Chen, J. and Yuan, X. and Ma, X.},
  title = {Complexity-driven transitions in quantum observation},
  year = {2026},
  eprint = {2606.09765},
  archivePrefix = {arXiv},
  primaryClass = {quant-ph}
}

@misc{DLMF,
  title = {{NIST Digital Library of Mathematical Functions}},
  url = {https://dlmf.nist.gov/18},
  year = {2026},
  note = {Chapter~18, Version~1.2.8}
}

@article{Tikhomirov2018,
  author = {Tikhomirov, K.},
  title = {Sample covariance matrices of heavy-tailed distributions},
  journal = {Int. Math. Res. Not.},
  volume = {2018},
  pages = {6254--6289},
  year = {2018},
  doi = {10.1093/imrn/rnx067}
}

@article{BrussMacchiavello1999,
  author = {Bru{\ss}, D. and Macchiavello, C.},
  title = {Optimal state estimation for $d$-dimensional quantum systems},
  journal = {Phys. Lett. A},
  volume = {253},
  pages = {249--251},
  year = {1999},
  doi = {10.1016/S0375-9601(99)00099-7}
}

@misc{KuengZhuGross2016,
  author = {Kueng, R. and Zhu, H. and Gross, D.},
  title = {Distinguishing quantum states using {Clifford} orbits},
  year = {2016},
  eprint = {1609.08595},
  archivePrefix = {arXiv},
  primaryClass = {quant-ph},
  url = {https://arxiv.org/abs/1609.08595}
}

@article{BalanCasazzaEdidin2006,
  author = {Balan, R. and Casazza, P. and Edidin, D.},
  title = {On signal reconstruction without phase},
  journal = {Appl. Comput. Harmon. Anal.},
  volume = {20},
  pages = {345--356},
  year = {2006},
  doi = {10.1016/j.acha.2005.07.001}
}

@article{BandeiraEtAl2014,
  author = {Bandeira, A. S. and Cahill, J. and Mixon, D. G. and Nelson, A. A.},
  title = {Saving phase: Injectivity and stability for phase retrieval},
  journal = {Appl. Comput. Harmon. Anal.},
  volume = {37},
  pages = {106--125},
  year = {2014},
  doi = {10.1016/j.acha.2013.10.002}
}

@misc{KuengGross2015,
  author = {Kueng, R. and Gross, D.},
  title = {Qubit stabilizer states are complex projective $3$-designs},
  year = {2015},
  eprint = {1510.02767},
  archivePrefix = {arXiv},
  primaryClass = {quant-ph},
  url = {https://arxiv.org/abs/1510.02767}
}

@article{Webb2016,
  author = {Webb, Z.},
  title = {The {Clifford} group forms a unitary $3$-design},
  journal = {Quantum Inf. Comput.},
  volume = {16},
  pages = {1379--1400},
  year = {2016},
  doi = {10.26421/QIC16.15-16-8}
}

@article{ChengEtAl2025,
  author = {Cheng, H.-C. and Datta, N. and Liu, N. and Nuradha, T. and Salzmann, R. and Wilde, M. M.},
  title = {An invitation to the sample complexity of quantum hypothesis testing},
  journal = {npj Quantum Inf.},
  volume = {11},
  pages = {94},
  year = {2025},
  doi = {10.1038/s41534-025-00980-8}
}

@article{HuangKuengPreskill2020,
  author = {Huang, H.-Y. and Kueng, R. and Preskill, J.},
  title = {Predicting many properties of a quantum system from very few measurements},
  journal = {Nat. Phys.},
  volume = {16},
  pages = {1050--1057},
  year = {2020},
  doi = {10.1038/s41567-020-0932-7}
}

@article{Wootters1981,
  author = {Wootters, W. K.},
  title = {Statistical distance and {Hilbert} space},
  journal = {Phys. Rev. D},
  volume = {23},
  pages = {357--362},
  year = {1981},
  doi = {10.1103/PhysRevD.23.357}
}

@article{RenesEtAl2004,
  author = {Renes, J. M. and Blume-Kohout, R. and Scott, A. J. and Caves, C. M.},
  title = {Symmetric informationally complete quantum measurements},
  journal = {J. Math. Phys.},
  volume = {45},
  pages = {2171--2180},
  year = {2004},
  doi = {10.1063/1.1737053}
}

@article{YangChenZhu2026,
  author = {Yang, Zhiyao and Chen, Datong and Zhu, Huangjun},
  title = {Optimal Shadow Estimation with Minimal Measurement Settings},
  journal = {Phys. Rev. Lett.},
  volume = {137},
  pages = {140804},
  year = {2026},
  doi = {10.1103/g4lh-56r1}
}

@article{Scott2006,
  author = {Scott, A. J.},
  title = {Tight informationally complete quantum measurements},
  journal = {J. Phys. A: Math. Gen.},
  volume = {39},
  pages = {13507--13530},
  year = {2006},
  doi = {10.1088/0305-4470/39/43/009}
}

@article{FlammiaEtAl2005,
  author = {Flammia, S. T. and Silberfarb, A. and Caves, C. M.},
  title = {Minimal informationally complete measurements for pure states},
  journal = {Found. Phys.},
  volume = {35},
  pages = {1985--2006},
  year = {2005},
  doi = {10.1007/s10701-005-8658-z}
}

@article{Finkelstein2004,
  author = {Finkelstein, J.},
  title = {Pure-state informationally complete and ``really'' complete measurements},
  journal = {Phys. Rev. A},
  volume = {70},
  pages = {052107},
  year = {2004},
  doi = {10.1103/PhysRevA.70.052107}
}

@misc{HuangLi2026Recovery,
  author = {Huang, Gao and Li, Song},
  title = {Low-Rank Matrix Recovery via Heavy-Tailed Quadratic Sampling},
  year = {2026},
  eprint = {2607.08671},
  archivePrefix = {arXiv},
  primaryClass = {math.ST},
  url = {https://arxiv.org/abs/2607.08671}
}

@article{Matsumoto2002,
  author = {Matsumoto, K.},
  title = {A new approach to the {Cram\'{e}r--Rao}-type bound of the pure-state model},
  journal = {J. Phys. A: Math. Gen.},
  volume = {35},
  pages = {3111--3123},
  year = {2002},
  doi = {10.1088/0305-4470/35/13/307}
}

@article{ElbenEtAl2023,
  author = {Elben, A. and Flammia, S. T. and Huang, H.-Y. and Kueng, R. and Preskill, J. and Vermersch, B. and Zoller, P.},
  title = {The randomized measurement toolbox},
  journal = {Nat. Rev. Phys.},
  volume = {5},
  pages = {9--24},
  year = {2023},
  doi = {10.1038/s42254-022-00535-2}
}

@article{ChenEtAl2021Robust,
  author = {Chen, S. and Yu, W. and Zeng, P. and Flammia, S. T.},
  title = {Robust shadow estimation},
  journal = {PRX Quantum},
  volume = {2},
  pages = {030348},
  year = {2021},
  doi = {10.1103/PRXQuantum.2.030348}
}

@article{MaoYiZhu2025,
  author = {Mao, Chengsi and Yi, Changhao and Zhu, Huangjun},
  title = {Qudit Shadow Estimation Based on the {Clifford} Group and the Power of a Single Magic Gate},
  journal = {Phys. Rev. Lett.},
  volume = {134},
  pages = {160801},
  year = {2025},
  doi = {10.1103/PhysRevLett.134.160801}
}

@article{ZhuMaoYi2024,
  author = {Zhu, Huangjun and Mao, Chengsi and Yi, Changhao},
  title = {Third moments of qudit {Clifford} orbits and 3-designs based on magic orbits},
  journal = {Commun. Math. Phys.},
  volume = {407},
  pages = {242},
  year = {2026},
  doi = {10.1007/s00220-026-05583-8}
}

@article{RoyScott2007,
  author = {Roy, Aidan and Scott, A. J.},
  title = {Weighted complex projective 2-designs from bases: optimal state determination by orthogonal measurements},
  journal = {J. Math. Phys.},
  volume = {48},
  pages = {072110},
  year = {2007},
  doi = {10.1063/1.2748617}
}

@phdthesis{Gottesman1997,
  author = {Gottesman, Daniel},
  title = {Stabilizer Codes and Quantum Error Correction},
  school = {California Institute of Technology},
  year = {1997},
  eprint = {quant-ph/9705052},
  archivePrefix = {arXiv},
  url = {https://arxiv.org/abs/quant-ph/9705052}
}

@book{Helstrom1976,
  author = {Helstrom, Carl W.},
  title = {Quantum Detection and Estimation Theory},
  publisher = {Academic Press},
  address = {New York},
  year = {1976}
}

@book{VanDerVaart1998,
  author = {van der Vaart, A. W.},
  title = {Asymptotic Statistics},
  publisher = {Cambridge University Press},
  address = {Cambridge},
  year = {1998},
  doi = {10.1017/CBO9780511802256}
}

@article{DeckerEtAl2003,
  author = {Decker, Thomas and Janzing, Dominik and Beth, Thomas},
  title = {Quantum circuits for single-qubit measurements corresponding to {Platonic} solids},
  journal = {Int. J. Quantum Inf.},
  volume = {2},
  pages = {353},
  year = {2004},
  doi = {10.1142/S0219749904000298}
}

@article{DeBurghEtAl2008,
  author = {de Burgh, Mark D. and Langford, Nathan K. and Doherty, Andrew C. and Gilchrist, Alexei},
  title = {Choice of measurement sets in qubit tomography},
  journal = {Phys. Rev. A},
  volume = {78},
  pages = {052122},
  year = {2008},
  doi = {10.1103/PhysRevA.78.052122}
}

@misc{YangEtAl2025CRM,
  author = {Yang, Zhongyi and Chen, Datong and Li, Zihao and Zhu, Huangjun},
  title = {High-Precision Fidelity Estimation with Common Randomized Measurements},
  year = {2025},
  eprint = {2511.22509},
  archivePrefix = {arXiv},
  primaryClass = {quant-ph},
  url = {https://arxiv.org/abs/2511.22509}
}

@article{Hradil1997,
  author = {Hradil, Zdenek},
  title = {Quantum-state estimation},
  journal = {Phys. Rev. A},
  volume = {55},
  pages = {R1561--R1564},
  year = {1997},
  doi = {10.1103/PhysRevA.55.R1561}
}

@article{Zhu2014Overcomplete,
  author = {Zhu, Huangjun},
  title = {Quantum state estimation with informationally overcomplete measurements},
  journal = {Phys. Rev. A},
  volume = {90},
  pages = {012115},
  year = {2014},
  doi = {10.1103/PhysRevA.90.012115}
}

@article{BertoniEtAl2024,
  author = {Bertoni, Christian and Haferkamp, Jonas and Hinsche, Marcel and Ioannou, Marios and Eisert, Jens and Pashayan, Hakop},
  title = {Shallow Shadows: Expectation Estimation Using Low-Depth Random {Clifford} Circuits},
  journal = {Phys. Rev. Lett.},
  volume = {133},
  pages = {020602},
  year = {2024},
  doi = {10.1103/PhysRevLett.133.020602}
}

@article{VermerschEtAl2024CRM,
  author = {Vermersch, Benoit and Rath, Aniket and Sundar, Bharathan and Branciard, Cyril and Preskill, John and Elben, Andreas},
  title = {Enhanced Estimation of Quantum Properties with Common Randomized Measurements},
  journal = {PRX Quantum},
  volume = {5},
  pages = {010352},
  year = {2024},
  doi = {10.1103/PRXQuantum.5.010352}
}

@article{InnocentiEtAl2023,
  author = {Innocenti, Luca and Lorenzo, Salvatore and Palmisano, Ivan and Albarelli, Francesco and Ferraro, Alessandro and Paternostro, Mauro and Palma, G. Massimo},
  title = {Shadow Tomography on General Measurement Frames},
  journal = {PRX Quantum},
  volume = {4},
  pages = {040328},
  year = {2023},
  doi = {10.1103/PRXQuantum.4.040328}
}

@article{ParkTeoJeong2025,
  author = {Park, Guedong and Teo, Yong Siah and Jeong, Hyunseok},
  title = {Resource-efficient shadow tomography using equatorial stabilizer measurements},
  journal = {Phys. Rev. Research},
  volume = {7},
  pages = {033097},
  year = {2025},
  doi = {10.1103/9pbp-jzr9}
}

@misc{Zhu2023Projective,
  author = {Zhu, Huangjun},
  title = {Information Theoretic Significance of Projective Measurements},
  year = {2023},
  eprint = {2305.14110},
  archivePrefix = {arXiv},
  primaryClass = {quant-ph},
  url = {https://arxiv.org/abs/2305.14110}
}

@CONTROL{apsrev42Control,
  title = {0}
}

@book{Gautschi2004,
  author = {Gautschi, Walter},
  title = {Orthogonal Polynomials: Computation and Approximation},
  publisher = {Oxford University Press},
  address = {Oxford},
  year = {2004},
  doi = {10.1093/oso/9780198506720.001.0001}
}

@article{KoenigSmolin2014,
  author = {Koenig, Robert and Smolin, John A.},
  title = {How to efficiently select an arbitrary {Clifford} group element},
  journal = {J. Math. Phys.},
  volume = {55},
  pages = {122202},
  year = {2014},
  doi = {10.1063/1.4903507}
}

@article{ByrdEtAl1995,
  author = {Byrd, Richard H. and Lu, Peihuang and Nocedal, Jorge and Zhu, Ciyou},
  title = {A limited memory algorithm for bound constrained optimization},
  journal = {SIAM J. Sci. Comput.},
  volume = {16},
  pages = {1190--1208},
  year = {1995},
  doi = {10.1137/0916069}
}

@article{Hayashi1998,
  author = {Hayashi, M.},
  title = {Asymptotic estimation theory for a finite-dimensional pure state model},
  journal = {J. Phys. A: Math. Gen.},
  volume = {31},
  pages = {4633--4655},
  year = {1998},
  doi = {10.1088/0305-4470/31/20/006}
}

@misc{LuSha2025,
  author = {Lu, J. and Sha, K.},
  title = {Quantum {Fisher} information matrix via its classical counterpart from random measurements},
  year = {2025},
  eprint = {2509.08196},
  archivePrefix = {arXiv},
  primaryClass = {quant-ph},
  url = {https://arxiv.org/abs/2509.08196}
}

@article{HuChoiYou2023,
  author = {Hu, Hong-Ye and Choi, Soonwon and You, Yi-Zhuang},
  title = {Classical shadow tomography with locally scrambled quantum dynamics},
  journal = {Phys. Rev. Res.},
  volume = {5},
  pages = {023027},
  year = {2023},
  doi = {10.1103/PhysRevResearch.5.023027}
}

@article{HuEtAl2025RobustShallow,
  author = {Hu, Hong-Ye and Gu, Andi and Majumder, Swarnadeep and Ren, Hang and Zhang, Yipei and Wang, Derek S. and You, Yi-Zhuang and Minev, Zlatko and Yelin, Susanne F. and Seif, Alireza},
  title = {Demonstration of robust and efficient quantum property learning with shallow shadows},
  journal = {Nat. Commun.},
  volume = {16},
  pages = {2943},
  year = {2025},
  doi = {10.1038/s41467-025-57349-w}
}

@article{AcharyaKypraiosGuta2016,
  author = {Acharya, A. and Kypraios, T. and Gu{\c t}{\u a}, M.},
  title = {Statistically efficient tomography of low rank states with incomplete measurements},
  journal = {New J. Phys.},
  volume = {18},
  pages = {043018},
  year = {2016},
  doi = {10.1088/1367-2630/18/4/043018}
}

@article{AcharyaGuta2017,
  author = {Acharya, A. and Gu{\c t}{\u a}, M.},
  title = {Statistical analysis of compressive low rank tomography with random measurements},
  journal = {J. Phys. A: Math. Theor.},
  volume = {50},
  pages = {195301},
  year = {2017},
  doi = {10.1088/1751-8121/aa682e}
}

@inproceedings{ChenGongZhou2026,
  author = {Chen, S. and Gong, W. and Zhou, S.},
  title = {Instance-optimal high-precision shadow tomography with few-copy measurements: A metrological approach},
  booktitle = {Proceedings of Thirty Ninth Conference on Learning Theory},
  series = {Proceedings of Machine Learning Research},
  volume = {336},
  pages = {1115--1185},
  year = {2026},
  url = {https://proceedings.mlr.press/v336/chen26c.html}
}

@article{Hoeffding1963,
  author = {Hoeffding, W.},
  title = {Probability inequalities for sums of bounded random variables},
  journal = {J. Am. Stat. Assoc.},
  volume = {58},
  number = {301},
  pages = {13--30},
  year = {1963},
  doi = {10.1080/01621459.1963.10500830}
}

@article{GiovannettiLloydMaccone2011,
  author = {Giovannetti, V. and Lloyd, S. and Maccone, L.},
  title = {Advances in quantum metrology},
  journal = {Nat. Photonics},
  volume = {5},
  pages = {222--229},
  year = {2011},
  doi = {10.1038/nphoton.2011.35}
}

@article{SzczykulskaBaumgratzDatta2016,
  author = {Szczykulska, M. and Baumgratz, T. and Datta, A.},
  title = {Multi-parameter quantum metrology},
  journal = {Adv. Phys. X},
  volume = {1},
  pages = {621--639},
  year = {2016},
  doi = {10.1080/23746149.2016.1230476}
}

@article{BayerTeichmann2006,
 author = {Bayer, Christian and Teichmann, Josef},
 title = {The proof of {Tchakaloff's} theorem},
 journal = {Proceedings of the American Mathematical Society},
 volume = {134},
 number = {10},
 pages = {3035--3040},
 year = {2006},
 eprint = {math/0502473},
 archivePrefix = {arXiv}
}

@article{KwonLieJiang2026,
  author = {Kwon, Hyukgun and Lie, Seok Hyung and Jiang, Liang},
  title = {Universal Sample Complexity Bounds in Quantum Learning Theory via {Fisher} Information Matrix},
  journal = {PRX Quantum},
  volume = {7},
  pages = {033063},
  year = {2026},
  doi = {10.1103/8k5v-ddtw}
}

@article{WangChenYuan2026,
  author = {Wang, Lingna and Chen, Hongzhen and Yuan, Haidong},
  title = {Minimal Trade-Off and Optimal Measurement for Multiparameter Quantum Estimation},
  journal = {Phys. Rev. Lett.},
  volume = {137},
  pages = {020804},
  year = {2026},
  doi = {10.1103/dvj7-9thf}
}

@article{Hoggar1982,
  author = {Hoggar, S. G.},
  title = {$t$-Designs in Projective Spaces},
  journal = {Eur. J. Combin.},
  volume = {3},
  number = {3},
  pages = {233--254},
  year = {1982},
  doi = {10.1016/S0195-6698(82)80035-8}
}

@article{Ivanovic1981,
  author = {Ivanovi{\'c}, I. D.},
  title = {Geometrical description of quantal state determination},
  journal = {J. Phys. A: Math. Gen.},
  volume = {14},
  number = {12},
  pages = {3241--3245},
  year = {1981},
  doi = {10.1088/0305-4470/14/12/019}
}

@article{WoottersFields1989,
  author = {Wootters, W. K. and Fields, B. D.},
  title = {Optimal state-determination by mutually unbiased measurements},
  journal = {Ann. Phys.},
  volume = {191},
  number = {2},
  pages = {363--381},
  year = {1989},
  doi = {10.1016/0003-4916(89)90322-9}
}

@article{Zauner2011,
  author = {Zauner, Gerhard},
  title = {Quantum designs: Foundations of a noncommutative design theory},
  journal = {Int. J. Quantum Inf.},
  volume = {9},
  number = {1},
  pages = {445--507},
  year = {2011},
  doi = {10.1142/S0219749911006776}
}

@article{DurtEtAl2010,
  author = {Durt, Thomas and Englert, Berthold-Georg and Bengtsson, Ingemar and {\.{Z}}yczkowski, Karol},
  title = {On mutually unbiased bases},
  journal = {Int. J. Quantum Inf.},
  volume = {8},
  number = {4},
  pages = {535--640},
  year = {2010},
  doi = {10.1142/S0219749910006502}
}

@article{FuchsHoangStacey2017,
  author = {Fuchs, Christopher A. and Hoang, Michael C. and Stacey, Blake C.},
  title = {The {SIC} Question: History and State of Play},
  journal = {Axioms},
  volume = {6},
  number = {3},
  pages = {21},
  year = {2017},
  doi = {10.3390/axioms6030021}
}

\clearpage
\onecolumngrid
\appendix

\pdfbookmark[1]{Notation for the appendices}{appendix-notation}
\phantomsection\label{app:notation}
\addcontentsline{toc}{section}{Notation for the appendices}
\begin{center}
\textbf{Notation for the appendices}
\end{center}
In addition to the main-text notation, $\rk(O)$ denotes the rank of an operator. We use $\norm{O}=\norm{O}_\infty$ for the operator norm, $\norm{O}_2$ for the Hilbert--Schmidt norm, and $\norm{x}_2$ for the Euclidean norm of a complex vector or real coordinate vector. For Hermitian $O$, $|O|=\left(O^2\right)^{1/2}$. We use $\IdMat{k}$ for the identity matrix on $\bbR^k$, $\IdTan{\psi}$ for the identity on $\caH_\psi$ viewed as a real space, and $\bbone$ for the identity on the complex space $\caH$. The symbol $\delta_{jk}$ is the Kronecker delta. State vectors are normalized; a horizontal vector $\nu\in\caH_\psi$ may have arbitrary norm unless $\braket\nu\nu=1$ is stated.

We write $\bbP$ for probability and $\bbE$ for expectation; subscripts specify the state, ensemble, or probability measure. In particular, $\bbE_\pi$ uses the design weights $\pi_i$, and $\bbE_\Ha$ uses Haar measure on pure states. The symbol $\bfone_A$ is the indicator of an event $A$, $\supp p=\{i\colon p_i>0\}$ is the support of a probability vector, and $|\supp p|$ is its cardinality. An $\varepsilon$-net is a finite subset of a metric space whose radius-$\varepsilon$ balls cover that space.

For a probability measure $\mu$, $L^2(\mu)$ is the real Hilbert space of measurable square-integrable functions, identified up to equality $\mu$-almost everywhere. Its inner product is $\inner{f}{g}_\mu=\int fg\,\rmd\mu$, and $\norm{f}_{L^2(\mu)}^2=\inner{f}{f}_\mu$. The space $\scrP_k$ consists of real polynomials of degree at most $k$, and $\deg f$ denotes polynomial degree. The rising factorial is $(g)_m=g(g+1)\cdots(g+m-1)$ for integers $m\ge1$, with $(g)_0=1$.

\section{Proofs of Fisher-geometry results}\label{app:geometry}

This appendix proves Propositions~\ref{prop:geometry}--\ref{prop:pure-ic} on Fisher geometry and pure-state identifiability.

\subsection{Proofs of Propositions~\ref{prop:geometry}--\ref{prop:pure-ic}: trace identity, spectral pairing, and Fisher floors}\label{app:trace-pairing}

\begin{proof}[Proof of \pref{prop:geometry}]
Fix $\rho=\proj\psi$ and write $G=G_{\psi,\caM}$. For every unit ket $\ket\nu\in\caH_\psi$, adding \eref{eq:hor-fi} in the directions $\ket\nu$ and $\rmi\ket\nu$ gives
\begin{equation}
 \begin{aligned}
  \inner{\nu}{G\nu}_{\bbR}+\inner{\rmi\nu}{G(\rmi\nu)}_{\bbR}
  &=\sum_{i\colon p_i(\rho)>0}w_i
 \frac{[\Rea(\ba_i\braket{\phi_i}{\nu})]^2+[\Ima(\ba_i\braket{\phi_i}{\nu})]^2}{|a_i|^2}\\
  &=\sum_{i\colon p_i(\rho)>0}w_i
 \frac{|\ba_i\braket{\phi_i}{\nu}|^2}{|a_i|^2}
 =\sum_{i\colon p_i(\rho)>0}w_i|\braket{\phi_i}{\nu}|^2=1-\bra\nu Z_\psi\ket\nu.
 \end{aligned}
 \label{eq:pairing-form}
\end{equation}
Here $a_i=\braket{\phi_i}{\psi}$, as in \eref{eq:hor-fi}; the last equality follows from \eref{eq:povm} and \eref{eq:zero-op}. The complex structure $\caJ$ is real orthogonal, with $\caJ^{-1}=-\caJ$. Hence the term $\inner{\rmi\nu}{G(\rmi\nu)}_{\bbR}$ in \eref{eq:pairing-form} satisfies
\begin{equation}
 \inner{\rmi\nu}{G(\rmi\nu)}_{\bbR}
 =\inner*{\nu}{\caJ^{-1}G\caJ\nu}_{\bbR}
 =\inner*{\nu}{\caJ G\caJ^{-1}\nu}_{\bbR}.
\end{equation}
Together with \eref{eq:pairing-form}, this identity shows that $G+\caJ G\caJ^{-1}$ and $\IdTan\psi-(Z_\psi)_{\bbR}$ have the same quadratic form on every unit ket $\ket\nu\in\caH_\psi$. Both operators are real self-adjoint and are therefore determined by their quadratic forms on the unit sphere, proving \eref{eq:pairing}.

Taking real traces in \eref{eq:pairing}, we have $\tr_{\bbR}\left(\caJ G\caJ^{-1}\right)=\tr_{\bbR}G$, $\tr_{\bbR}\IdTan\psi=2d-2$, and $\tr_{\bbR}(Z_\psi)_{\bbR}=2\tr Z_\psi$. Since each projector in \eref{eq:zero-op} has complex trace one, this yields
\begin{equation}
 2\tr_{\bbR}G=2d-2-2\sum_{i\colon p_i(\rho)=0}w_i,
\end{equation}
which proves \eref{eq:trace-defect}. At a regular state, $Z_\psi=0$, and \eref{eq:pairing} implies $\caJ(G-\IdTan\psi/2)\caJ^{-1}=-(G-\IdTan\psi/2)$. Hence the eigenvalues of $G-\IdTan\psi/2$ occur in opposite pairs. Writing these pairs as $\pm s_j$ with $s_j\ge0$, the corresponding eigenvalues of $G$ are $1/2\pm s_j$. Since $G\succeq0$, $1/2-s_j\ge0$ and hence $s_j\le1/2$.
\end{proof}

\begin{proof}[Proof of \pref{prop:reg-eq}]
\emph{Attainment.} Let $\ket\psi,\ket\nu$ range over orthonormal pairs and put $\rho=\proj\psi$. This pair space is compact. For each outcome $i$, evaluate its summand in \eref{eq:score-fi} at $\Delta=\Delta_{\psi,\nu}$ and divide by four. This function is continuous when $p_i(\rho)>0$. At a pair with $p_i(\rho)=0$, the summand is assigned the value zero. All its values are nonnegative, so its lower limit along any sequence approaching that pair is at least its value zero; thus it is lower semicontinuous there. The finite sum is therefore lower semicontinuous and attains its minimum. For fixed $\ket\psi$, minimizing over unit horizontal $\ket\nu$ gives $\lambda_{\min}(G_{\psi,\caM})$ by \eref{eq:normalized-fi-form}; the minimum over pairs is thus $\eta(\caM)$. Consequently,
\begin{equation}
 \eta(\caM)=\min_{\ket\psi}\lambda_{\min}(G_{\psi,\caM}).
 \label{eq:floor-attainment}
\end{equation}

\emph{Equality of Fisher floors.} Fix $\rho=\proj\psi$. If $p_i(\rho)=0$, then $\bigl\|E_i^{1/2}\ket\psi\bigr\|_2^2=0$, so $E_i\ket\psi=0$. For each such index, $\ker E_i\cap\caH_\psi$ is a proper complex subspace of $\caH_\psi$: otherwise $E_i$ would annihilate both $\ket\psi$ and its orthogonal complement, contradicting $E_i\ne0$. The complement of the union of these finitely many subspaces is dense. We may therefore first choose a unit horizontal ket $\ket\nu$ satisfying $\bra\nu E_i\ket\nu>0$ for every $i$ with $p_i(\rho)=0$.

For $0<\varepsilon<\pi/2$, define
\begin{equation}
 \ket{\psi_\varepsilon}=\cos\varepsilon\ket\psi+\rmi\sin\varepsilon\ket\nu,\qquad
 \ket{\nu_\varepsilon}=\cos\varepsilon\ket\nu+\rmi\sin\varepsilon\ket\psi.
 \label{eq:reg-path}
\end{equation}
These kets are orthonormal. Put $\rho_\varepsilon=\proj{\psi_\varepsilon}$. Substitution into \eref{eq:tan}, using $E_i\ket\psi=0$, gives
\begin{equation}
 p_i(\rho_\varepsilon)=\sin^2\varepsilon\bra\nu E_i\ket\nu>0,\qquad
 \tr(E_i\Delta_{\psi_\varepsilon,\nu_\varepsilon})=0,
 \label{eq:reg-zero-score}
\end{equation}
for every $i$ with $p_i(\rho)=0$. If $p_i(\rho)>0$, continuity of $\varepsilon\mapsto p_i(\rho_\varepsilon)$ at zero gives $p_i(\rho_\varepsilon)>0$ for all sufficiently small $\varepsilon$. There are finitely many outcomes, so $\rho_\varepsilon$ is regular for all sufficiently small $\varepsilon>0$. At each such state, \eref{eq:reg-eta} gives
\begin{equation}
 \eta_{\reg}(\caM)
 \le\lambda_{\min}(G_{\psi_\varepsilon,\caM})
 \le\inner{\nu_\varepsilon}{G_{\psi_\varepsilon,\caM}\nu_\varepsilon}_{\bbR}.
\end{equation}
The summands in \eref{eq:score-fi} with $p_i(\rho)=0$ vanish along this path by \eref{eq:reg-zero-score}; those with $p_i(\rho)>0$ converge to their values at $\rho$. Taking $\varepsilon\to0^+$ and using \eref{eq:normalized-fi-form} therefore yields
\begin{equation}
 \eta_{\reg}(\caM)
 \le\lim_{\varepsilon\to0^+}\inner{\nu_\varepsilon}{G_{\psi_\varepsilon,\caM}\nu_\varepsilon}_{\bbR}
 =\inner{\nu}{G_{\psi,\caM}\nu}_{\bbR}.
 \label{eq:reg-limit}
\end{equation}
For fixed $\psi$, the right-hand side is continuous in $\ket\nu$. Since the admissible directions chosen above are dense in the unit sphere of $\caH_\psi$, \eref{eq:reg-limit} holds for every unit ket $\ket\nu\in\caH_\psi$. Taking the infimum over $\psi$ and $\nu$ gives
\begin{equation}
 \eta_{\reg}(\caM)
 \le\inf_{\substack{\braket{\psi}{\psi}=\braket{\nu}{\nu}=1\\\braket{\psi}{\nu}=0}}
       \inner{\nu}{G_{\psi,\caM}\nu}_{\bbR}
 =\eta(\caM).
\end{equation}
Since regular states are a subset of all pure states, their defining infimum satisfies $\eta(\caM)\le\eta_{\reg}(\caM)$. Hence
\begin{equation}
 \eta_{\reg}(\caM)=\eta(\caM).
 \label{eq:floor-equality-proof}
\end{equation}

\emph{Upper bound.} Write $E_i=\sum_j E_{ij}$ as a sum of its nonzero rank-one spectral components, and let $\widetilde{\caM}=\{E_{ij}\}_{i,j}$ be the resulting refinement. For fixed $\rho=\proj\psi$ and $\Delta\in T_\rho$, put $p_{ij}=\tr(E_{ij}\rho)$, so $p_i(\rho)=\sum_jp_{ij}$. If $p_{ij}=0$, the positive semidefiniteness of $E_{ij}$ implies $E_{ij}^{1/2}\ket\psi=0$ and hence $E_{ij}\ket\psi=0$. Equation~\eqref{eq:tan} then gives $\tr(E_{ij}\Delta)=0$. Thus, for each $i$ with $p_i(\rho)>0$, Cauchy--Schwarz yields
\begin{equation}
 \frac{[\tr(E_i\Delta)]^2}{p_i(\rho)}
 =\frac{\bigl[\sum_{j\colon p_{ij}>0}\tr(E_{ij}\Delta)\bigr]^2}
         {\sum_{j\colon p_{ij}>0}p_{ij}}
 \le\sum_{j\colon p_{ij}>0}\frac{[\tr(E_{ij}\Delta)]^2}{p_{ij}}.
\end{equation}
Summing over $i$ in \eref{eq:score-fi}, and then using \eref{eq:normalized-fi-form}, gives
\begin{equation}
 I_{\rho,\caM}\preceq I_{\rho,\widetilde{\caM}},
 \qquad \eta(\caM)\le\eta(\widetilde{\caM}).
\end{equation}
For every pure state $\rho=\proj\psi$, \eref{eq:trace-defect} in \pref{prop:geometry} bounds the trace of $G_{\psi,\widetilde{\caM}}$ by $d-1$. Since this operator acts on a real space of dimension $2d-2$,
\begin{equation}
 \eta(\widetilde{\caM})
 \le\lambda_{\min}(G_{\psi,\widetilde{\caM}})
 \le\frac{\tr_{\bbR}G_{\psi,\widetilde{\caM}}}{2d-2}
 \le\frac{d-1}{2d-2}=\frac12.
\end{equation}
Each summand in \eref{eq:score-fi} is a square divided by a positive probability, so $I_{\rho,\caM}(\Delta,\Delta)\ge0$ for every tangent $\Delta$. Equation~\eqref{eq:normalized-fi-form} therefore gives $G_{\psi,\caM}\succeq0$ at every state and $\eta(\caM)\ge0$. We have therefore proved
\begin{equation}
 0\le\eta(\caM)\le\frac12.
 \label{eq:floor-range-proof}
\end{equation}
Combining \eref{eq:floor-attainment}, \eref{eq:floor-equality-proof}, and \eref{eq:floor-range-proof} proves \eref{eq:reg-eq}.
\end{proof}

\begin{proof}[Proof of \pref{prop:pure-ic}]
\emph{(\ref{cond:pure-identifiability})$\Leftrightarrow$(\ref{cond:pointwise-fisher}).} Fix $\rho=\proj\psi$ and $\Delta\in T_\rho$. If $p_i(\rho)=0$, the positive semidefiniteness of $E_i$ implies $E_i^{1/2}\ket\psi=0$ and hence $E_i\ket\psi=0$. Equation~\eqref{eq:tan} then gives $\tr(E_i\Delta)=0$. All remaining terms in \eref{eq:score-fi} are nonnegative with positive denominators. Therefore,
\begin{equation}
 I_{\rho,\caM}(\Delta,\Delta)=0
 \quad\Longleftrightarrow\quad
 \tr(E_i\Delta)=0\ \text{for every }i.
 \label{eq:tangent-score-kernel}
\end{equation}

Assume (\ref{cond:pure-identifiability}). If (\ref{cond:pointwise-fisher}) failed, there would be a pure state $\rho$ and a nonzero $\Delta\in T_\rho$ with $I_{\rho,\caM}(\Delta,\Delta)=0$. By \eref{eq:tan}, $\Delta$ is traceless Hermitian of rank two, so its spectral decomposition has the form
\begin{equation}
 \Delta=\lambda(\proj{e_+}-\proj{e_-}),
 \qquad \lambda>0,\qquad \braket{e_\pm}{e_\pm}=1,\quad\braket{e_+}{e_-}=0.
\end{equation}
Equation~\eqref{eq:tangent-score-kernel} then gives
\begin{equation}
 p_i(\proj{e_+})-p_i(\proj{e_-})
 =\frac{\tr(E_i\Delta)}{\lambda}=0
 \quad\text{for every }i.
\end{equation}
The distinct pure states $\proj{e_+}$ and $\proj{e_-}$ would have identical outcome probabilities, contradicting (\ref{cond:pure-identifiability}).

Conversely, assume (\ref{cond:pointwise-fisher}). If two distinct pure states $\rho_1,\rho_2$ had identical outcome probabilities, their difference $\Delta=\rho_1-\rho_2$ would be nonzero, traceless Hermitian of rank two, with $\tr(E_i\Delta)=0$ for every $i$. Write its spectral decomposition as above and define the orthonormal pair
\begin{equation}
 \ket\psi=\frac{\ket{e_+}+\ket{e_-}}{\sqrt{2}\lsp},
 \qquad\ket\nu=\frac{\ket{e_+}-\ket{e_-}}{\sqrt{2}\lsp}.
\end{equation}
Put $\rho=\proj\psi$. Direct expansion using \eref{eq:tan} gives
\begin{equation}
 \Delta=\lambda\Delta_{\psi,\nu}\in T_\rho.
\end{equation}
Equation~\eqref{eq:tangent-score-kernel} now implies $I_{\rho,\caM}(\Delta,\Delta)=0$, contradicting (\ref{cond:pointwise-fisher}). This proves (\ref{cond:pure-identifiability})$\Leftrightarrow$(\ref{cond:pointwise-fisher}) by the rank-two kernel argument of Proposition~4 and Corollary~1 in \rcite{HeinosaariEtAl2013}.

\emph{(\ref{cond:pointwise-fisher})$\Leftrightarrow$(\ref{cond:uniform-fisher}).} Assume (\ref{cond:pointwise-fisher}). By \eref{eq:normalized-fi-form}, every $G_{\psi,\caM}$ is positive definite. Proposition~\ref{prop:reg-eq} ensures that its least eigenvalue attains its minimum at some unit ket $\ket{\psi_*}$. Consequently,
\begin{equation}
 \eta(\caM)=\lambda_{\min}(G_{\psi_*,\caM})>0.
\end{equation}

Conversely, assume (\ref{cond:uniform-fisher}). Then Eqs.~\eqref{eq:qfi} and~\eqref{eq:normalized-fi-form} give, for every pure state $\rho$ and nonzero $\Delta\in T_\rho$,
\begin{equation}
 I_{\rho,\caM}(\Delta,\Delta)
 \ge\eta(\caM)J_\rho(\Delta,\Delta)
 =2\eta(\caM)\tr\left(\Delta^2\right)>0.
\end{equation}
Thus (\ref{cond:pure-identifiability}), (\ref{cond:pointwise-fisher}), and (\ref{cond:uniform-fisher}) are equivalent.
\end{proof}

\section{Proof of the shadow-variance criterion}
\label{app:shadow-fi}

This appendix proves \pref{prop:shadow-fi} and the variance-based bounds in Table~\ref{tab:shadow-fi-ens}.

\subsection{Proof of \pref{prop:shadow-fi}}\label{app:shadow-crit}

\begin{proof}[Proof of \pref{prop:shadow-fi}]
Fix a pure state $\rho=\proj\psi$ and set $O=\Delta\in T_\rho$, with $\Delta\ne0$. By \eref{eq:tan}, $\Delta\in\caL^{\rmH}_0(\caH)$ and $\norm{\Delta}_2^2=\tr\left(\Delta^2\right)$, so \eref{eq:shadow-var} gives $\Var_\rho[\tr(\Delta\hrho_i)]\le C\tr\left(\Delta^2\right)$. Put $f_i=\tr(\Delta\hrho_i)$. Equation~\eqref{eq:tan} gives $\tr(\rho\Delta)=0$, so \eref{eq:shadow-mean} implies $\sum_i p_i(\rho)f_i=0$. By linearity and spectral decomposition, the reconstruction identity \eref{eq:shadow-mean} extends from density operators to every Hermitian operator. Applying it to $\Delta$ and taking its trace against $\Delta$ gives
\begin{equation}
 \sum_i\tr(E_i\Delta)\hrho_i=\Delta,\qquad
 \sum_i\tr(E_i\Delta)f_i=\tr\left(\Delta^2\right).
 \label{eq:shadow-tangent-reconstruction}
\end{equation}
If $p_i(\rho)=0$, the positive semidefiniteness of $E_i$ implies $E_i^{1/2}\ket\psi=0$ and hence $E_i\ket\psi=0$. Equation~\eqref{eq:tan} then gives $\tr(E_i\Delta)=0$. Applying Cauchy--Schwarz to the positive-probability outcomes in \eref{eq:shadow-tangent-reconstruction}, and using \eref{eq:score-fi} and the zero mean of $f_i$, we obtain
\begin{equation}
 \left[\tr\left(\Delta^2\right)\right]^2
 \le I_{\rho,\caM}(\Delta,\Delta)\sum_{i\colon p_i(\rho)>0}p_i(\rho)f_i^2
 =I_{\rho,\caM}(\Delta,\Delta)\Var_\rho(f_i)
 \le C\tr\left(\Delta^2\right)I_{\rho,\caM}(\Delta,\Delta).
 \label{eq:shadow-tangent-cs}
\end{equation}
Dividing by $C\tr\left(\Delta^2\right)>0$ and using \eref{eq:qfi} gives
\begin{equation}
 I_{\rho,\caM}(\Delta,\Delta)
 \ge\frac{\tr(\Delta^2)}{C}
 =\frac{J_\rho(\Delta,\Delta)}{2C}.
\end{equation}
The inequality also holds at $\Delta=0$. Taking the infimum over pure states and unit tangent directions gives \eref{eq:shadow-fi-bnd}. The argument only uses the variance bound for $\sigma=\rho$ and $O=\Delta\in T_\rho$.
\end{proof}

\subsection{Reconstruction variance bounds and proof of Table~\ref{tab:shadow-fi-ens}}\label{app:shadow-apps}

We consider exact projective $3$-designs~\cite{HuangKuengPreskill2020,InnocentiEtAl2023}, phase $3$-designs over mutually unbiased bases~\cite{YangChenZhu2026}, complete odd-prime qudit stabilizer measurements~\cite{MaoYiZhu2025,ZhuMaoYi2024}, and the equal mixture of equatorial-stabilizer and computational-basis measurements~\cite{ParkTeoJeong2025}. Write each rank-one POVM as $\caM=\{E_i=w_i\phi_i\}_i$. We use the canonical reconstruction obtained from the measurement channel~\cite{HuangKuengPreskill2020,InnocentiEtAl2023}:
\begin{equation}
 \mathcal C_{\caM}(X)=\sum_i w_i\tr(\phi_i X)\phi_i,
 \qquad \hrho_i=\mathcal C_{\caM}^{-1}(\phi_i),
 \qquad X\in\caL(\caH).
 \label{eq:canonical-shadow}
\end{equation}
In each case, the projectors $\phi_i$ span $\caL(\caH)$. Thus $\mathcal C_{\caM}$ is invertible: for $X\ne0$, the sum $\tr\left[X^\dagger\mathcal C_{\caM}(X)\right]=\sum_i w_i|\tr(\phi_iX)|^2$ is positive. The reconstruction therefore satisfies \eref{eq:shadow-mean}. Here $p_i(\sigma)=\tr(E_i\sigma)$ is the probability of outcome $i$.

\begin{lemma}[Reconstruction variance formulas and bounds]\label{lem:ensemble-variance}
For the following measurements, let $\hrho_i$ be the canonical reconstruction operators in \eref{eq:canonical-shadow}. For every $\sigma\in\caD(\caH)$ and $O\in\caL^{\rmH}_0(\caH)$, the following statements hold.
\begin{enumerate}
\renewcommand{\labelenumi}{(\arabic{enumi})}
\item\label{part:var-three-design} For an exact projective $3$-design [Eq.~(36) of \rcite{InnocentiEtAl2023}],
\begin{equation}
 \Var_\sigma[\tr(O\hrho_i)]
 =\frac{d+1}{d+2}\left[\tr\left(O^2\right)+2\tr\left(\sigma O^2\right)\right]-[\tr(\sigma O)]^2.
 \label{eq:3-des-rec-var}
\end{equation}
\item\label{part:var-phase} For phase $3$-designs over $L$ mutually unbiased reference bases [Theorem~2 of \rcite{YangChenZhu2026}], with $2\le L\le d+1$,
\begin{equation}
 \Var_\sigma[\tr(O\hrho_i)]\le\frac{L(3L+1)}{(L-1)^2}\norm{O}_2^2.
 \label{eq:var-phase}
\end{equation}
\item\label{part:var-qudit} For the complete qudit stabilizer measurement [Theorem~2 of \rcite{MaoYiZhu2025}] in dimension $d=q^n$, where $q$ is an odd prime and $n\ge1$,
\begin{equation}
 \Var_\sigma[\tr(O\hrho_i)]\le(2q-3)\norm{O}_2^2+2\norm{O}_\infty^2\le(2q-1)\norm{O}_2^2.
 \label{eq:var-qudit}
\end{equation}
\item\label{part:var-equatorial} For the equal-probability mixture of the complex equatorial-stabilizer and computational-basis measurements [Eqs.~(B17) and~(B18) of \rcite{ParkTeoJeong2025}] on $n$ qubits, with $n\ge1$ and $d=2^n$,
\begin{equation}
 \Var_\sigma[\tr(O\hrho_i)]\le30\norm{O}_2^2.
 \label{eq:var-equatorial}
\end{equation}
\end{enumerate}
\end{lemma}

\begin{proof}[Proof of the bounds in Table~\ref{tab:shadow-fi-ens}]
For the mixture in part~(\ref{part:var-equatorial}), the canonical reconstruction values for traceless $O$ are $2d\bra\phi O\ket\phi$ for an equatorial outcome and $2\bra xO\ket x$ for a computational-basis outcome. With setting probabilities $1/2$, the second-moment bounds in the cited equations give
\begin{equation*}
 \Var_\sigma[\tr(O\hrho_i)]
 \le\left(\frac12\cdot4\cdot14+\frac12\cdot4\right)\norm{O}_2^2
 =30\norm{O}_2^2.
\end{equation*}

Fix a pure state $\rho$ and $\Delta\in T_\rho$. Equation~\eqref{eq:tan} gives $\tr(\rho\Delta)=0$ and $\rho\Delta+\Delta\rho=\Delta$. Multiplying the latter identity by $\Delta$ and taking the trace gives $2\tr\left(\rho\Delta^2\right)=\tr\left(\Delta^2\right)$. Thus, for an exact projective $3$-design, substituting $O=\Delta$ and $\sigma=\rho$ into \eref{eq:3-des-rec-var} in \lref{lem:ensemble-variance} gives
\begin{equation}
 \Var_\rho[\tr(\Delta\hrho_i)]=\frac{2(d+1)}{d+2}\tr\left(\Delta^2\right)\le2\tr\left(\Delta^2\right).
 \label{eq:var-three-tangent}
\end{equation}
The tangent-space version of \pref{prop:shadow-fi}, noted after \eref{eq:shadow-fi-bnd} and proved in \aref{app:shadow-crit}, applies with variance constant $2(d+1)/(d+2)$. For phase $3$-designs and the equatorial-stabilizer mixture, parts~(\ref{part:var-phase}) and~(\ref{part:var-equatorial}) of \lref{lem:ensemble-variance} give variance constants $L(3L+1)/(L-1)^2$ and $30$. Applying \pref{prop:shadow-fi}, the corresponding Fisher guarantees for these three measurements are
\begin{equation}
 \frac{d+2}{4(d+1)},\qquad
 \frac{(L-1)^2}{2L(3L+1)},\qquad
 \frac1{60}.
\end{equation}
For the qudit stabilizer measurement, the rank-two tangent form in \eref{eq:tan} gives $2\norm{\Delta}_\infty^2=\tr\left(\Delta^2\right)$. Substituting $O=\Delta$ and $\sigma=\rho$ into \eref{eq:var-qudit} in \lref{lem:ensemble-variance}, we obtain
\begin{equation}
 \Var_\rho[\tr(\Delta\hrho_i)]
 \le(2q-3)\tr\left(\Delta^2\right)+2\norm{\Delta}_\infty^2
 =2(q-1)\tr\left(\Delta^2\right).
 \label{eq:var-qudit-tangent}
\end{equation}
The tangent-space version of \pref{prop:shadow-fi} therefore gives
\begin{equation}
 I_{\rho,\caM_{\rmstab(q,n)}}\succeq\frac{1}{4(q-1)}J_\rho.
\end{equation}
This also proves the qudit bound in \pref{prop:stab-fi}. At $L=2$, part~(\ref{part:var-phase}) of \lref{lem:ensemble-variance} gives variance constant $14$ and Fisher guarantee $1/28$. The exact multiqubit stabilizer entry follows from the qubit case of \pref{prop:stab-fi}, whose lower bound and attainment are proved in \aref{app:stab-fi}. Together, these establish all entries of Table~\ref{tab:shadow-fi-ens}.
\end{proof}

\section{Proofs of design-order guarantees}\label{app:hier}

In this appendix, we prove \thref{thm:hier} and \coref{cor:zero-wt}. The design bounds rely on polynomial moment inequalities, and finite radial constructions establish sharpness at second and odd orders.

The odd-order lower bound combines the endpoint kernel, Fisher-kernel identity, and polynomial minorant in Lemmas~\ref{lem:end}, \ref{lem:fi-kernel}, and~\ref{lem:minorant}. Attainment follows from the radial quadrature, finite lifting, and saturation argument in Lemmas~\ref{lem:rad-quad}, \ref{lem:fin-lift}, and~\ref{lem:sharp-sat}. At even orders, Cauchy--Schwarz retains a positive contribution from the polynomial remainder that was discarded at odd orders; Lemmas~\ref{lem:eq-nodes}, \ref{lem:even-int}, and~\ref{lem:even-fi} give the improved lower bound, while \lref{lem:even-zero} supplies the upper bound.

\subsection{Polynomial and moment tools}
\label{app:poly-tools}

\emph{Polynomial reproducing kernels.} Let $\mu$ be a probability measure on $[0,1]$ with positive density on $(0,1)$. For a nonnegative integer $k$, let $\scrP_k$ be the real polynomials of degree at most $k$, equipped with the inner product and norm
\begin{equation}
 \inner{p}{f}_{\mu}=\int_0^1 p(u)f(u)\,\mu(\rmd u),
 \qquad \norm{p}_{L^2(\mu)}^2=\inner{p}{p}_{\mu}.
 \label{eq:poly-inner-product}
\end{equation}
For any orthogonal basis $B_0,\ldots,B_k$ of $\scrP_k$, the reproducing kernel is
\begin{equation}
 K_k^{\mu}(u,z)=\sum_{\ell=0}^k
     \frac{B_\ell(u)B_\ell(z)}{\norm{B_\ell}_{L^2(\mu)}^2},
 \qquad
 \inner*{p}{K_k^{\mu}(\cdot,z)}_{\mu}=p(z)
 \quad(p\in\scrP_k).
 \label{eq:poly-kernel-def}
\end{equation}
The second identity follows by expanding $p$ in the orthogonal basis, and shows that the kernel does not depend on that basis. In particular,
\begin{equation}
 \norm*{K_k^{\mu}(\cdot,0)}_{L^2(\mu)}^2=K_k^{\mu}(0,0),
 \qquad
 \inner*{1}{K_k^{\mu}(\cdot,0)}_{\mu}=1.
 \label{eq:kernel-eval-ids}
\end{equation}
For every nonnegative integer $n$, define
\begin{equation}
 q_n^\mu(u):=\frac{K_n^\mu(u,0)}{K_n^\mu(0,0)}.
 \label{eq:normalized-poly-def}
\end{equation}
\emph{Shifted Jacobi polynomials.} For $a,b>0$, the beta function is
\begin{equation}
 \rmB(a,b)=\int_0^1 u^{a-1}(1-u)^{b-1}\rmd u
 =\frac{\Gamma(a)\Gamma(b)}{\Gamma(a+b)},
\end{equation}
where $\Gamma$ is the Euler gamma function. The $\Beta(a,b)$ distribution has density $u^{a-1}(1-u)^{b-1}/\rmB(a,b)$ on $[0,1]$. For $\alpha,\beta\ge0$ and a nonnegative integer $\ell$, define the Jacobi polynomial by Rodrigues' formula~\cite[Eq.~(18.5.5) and Table~18.5.1]{DLMF}:
\begin{equation}
 P_\ell^{(\alpha,\beta)}(x)
 =\frac{(-1)^\ell}{2^\ell\ell!(1-x)^\alpha(1+x)^\beta}
 \frac{\rmd^\ell}{\rmd x^\ell}
 \left[(1-x)^{\alpha+\ell}(1+x)^{\beta+\ell}\right],
 \qquad -1<x<1.
\end{equation}
The endpoint values are obtained by continuity. Set
\begin{equation}
 \mu_{\alpha,\beta}(\rmd u)
 =\frac{u^{\alpha}(1-u)^{\beta}}{\rmB(\alpha+1,\beta+1)}\rmd u,
 \qquad Q_\ell^{(\alpha,\beta)}(u)=P_\ell^{(\alpha,\beta)}(1-2u).
 \label{eq:jacobi-prob-wt}
\end{equation}

The endpoint value and squared norm are
\begin{align}
 Q_\ell^{(\alpha,\beta)}(0)&=\frac{(\alpha+1)_\ell}{\ell!},
 \label{eq:jacobi-end-val}\\
 \norm*{Q_\ell^{(\alpha,\beta)}}_{L^2(\mu_{\alpha,\beta})}^2
 &=\frac{\Gamma(\alpha+\beta+2)\Gamma(\ell+\alpha+1)\Gamma(\ell+\beta+1)}
 {\Gamma(\alpha+1)\Gamma(\beta+1)(2\ell+\alpha+\beta+1)\ell!\Gamma(\ell+\alpha+\beta+1)}.
 \label{eq:jacobi-end-norm}
\end{align}
These follow from the Jacobi normalization and orthogonality formulas in \rcite{DLMF} after the change of variable $x=1-2u$. A degree-$k$ orthogonal polynomial for a positive weight on $(0,1)$ has $k$ simple roots there; a sign-change proof is given in \lref{lem:eq-nodes}. Those roots, together with $0$ and $1$, will form the quadrature nodes.

For a positive integer $m$, we also write
\begin{equation}
 \mu_m:=\mu_{0,m-1},\qquad
 \mu_m(\rmd u)=m(1-u)^{m-1}\rmd u.
 \label{eq:radial-measure-family}
\end{equation}

\Needspace{5\baselineskip}
\begin{lemma}[Polynomial norm minimization]\label{lem:end}
Let $\mu$ be a probability measure on $[0,1]$ with positive density on $(0,1)$, and let $k\ge1$. The polynomial $q_k^\mu$ in \eref{eq:normalized-poly-def} uniquely attains the minimum
\begin{equation}
 \min_{\substack{q\in\scrP_k\\q(0)=1}}\bbE_\mu q^2=\frac1{K_k^\mu(0,0)},
 \label{eq:end-min}
\end{equation}
and
\begin{equation}
 \bbE_\mu q_k^\mu=\bbE_\mu \left(q_k^\mu\right)^2=\frac1{K_k^\mu(0,0)}.
 \label{eq:opt-mom}
\end{equation}
\end{lemma}
\begin{proof}[Proof of \lref{lem:end}]
For every $q\in\scrP_k$ with $q(0)=1$, the reproducing identity in \eref{eq:poly-kernel-def}, Cauchy--Schwarz, and \eref{eq:kernel-eval-ids} give
\begin{equation}
 1=\inner*{q}{K_k^\mu(\cdot,0)}_\mu^2
 \le\bbE_\mu q^2\,K_k^\mu(0,0).
\end{equation}
Equality holds only when $q$ is proportional to $K_k^\mu(\cdot,0)$; the constraint $q(0)=1$ fixes that polynomial as $q_k^\mu$. This proves \eref{eq:end-min}. Substituting \eref{eq:normalized-poly-def} into \eref{eq:kernel-eval-ids} gives both equalities in \eref{eq:opt-mom}.
\end{proof}

\begin{lemma}[Kernel for $\mu_d$]\label{lem:fi-kernel}
Let $d\ge1$ and $k\ge0$ be integers, and write $K_k=K_k^{\mu_d}$ for the kernel in \eref{eq:poly-kernel-def}. Set $Q_\ell(u)=Q_\ell^{(0,d-1)}(u)$. Then
\begin{equation}
 K_k(u,z)=\sum_{\ell=0}^k\frac{d+2\ell}{d}Q_\ell(u)Q_\ell(z),\qquad
 K_k(0,0)=\frac{(k+1)(d+k)}d.
 \label{eq:expl-fi-kernel}
\end{equation}
\end{lemma}
\begin{proof}[Proof of \lref{lem:fi-kernel}]
Equations~\eqref{eq:jacobi-end-val} and \eqref{eq:jacobi-end-norm} give $Q_\ell(0)=1$ and $\norm{Q_\ell}_{L^2(\mu_d)}^2=d/(d+2\ell)$. Substituting these squared norms into \eref{eq:poly-kernel-def} gives the kernel formula. Evaluating it at $u=z=0$ yields
\begin{equation}
 K_k(0,0)=\sum_{\ell=0}^k\frac{d+2\ell}{d}
 =\frac{(k+1)d+k(k+1)}d=\frac{(k+1)(d+k)}d.
\end{equation}
\end{proof}

\emph{Exact quadrature and finite moments.} A positive quadrature formula of degree $t$ is a finite set of nodes $u_j$ and weights $\alpha_j>0$ satisfying $\sum_j\alpha_jF(u_j)=\int F\,\rmd\mu$ for every $F\in\scrP_t$. The quadrature formula containing both $0$ and $1$ in \lref{lem:rad-quad} follows from polynomial division and interpolation; strict positivity of all its quadrature weights is established in the same proof. In every finite-dimensional complex Hilbert space and at every finite order $t$, a finite weighted exact projective $t$-design with positive weights exists by Tchakaloff's theorem~\cite{BayerTeichmann2006}.

\subsection{Auxiliary results on the design-order bounds}\label{app:univ-lb-bnd}

We establish the polynomial inequalities and finite design constructions used in \thref{thm:hier}. The complete proof of the theorem is collected in \aref{app:theorem-one-proof}.

For the design-order bounds and constructions below, $\caE$ denotes the design under consideration. For orthonormal kets $\ket\psi,\ket\nu$, write $\rho=\proj\psi$, $\Delta=\Delta_{\psi,\nu}$, and $G=G_{\psi,\caM_\caE}$. For each projector $\phi$ in $\caE$, use the quantities in \eref{eq:m-ov-fi}:
\begin{equation}
 u=\tr(\rho\phi),\qquad x=\tr(\Delta\phi).
 \label{eq:px-def}
\end{equation}

\Needspace{5\baselineskip}
For $k\ge2$, write $q_{k-1}=q_{k-1}^{\mu_d}$, with $q_n^\mu$ defined in \eref{eq:normalized-poly-def}.

\subsubsection{Auxiliary results on \texorpdfstring{$2$}{2}-designs}

\begin{lemma}\label{lem:2des-cert}
Let $d\ge2$, $\chi=\sqrt{d+1}$, and
\begin{equation}
 A=\frac{\chi(\chi+1)}2,\qquad B=\frac{\chi(\chi-1)}2,\qquad F(u,s)=Au+Bs-1.
 \label{eq:2des-cert-const}
\end{equation}
For $u,s\ge0$ with $u+s\le1$ and $0\le y\le s$,
\begin{equation}
 dy\ge\frac{c_{d,2}}{2}\left[(\chi+3)d(d+1)uy-(\chi-1)F(u,s)^2\right].
 \label{eq:2des-score-cert}
\end{equation}
\end{lemma}
\begin{proof}[Proof of \lref{lem:2des-cert}]
The certificate is affine in $y$ and uses only $uy$ and $F(u,s)^2$ on the right, whose averages are fixed by the second design moment; the four layers of Table~\ref{tab:2ext-layers} saturate it.
Put $c=(\chi-1)c_{d,2}/2$, $h=(\chi+3)c_{d,2}/2$, and $L=\chi(\chi+3)/2$.
Expanding $F(u,s)^2=(Au+Bs-1)^2$ and $(Au-Ls-1)^2$, using $A,B$ from \eref{eq:2des-cert-const} and $d=\chi^2-1$, gives
\begin{equation}
 ds+cF(u,s)^2-hd(d+1)us
 =c(Au-Ls-1)^2+2Bs(1-u-s)\ge0.
 \label{eq:2des-pos-cert}
\end{equation}
The right-hand side of \eref{eq:2des-pos-cert} is nonnegative because $c,B>0$ and $s,1-u-s\ge0$. For fixed $u,s$, $dy+cF(u,s)^2-hd(d+1)uy$ is affine in $y$ and nonnegative at both $y=0$ and $y=s$, by $cF(u,s)^2\ge0$ and \eref{eq:2des-pos-cert}, respectively. Hence \eref{eq:2des-score-cert} holds for every $0\le y\le s$.
\end{proof}

\begin{lemma}[Attainment of the $2$-design bound]\label{lem:2des-attain}
For every $d\ge2$, there exist a finite weighted exact projective $2$-design $\caE_*$ in $\caH$ with positive weights and orthonormal kets $\ket\psi,\ket\nu$ such that
\begin{equation}
 \inner{\nu}{G_{\psi,\caM_{\caE_*}}\nu}_{\bbR}=c_{d,2}.
 \label{eq:2des-attain-statement}
\end{equation}
\end{lemma}
\begin{proof}[Proof of \lref{lem:2des-attain}]
For $d\ge3$, fix orthonormal kets $\ket\psi,\ket\nu$. We construct a finite ensemble with four layers and verify its second tensor moment. The layer weights require $\chi=\sqrt{d+1}\lsp\ge2$, so we treat $d=2$ separately below.

\emph{The finite ensemble.} Set $\caK=\spanop\{\ket\psi,\ket \nu\}^{\perp}$, and write $r=d-2$ and $\chi=\sqrt{d+1}$. Choose a finite weighted exact projective $2$-design $\{(\gamma_a,\zeta_a)\}_a$ in $\caK$, with unit vectors $\ket{\zeta_a}$. Its existence follows from the finite-design existence result cited in \aref{app:poly-tools}; when $r=1$, a single projector suffices. For each row $j$ of Table~\ref{tab:2ext-layers}, the construction below specifies projectors with total ensemble weight $\alpha_j$ and squared overlaps $u_j,s_j$ with $\ket\psi,\ket\nu$, respectively. The individual projectors and their weights are given after the table.
\begin{table}[!ht]
\caption{Weights and squared overlaps for the extremal $2$-design construction in dimensions $d\ge3$, with $\chi=\sqrt{d+1}$.}
\label{tab:2ext-layers}
\centering
\begingroup
\setlength{\tabcolsep}{14pt}
\renewcommand{\arraystretch}{1.8}
\begin{tabular}{@{}cccc@{}}
\toprule
$j$ & $\alpha_j$ & $u_j$ & $s_j$\\
\midrule
$1$ & $\dfrac2{d^2}$ & $\dfrac{1+\chi^{-1}}2$ & $\dfrac{1-\chi^{-1}}2$\\
\addlinespace[4pt]
$2$ & $\dfrac{d+1}{2d}$ & $\dfrac1{d+1}$ & $\dfrac1{d+1}$\\
\addlinespace[4pt]
$3$ & $\dfrac{(\chi-2)(\chi+1)}{4(\chi-1)^2}$ & $\dfrac2{\chi(\chi+1)}$ & $0$\\
\addlinespace[4pt]
$4$ & $\dfrac{(\chi+2)(\chi-1)}{4(\chi+1)^2}$ & $0$ & $\dfrac2{\chi(\chi-1)}$\\
\bottomrule
\end{tabular}
\endgroup
\end{table}

For $d\ge3$, $\chi\ge2$, so $\alpha_j\ge0$, $\sum_j\alpha_j=1$, and $u_j+s_j\le1$. At $d=3$, $\alpha_3=0$, and the third row is omitted from the construction.

The first layer consists of
\begin{equation}
\ket{\phi_{1,\tau}}=\sqrt{u_1}\lsp\ket\psi+\tau\sqrt{s_1}\lsp\ket \nu,\qquad \pi_{1,\tau}=\frac{\alpha_1}2=\frac1{d^2},\qquad \tau\in\{-1,1\}.
 \label{eq:2ext-real}
\end{equation}
For $j=2,3,4$, let $\omega_p=\rme^{2\pi\rmi p/3}$, $p=0,1,2$, and take
\begin{equation}
\ket{\phi_{j,a,\tau,p}}=\sqrt{u_j}\lsp\ket\psi+\rmi\tau\sqrt{s_j}\lsp\ket \nu+\omega_p\sqrt{1-u_j-s_j}\lsp\ket{\zeta_a},\qquad \pi_{j,a,\tau,p}=\frac{\alpha_j\gamma_a}{6}.
 \label{eq:2ext-other}
\end{equation}
Only positive weights are retained. Coincident projectors may be merged by adding their weights. Equations~\eqref{eq:2ext-real} and~\eqref{eq:2ext-other} therefore specify a finite ensemble with positive weights.

\emph{Verification of the design identity.} For a random ket from this ensemble, write $u=|\braket{\psi}{\phi}|^2$, $s=|\braket{\nu}{\phi}|^2$, and $h=1-u-s$. Substitution of the four rows of Table~\ref{tab:2ext-layers} gives
\begin{equation}
\bbE_\pi u=\bbE_\pi s=\frac1d,\qquad \bbE_\pi u^2=\bbE_\pi s^2=\frac2{d(d+1)},\qquad \bbE_\pi us=\frac1{d(d+1)}.
 \label{eq:2ext-rad-mom}
\end{equation}
These identities also imply
\begin{equation}
\bbE_\pi h=\frac rd,\qquad \bbE_\pi uh=\bbE_\pi sh=\frac r{d(d+1)},\qquad \bbE_\pi h^2=\frac{r(r+1)}{d(d+1)}.
 \label{eq:2ext-comp-mom}
\end{equation}
To check the tensor entries, expand $\ket\phi^{\otimes2}$ according to the number of factors in $\caK$. Averaging $\omega_p$ removes blocks with different numbers of factors in $\caK$ on the ket and bra sides, because their difference is an integer of magnitude at most two. Within the remaining blocks, the sign average removes terms containing an odd total number of $\ket \nu$ and $\bra \nu$ factors. In the block with no factors in $\caK$, the remaining off-diagonal terms are proportional to $\ket\psi^{\otimes2}\bra\nu^{\otimes2}$ and its Hermitian conjugate. The contributions from the first and second layers cancel because
\begin{equation}
\alpha_1u_1s_1=\frac1{2d(d+1)}=\alpha_2u_2s_2,
 \label{eq:2ext-cancel}
\end{equation}
and the squared relative phases in these layers are $+1$ and $-1$, respectively. The third and fourth layers have $u_js_j=0$.

The remaining diagonal blocks are fixed by \eref{eq:2ext-rad-mom}, \eref{eq:2ext-comp-mom}, and the first two angular design moments. In particular, the coefficients of the normalized symmetric vectors $\ket\psi^{\otimes2}$, $\ket \nu^{\otimes2}$, and $(\ket\psi\ket \nu+\ket \nu\ket\psi)/\sqrt{2}\lsp$ are all $2/[d(d+1)]$. Each mixed block with one factor in $\caK$ has coefficient $2\bbE_\pi(uh)/r$ or $2\bbE_\pi(sh)/r$, also equal to $2/[d(d+1)]$. The block on $\Sym^2(\caK)$ has coefficient $\bbE_\pi h^2/[r(r+1)/2]$, with the same value. Thus
\begin{equation}
\bbE_\pi\phi^{\otimes2}=\frac2{d(d+1)}P_{[2]}=\frac{P_{[2]}}{D_{[2]}}.
 \label{eq:2ext-design}
\end{equation}
The first moment follows by partial trace, so the ensemble is an exact projective $2$-design.

\emph{Attainment.} At the state $\proj\psi$ and along $\ket \nu$, only the first layer contributes to the Fisher information. The second layer has imaginary relative phase, the third has zero $\nu$ component, and the fourth has zero probability at $\rho=\proj\psi$. \eref{eq:hor-fi} therefore gives
\begin{equation}
\inner{\nu}{G_{\psi,\caM_{\caE_*}}\nu}_{\bbR}=d\alpha_1s_1=\frac{1-(d+1)^{-1/2}}d=\frac1{d+1+\sqrt{d+1}\lsp}=c_{d,2}.
 \label{eq:2ext-attain}
\end{equation}

\emph{Dimension two.} At $d=2$, take the tetrahedral SIC POVM $\caM_{\rm tet}=\{E_i=\phi_i/2\}_{i=1}^4$. Choose unit vectors $\ket{\phi_1},\ket{\phi_2}$ for $\phi_1,\phi_2$, with $\braket{\phi_1}{\phi_2}=\gamma=1/\sqrt{3}$, and set
\begin{equation}
 \ket\psi=\frac{\ket{\phi_1}+\ket{\phi_2}}{\sqrt{2(1+\gamma)}\lsp},\qquad
 \ket\nu=\frac{\ket{\phi_1}-\ket{\phi_2}}{\sqrt{2(1-\gamma)}\lsp}.
\end{equation}
These kets are orthonormal. At $\rho=\proj\psi$, the direction $\Delta=\Delta_{\psi,\nu}$ satisfies
\begin{equation}
 p_1=p_2=\frac{1+\gamma}4,\qquad
 \tr(E_1\Delta)=-\tr(E_2\Delta)=\frac{\sqrt{1-\gamma^2}\lsp}2.
\end{equation}
The SIC overlaps in \eref{eq:sic-overlap} give $\tr(E_i\Delta)=0$ for $i=3,4$. Thus \eref{eq:normalized-fi-form} gives
\begin{equation}
 \inner{\nu}{G_{\psi,\caM_{\rm tet}}\nu}_{\bbR}
 =\frac14\sum_{i=1}^2\frac{[\tr(E_i\Delta)]^2}{p_i}
 =\frac{1-\gamma}2=c_{2,2}.
\end{equation}
This proves the claimed equality also at $d=2$.
\end{proof}

\subsubsection{Higher-order auxiliary lemmas}\label{app:hi-ord-tools}

\begin{lemma}[A Fisher bound from design moments]\label{lem:minorant}
Let $d\ge2$ and $t\ge2$ be integers, let $\caE=\{(\pi_i,\phi_i)\}_i$ be a finite weighted exact projective $t$-design in $\caH$, and let $\ket\psi,\ket \nu$ be orthonormal. Set $\rho=\proj\psi$, $\Delta=\Delta_{\psi,\nu}$, and $G=G_{\psi,\caM_\caE}$. Recall the measure $\mu_d$ in \eref{eq:m-rad-mu}.
Suppose that $r$ is a real polynomial with $r(0)=1$, $r(u)\ge0$ for $0\le u\le1$, and $\deg r\le t-1$. Then
\begin{equation}
\inner{\nu}{G\nu}_{\bbR}\ge\frac12\left[1-\bbE_{\mu_d}r(u)\right].
 \label{eq:min-bnd}
\end{equation}
More precisely, write $u_i=\tr(\rho\phi_i)$ and $x_i=\tr(\Delta\phi_i)$ for each projector $\phi_i$ in $\caE$. Then
\begin{equation}
\inner{\nu}{G\nu}_{\bbR}=\frac12\left[1-\bbE_{\mu_d}r(u)\right]+\frac d4\sum_{i\colon u_i>0}\pi_i\frac{x_i^2r(u_i)}{u_i}.
 \label{eq:min-rem}
\end{equation}
For every real polynomial $h$ of degree at most $t-2$, we also have
\begin{equation}
 \bbE_\pi\left[x^2h(u)\right]=\frac2d\int_0^1u h(u)\,\mu_d(\rmd u).
 \label{eq:design-radial-moment}
\end{equation}
\end{lemma}

\begin{lemma}[Jacobi expressions and roots]\label{lem:eq-nodes}
Let $m\ge1$ and $n\ge1$ be integers. For $q_n^{\mu_m}$ and $\mu_m$ defined in Eqs.~\eqref{eq:normalized-poly-def} and~\eqref{eq:radial-measure-family},
\begin{equation}
 q_n^{\mu_m}(u)=\frac{P_n^{(1,m-1)}(1-2u)}{n+1}.
 \label{eq:expl-eq-poly}
\end{equation}
This polynomial has $n$ simple roots in $(0,1)$ and satisfies
\begin{equation}
 \int_0^1u q_n^{\mu_m}(u)p(u)\,\mu_m(\rmd u)=0
 \qquad(p\in\scrP_{n-1}).
 \label{eq:qn-orth}
\end{equation}
In particular, for $d,k\ge2$, the two choices used at odd and even orders are
\begin{equation}
 q_{k-1}^{\mu_d}(u)=\frac{P_{k-1}^{(1,d-1)}(1-2u)}k,\qquad
 q_{k-1}^{\mu_{d+1}}(u)=\frac{P_{k-1}^{(1,d)}(1-2u)}k.
 \label{eq:odd-even-jacobi}
\end{equation}
\end{lemma}

\begin{lemma}[Even-order polynomial integrals]\label{lem:even-int}
Let $d\ge2$ and $k\ge2$ be integers, and write $q=q_{k-1}^{\mu_{d+1}}$, with $q_n^\mu$ and $\mu_m$ defined in Eqs.~\eqref{eq:normalized-poly-def} and~\eqref{eq:radial-measure-family}. Then
\begin{align}
\int_0^1(1-u)q(u)^2\,\mu_d(\rmd u)&=\frac d{k(d+k)},
 \label{eq:even-loss}\\
\int_0^1u(1-u)^{d-1}q(u)^2\,\rmd u&=\frac1{dk(d+k)},
 \label{eq:even-int-one}\\
\int_0^1u^2q(u)^2\,\mu_{d-1}(\rmd u)&=\frac2{d(d+1)}.
 \label{eq:even-int-two}
\end{align}
\end{lemma}

\begin{lemma}[Even-order Fisher lower bound]\label{lem:even-fi}
Let $d,k\ge2$, let $t\ge2k$, and let $\caE$ be a finite weighted exact projective $t$-design in $\caH$. At every pure state $\proj\psi$,
\begin{equation}
 G_{\psi,\caM_\caE}\succeq c_{d,2k}\IdTan\psi,
 \label{eq:even-fi-bound}
\end{equation}
where $c_{d,2k}$ is defined in \eref{eq:even-coef}.
\end{lemma}

\subsubsection{Proofs of the higher-order auxiliary lemmas}\label{app:hi-ord-aux-prfs}

\begin{proof}[Proof of \lref{lem:minorant}]
After fixing its global phase, a Haar ket has the decomposition
\begin{equation}
 \ket\phi=\sqrt{u}\lsp\ket\psi+\sqrt{1-u}\lsp\rme^{\rmi\theta}\ket\xi,
 \label{eq:m-haar-dec}
\end{equation}
where $u$ has the $\Beta(1,d-1)$ density $(d-1)(1-u)^{d-2}$ on $[0,1]$, $\theta$ is uniform on $[0,2\pi)$, and $\ket\xi$ is Haar distributed on the unit sphere of $\caH_\psi$, independently. At fixed $u$,
\begin{equation}
 \bbE_\Ha\left[x^2\mid u\right]=4u(1-u)\bbE_{\theta,\xi}\left\{\left[\Rea\left(\rme^{-\rmi\theta}\braket{\xi}{\nu}\right)\right]^2\right\}=\frac{2u(1-u)}{d-1}.
 \label{eq:m-haar-cond}
\end{equation}
Phase averaging contributes $1/2$, and the angular first moment gives $\bbE_\xi|\braket{\xi}{\nu}|^2=1/(d-1)$.

For a real polynomial $h$ of degree at most $t-2$, $x^2h(u)$ has degree at most $t$ in $\phi$. Its ensemble expectation therefore equals its Haar expectation by \eref{eq:des}. The conditional Haar moment in \eref{eq:m-haar-cond}, together with the density in \eref{eq:m-rad-mu}, gives
\begin{equation}
 \bbE_\pi\left[x^2h(u)\right]=\bbE_\Ha\left[x^2h(u)\right]
 =\frac2d\int_0^1u h(u)\,\mu_d(\rmd u),
\end{equation}
proving \eref{eq:design-radial-moment}.

Now take $h(u)=[1-r(u)]/u$, with its polynomial value at zero. Since $r(0)=1$, this polynomial has degree at most $t-2$. For $u>0$, substitute $1/u=h(u)+r(u)/u$ into \eref{eq:hor-fi}. At $u=0$, $x=0$, so $x^2h(u)=0$. Thus
\begin{equation}
 \inner{\nu}{G\nu}_{\bbR}=\frac d4\bbE_\pi\left[x^2h(u)\right]
 +\frac d4\sum_{i\colon u_i>0}\pi_i\frac{x_i^2r(u_i)}{u_i}.
 \label{eq:min-split}
\end{equation}
Substituting \eref{eq:design-radial-moment} and $uh(u)=1-r(u)$ gives \eref{eq:min-rem}. The sum in \eref{eq:min-rem} is nonnegative because $\pi_i>0$, $u_i>0$, and $r(u_i)\ge0$, which proves \eref{eq:min-bnd}.
\end{proof}

\begin{proof}[Proof of \lref{lem:eq-nodes}]
Write $q=q_n^{\mu_m}$. The reproducing identity in \eref{eq:poly-kernel-def} gives
\begin{equation}
 \inner{q}{f}_{\mu_m}=\frac{f(0)}{K_n^{\mu_m}(0,0)}
 \qquad(f\in\scrP_n).
 \label{eq:normalized-poly-reproduction}
\end{equation}
Taking $f(u)=up(u)$ proves \eref{eq:qn-orth}. The degree of $q$ is exactly $n$: otherwise $p=q$ would be admissible and \eref{eq:qn-orth} would give $\int_0^1u q(u)^2\,\mu_m(\rmd u)=0$, contradicting $q(0)=1$. The degree-$n$ orthogonal polynomial for the positive weight $u(1-u)^{m-1}$ is unique up to scale. Since $P_n^{(1,m-1)}(1)=n+1$ by \eref{eq:jacobi-end-val}, this proves \eref{eq:expl-eq-poly}.

To locate its roots, suppose $q$ has only $\ell<n$ sign changes in $(0,1)$, at $z_1,\ldots,z_\ell$. Choose the sign of $p(u)=\pm\prod_{j=1}^\ell(u-z_j)$ so that $q(u)p(u)\ge0$ on $(0,1)$. This product is not identically zero, so its integral against the positive weight $u\mu_m$ is strictly positive, contrary to \eref{eq:qn-orth}. Thus $q$ has $n$ sign changes, and its degree implies that these are exactly $n$ simple roots in $(0,1)$. Setting $(n,m)=(k-1,d)$ and $(k-1,d+1)$ in \eref{eq:expl-eq-poly} gives \eref{eq:odd-even-jacobi}.
\end{proof}

\begin{proof}[Proof of \lref{lem:even-int}]
For $q=q_{k-1}^{\mu_{d+1}}$, \eref{eq:opt-mom} in \lref{lem:end}, \eref{eq:expl-fi-kernel} in \lref{lem:fi-kernel}, and the densities in \eref{eq:radial-measure-family} give
\begin{equation}
 \int_0^1(1-u)q(u)^2\,\mu_d(\rmd u)
 =\frac d{d+1}\bbE_{\mu_{d+1}}q^2
 =\frac d{k(d+k)}.
 \label{eq:even-loss-proof}
\end{equation}
This proves \eref{eq:even-loss}. Equation~\eqref{eq:qn-orth} in \lref{lem:eq-nodes}, with $(n,m)=(k-1,d+1)$, also gives
\begin{equation}
\int_0^1u(1-u)^dq(u)p(u)\,\rmd u=0\qquad(\deg p\le k-2).
 \label{eq:even-orth}
\end{equation}
Put $p=q'$ in \eref{eq:even-orth} and integrate the derivative of $u(1-u)^dq(u)^2$. Its endpoint values vanish, so
\begin{equation}
0=\int_0^1(1-u)^dq(u)^2\,\rmd u-d\int_0^1u(1-u)^{d-1}q(u)^2\,\rmd u.
 \label{eq:even-ibp-one}
\end{equation}
\eref{eq:even-loss} now proves \eref{eq:even-int-one}.

For the final integral, differentiate the Jacobi expression for $q$ in \eref{eq:odd-even-jacobi} to obtain
\begin{equation}
q'(u)=-\frac{k+d+1}{k}P_{k-2}^{(2,d+1)}(1-2u),\qquad b:=\frac{q'(1)}{q(1)}=\frac{(k-1)(k+d+1)}{d+1}.
 \label{eq:even-end-der}
\end{equation}
The first identity follows from the Jacobi derivative formula [Eq.~(18.9.15) of \rcite{DLMF}] and the chain rule. The last equality follows by evaluating the Jacobi polynomials at $-1$, using the symmetry and special values in Table~18.6.1 of \rcite{DLMF}. The roots in \lref{lem:eq-nodes} imply $q(1)\ne0$. The quotient $[uq'(u)-bq(u)]/(1-u)$ is a polynomial of degree at most $k-2$: its numerator vanishes at one by the definition of $b$. Applying \eref{eq:even-orth} to this quotient yields
\begin{equation}
\int_0^1u^2(1-u)^{d-1}q(u)q'(u)\,\rmd u=b\int_0^1u(1-u)^{d-1}q(u)^2\,\rmd u.
 \label{eq:even-der-mom}
\end{equation}
Integrate the derivative of $u^2(1-u)^{d-1}q(u)^2$, whose endpoints again vanish since $d\ge2$, and use \eref{eq:even-der-mom}. The result is
\begin{equation}
(d-1)\int_0^1u^2(1-u)^{d-2}q(u)^2\,\rmd u=2(1+b)\int_0^1u(1-u)^{d-1}q(u)^2\,\rmd u=\frac2{d(d+1)}.
 \label{eq:even-ibp-two}
\end{equation}
The last equality uses \eref{eq:even-int-one} and $1+b=k(d+k)/(d+1)$. This proves \eref{eq:even-int-two}.
\end{proof}

\begin{proof}[Proof of \lref{lem:even-fi}]
Fix orthonormal kets $\ket\psi,\ket\nu$ and use the notation $G,u_i,x_i$ of \lref{lem:minorant}.
Set $e_{d,k}:=(1/2)\{1-d/[k(d+k)]\}$ and write $q=q_{k-1}^{\mu_{d+1}}$. The polynomial $(1-u)q(u)^2$ has degree $2k-1$, is nonnegative on $[0,1]$, and equals one at zero. Substituting $r(u)=(1-u)q(u)^2$ into \eref{eq:min-rem} and using \eref{eq:even-loss} gives
\begin{align}
 \frac12\left\{1-\bbE_{\mu_d}\left[(1-u)q(u)^2\right]\right\}
 &=\frac12\left[1-\frac d{k(d+k)}\right]=e_{d,k},
 \label{eq:even-constant}\\
 \inner{\nu}{G\nu}_{\bbR}&=e_{d,k}+\frac d4\sum_{i\colon u_i>0}\pi_i\frac{x_i^2(1-u_i)q(u_i)^2}{u_i}.
 \label{eq:even-fisher-expression}
\end{align}
Define
\begin{equation}
f=xq(u)\sqrt{\frac{1-u}{u}},\qquad g=xq(u)\sqrt{\frac{u}{1-u}}\lsp\quad(0<u<1),\qquad f=g=0\quad(u\in\{0,1\}).
 \label{eq:even-factor}
\end{equation}
At $u=0$ and $u=1$, $x=0$. Substituting the definition of $f$ in \eref{eq:even-factor} into the second term on the right-hand side of \eref{eq:min-rem} gives
\begin{equation}
\inner{\nu}{G\nu}_{\bbR}=e_{d,k}+\frac d4\bbE_\pi f^2.
 \label{eq:even-rem}
\end{equation}
For every projector $\phi$ in $\caE$, $fg=x^2q(u)^2$. With $s=|\braket{\phi}{\nu}|^2$, the orthonormality of $\ket\psi$ and $\ket\nu$ gives $u+s\le1$, while $x=2\Rea(\braket{\psi}{\phi}\braket{\phi}{\nu})$ gives $x^2\le4us\le4u(1-u)$. Hence $g^2\le4u^2q(u)^2$. The polynomials on the right have degree at most $2k$. Equation~\eqref{eq:design-radial-moment} in \lref{lem:minorant}, the $t$-design property in \eref{eq:des}, and \lref{lem:even-int} give
\begin{equation}
\bbE_\pi(fg)=2\int_0^1u(1-u)^{d-1}q(u)^2\,\rmd u=\frac2{dk(d+k)},\qquad \bbE_\pi g^2\le4\bbE_{\mu_{d-1}}\left[u^2q(u)^2\right]=\frac8{d(d+1)}.
 \label{eq:even-fixed-mom}
\end{equation}
Equation~\eqref{eq:even-fixed-mom} gives $\bbE_\pi(fg)>0$, so $\bbE_\pi g^2>0$. Cauchy--Schwarz applied to $f$ and $g$ gives
\begin{equation}
\bbE_\pi f^2\ge\frac{\left[\bbE_\pi(fg)\right]^2}{\bbE_\pi g^2}\ge\frac{d+1}{2dk^2(d+k)^2}.
 \label{eq:even-pos-gap}
\end{equation}
Substituting \eref{eq:even-pos-gap} into \eref{eq:even-rem} gives $\inner{\nu}{G\nu}_{\bbR}\ge c_{d,2k}$ for every unit horizontal ket, hence $G\succeq c_{d,2k}\IdTan\psi$.
This proves \eref{eq:even-fi-bound}.
\end{proof}

\subsubsection{Radial quadrature and finite-design constructions}\label{app:fin-ext}

We construct finite designs attaining the odd-order Fisher lower bound and the bounds on the zero-probability weight $\alpha_0(\rho)$, and establishing the even-order Fisher upper bound. We use positive quadrature formulas: nodes $u_j$ and weights $\alpha_j>0$ whose weighted sums reproduce the integrals of all polynomials through the specified degree, as defined in \aref{app:poly-tools}. We choose such formulas for $\mu_{d-1}$. A finite phase average removes terms with unequal numbers of $\ket\psi$ and $\bra\psi$ factors when the tensor power of each rank-one projector is expanded. A projective design in $\caH_\psi$ gives the tensor moments within that subspace.

\phantomsection\label{app:ext-rad-data}\label{app:ext-fin-lift}

\Needspace{14\baselineskip}
\begin{lemma}[Positive radial quadrature]\label{lem:rad-quad}
Let $d,k\ge2$, and let $u_1<\cdots<u_{k-1}$ be the roots of $q_{k-1}^{\mu_d}$ in $(0,1)$. There are positive weights $\alpha_0,\ldots,\alpha_k$ on the nodes
\begin{equation}
 0=u_0<u_1<\cdots<u_{k-1}<u_k=1,
 \label{eq:rad-nodes}
\end{equation}
such that
\begin{equation}
 \sum_{j=0}^{k}\alpha_jF(u_j)=\int_0^1F(u)\,\mu_{d-1}(\rmd u)
 \label{eq:rad-quad}
\end{equation}
for every polynomial $F$ of degree at most $2k-1$. The endpoint weight
at zero is
\begin{equation}
 \alpha_0=\frac{d-1}{k(d+k-1)}.
 \label{eq:a0-val}
\end{equation}
\end{lemma}

\begin{lemma}[Finite designs from a quadrature formula]\label{lem:fin-lift}
Let $d\ge2$ and $t\ge1$, fix a unit ket $\ket\psi$, and let $\{(\alpha_j,u_j)\}_j$ be a positive quadrature formula for $\mu_{d-1}$ exact through degree $t$, containing $u_0=0$. Choose a finite weighted projective $t$-design $\{(\gamma_a,\varphi_a)\}_a$ in $\caH_\psi$, with unit vectors $\ket{\varphi_a}$, and set $L=t+1$ and $\omega_p=\rme^{2\pi\rmi p/L}$. For $0<u_j<1$, take
\begin{equation}
 \ket{\phi_{j,a,p}}=\sqrt{u_j}\lsp\ket\psi+\sqrt{1-u_j}\lsp\omega_p\ket{\varphi_a},\qquad
 \pi_{j,a,p}=\frac{\alpha_j\gamma_a}{L},\qquad 0\le p<L.
 \label{eq:fin-lift-states}
\end{equation}
At $u=0$, take the projectors $\varphi_a$ with weights $\alpha_0\gamma_a$; at $u=1$, if present, take $\proj\psi$ with that node's weight. This ensemble is a finite weighted exact projective $t$-design in $\caH$, with total weight $\alpha_0$ on $\caH_\psi$.
\end{lemma}

\begin{lemma}[Equality conditions at odd orders]\label{lem:sharp-sat}
Let $d\ge2$ and $k\ge2$, let $\caE=\{(\pi_i,\phi_i)\}_i$ be a finite weighted exact projective $(2k-1)$-design in $\caH$ with positive weights, and let $\rho=\proj\psi$ be a pure state. Let $u_1^*<\cdots<u_{k-1}^*$ be the roots of $q_{k-1}^{\mu_d}$ in $(0,1)$. The following conditions are equivalent:
\begin{equation}
 \begin{aligned}
 \text{(i)}\quad &G_{\psi,\caM_\caE}=c_{d,2k-1}\IdTan\psi;\\
 \text{(ii)}\quad &\alpha_0(\rho)=\frac{d-1}{k(d+k-1)};\\
 \text{(iii)}\quad &\tr(\rho\phi_i)\in\left\{0,u_1^*,\ldots,u_{k-1}^*,1\right\}
                    \quad\text{for every }i.
 \end{aligned}
 \label{eq:odd-saturation-equivalence}
\end{equation}
\end{lemma}

\begin{lemma}[Sharp even-order zero-probability weight]\label{lem:even-zero}
Let $d\ge2$ and $k\ge1$ be integers. For every finite weighted exact projective $t$-design $\caE=\{(\pi_i,\phi_i)\}_i$ with $t\ge2k$ and every pure state $\rho$,
\begin{equation}
\alpha_0(\rho)\le\frac{d-1}{(k+1)(d+k-1)}.
 \label{eq:zero-even}
\end{equation}
At order $t=2k$, the bound is attained by a finite weighted exact projective $2k$-design and a pure state.
\end{lemma}

\subsubsection{Proofs of the radial auxiliary lemmas}\label{app:ext-aux-prfs}

\begin{proof}[Proof of \lref{lem:rad-quad}]
For the nodes $u_0,\ldots,u_k$ in \eref{eq:rad-nodes} and $0\le j\le k$, define
\begin{equation}
 h_j(u):=\prod_{\substack{0\le\ell\le k\\\ell\ne j}}\frac{u-u_\ell}{u_j-u_\ell},\qquad
 \alpha_j:=\int_0^1h_j(u)\,\mu_{d-1}(\rmd u).
 \label{eq:rad-interp-wts}
\end{equation}
Then $\deg h_j\le k$ and $h_j(u_\ell)=\delta_{j\ell}$ for $0\le j,\ell\le k$. The weights $\alpha_j$ are independent of the polynomial to be integrated. For $\deg F\le2k-1$, division by the degree-$(k+1)$ polynomial $u(1-u)q_{k-1}(u)$ gives
\begin{equation}
 F(u)=u(1-u)q_{k-1}(u)A(u)+B(u),\qquad
 \deg A\le k-2,\quad\deg B\le k.
\end{equation}
The first term has zero integral by \eref{eq:qn-orth} and $u(1-u)\mu_{d-1}=(d-1)u\mu_d/d$, and vanishes at every node. The remainder satisfies $B(u_j)=F(u_j)$, so interpolation gives $B(u)=\sum_j F(u_j)h_j(u)$. Integrating this identity proves \eref{eq:rad-quad}.

To prove that every quadrature weight $\alpha_j$ is strictly positive, consider the following polynomials:
\begin{align}
 L_0(u)&=(1-u)q_{k-1}(u)^2,\\
 L_k(u)&=\frac{u q_{k-1}(u)^2}{q_{k-1}(1)^2},\\
 L_j(u)&=\frac{u(1-u)}{u_j(1-u_j)}
 \left[\frac{q_{k-1}(u)}{(u-u_j)q_{k-1}'(u_j)}\right]^2,
 \qquad 1\le j\le k-1.
 \label{eq:pos-card}
\end{align}
The apparent quotient in $L_j$ is a polynomial because $u_j$ is a simple root. The endpoint polynomials have degree $2k-1$ and the internal ones have degree $2k-2$, so \eref{eq:rad-quad} applies to each $L_j$. Each is nonnegative on $[0,1]$ and has values $L_j(u_\ell)=\delta_{j\ell}$. It follows that $\alpha_j=\int L_j\,\rmd\mu_{d-1}>0$: the density is positive on $(0,1)$ and no $L_j$ is identically zero.

For the endpoint $u=0$, \eref{eq:opt-mom} and $(1-u)\mu_{d-1}=(d-1)\mu_d/d$ give
\begin{equation}
 \alpha_0=\int(1-u)q_{k-1}(u)^2\,\mu_{d-1}(\rmd u)
 =\frac{d-1}{d}\int q_{k-1}(u)^2\,\mu_d(\rmd u)
 =\frac{d-1}{dK_{k-1}(0,0)}
 =\frac{d-1}{k(d+k-1)}.
\end{equation}
This formula is the Gauss--Lobatto quadrature for the Haar radial measure, including both endpoints~\cite[Sec.~1.4.2]{Gautschi2004}.
\end{proof}

\begin{proof}[Proof of \lref{lem:fin-lift}]
A finite weighted exact projective $t$-design in $\caH_\psi\simeq\bbC^{d-1}$ exists by the finite-design existence result cited in \aref{app:poly-tools}; at $d=2$, a single projector suffices. The construction in \eref{eq:fin-lift-states}, together with the endpoint projectors in the lemma, has positive weights summing to one.

Fix $1\le s\le t$ and an interior node $u=u_j$. Using the symmetric projector $P_{[s]}$ in \eref{eq:des}, we expand
\begin{equation}
 \phi_{j,a,p}^{\otimes s}
 =\sum_{\ell,\ell'=0}^s\binom{s}{\ell}\binom{s}{\ell'}
 u^{(\ell+\ell')/2}(1-u)^{s-(\ell+\ell')/2}\omega_p^{\ell'-\ell}P_{[s]}
 \left(\ket\psi^{\otimes\ell}\otimes\ket{\varphi_a}^{\otimes(s-\ell)}\right)
 \left(\bra\psi^{\otimes\ell'}\otimes\bra{\varphi_a}^{\otimes(s-\ell')}\right)P_{[s]}.
 \label{eq:fin-lift-tensor-expansion}
\end{equation}
Since $0\le\ell,\ell'\le s<L$, averaging over $p$ gives
\begin{equation}
 \frac1L\sum_{p=0}^{L-1}\omega_p^{\ell'-\ell}
 =\delta_{\ell\ell'}\qquad(0\le\ell,\ell'\le s).
\end{equation}
Let $\Pi_\ell$ be the orthogonal projector onto the symmetric tensors with exactly $\ell$ factors along $\ket\psi$ and $s-\ell$ factors in $\caH_\psi$. Its dimension is $\binom{d+s-\ell-2}{s-\ell}$. The angular design identity applies at order $s-\ell\le t$, so averaging \eref{eq:fin-lift-tensor-expansion} over the phases and angular design gives
\begin{equation}
 \bbE_{a,p}\phi_{j,a,p}^{\otimes s}
 =\sum_{\ell=0}^s\frac{\binom{s}{\ell}u^\ell(1-u)^{s-\ell}}{\binom{d+s-\ell-2}{s-\ell}}\Pi_\ell.
 \label{eq:lift-angular-blocks}
\end{equation}
Here the factor $\binom{s}{\ell}$ follows from the normalization of the symmetric tensors: applying $P_{[s]}$ to a tensor with $\ell$ copies of $\ket\psi$ and $s-\ell$ copies of an orthogonal unit ket gives squared norm $1/\binom{s}{\ell}$. Equation~\eqref{eq:lift-angular-blocks} also holds for the endpoint projectors specified in the lemma. The radial integral is
\begin{equation}
 \int_0^1\binom{s}{\ell}u^\ell(1-u)^{s-\ell}\,\mu_{d-1}(\rmd u)
 =\frac{\binom{d+s-\ell-2}{s-\ell}}{\binom{d+s-1}{s}}
 \qquad(0\le\ell\le s).
 \label{eq:lift-sector-weight}
\end{equation}
The quadrature is exact for these degree-$s$ polynomials because $s\le t$. Substituting their integrals into \eref{eq:lift-angular-blocks} gives
\begin{equation*}
 \bbE_\pi\phi^{\otimes s}
 =\frac1{D_{[s]}}\sum_{\ell=0}^s\Pi_\ell
 =\frac{P_{[s]}}{D_{[s]}},
\end{equation*}
where the last equality uses the orthogonal decomposition of $\Sym^s(\caH)$ into the ranges of $\Pi_\ell$. This proves the design identity for every $s\le t$. Only the $u=0$ projectors are orthogonal to $\ket\psi$, and their total weight is $\alpha_0$.
\end{proof}

\begin{proof}[Proof of \lref{lem:sharp-sat}]
\emph{Trace condition.} Equations~\eqref{eq:opt-mom}, \eqref{eq:expl-fi-kernel}, and~\eqref{eq:min-bnd} in Lemmas~\ref{lem:end}, \ref{lem:fi-kernel}, and~\ref{lem:minorant}, respectively, together with the trace identity \eref{eq:trace-defect} in \pref{prop:geometry} give
\begin{equation}
 G_{\psi,\caM_\caE}-c_{d,2k-1}\IdTan\psi\succeq0,\qquad
 \tr_{\bbR}G_{\psi,\caM_\caE}=d-1-d\alpha_0(\rho).
 \label{eq:odd-sat-trace}
\end{equation}
Since $(2d-2)c_{d,2k-1}=d-1-d(d-1)/[k(d+k-1)]$, conditions (i) and (ii) are equivalent: the operator $G_{\psi,\caM_\caE}-c_{d,2k-1}\IdTan\psi$ is positive semidefinite by \eref{eq:odd-sat-trace}, and such an operator is zero if and only if its trace is zero.

\emph{Equivalence of (ii) and (iii).} Put $u_i=\tr(\rho\phi_i)$ and $p(u)=(1-u)q_{k-1}(u)^2$. This polynomial has degree $2k-1$, is nonnegative on $[0,1]$, and satisfies $p(0)=1$ and $p(1)=0$. The design property \eref{eq:des}, the densities in \eref{eq:radial-measure-family}, and Lemmas~\ref{lem:end} and~\ref{lem:fi-kernel} give
\begin{equation}
 \sum_i\pi_i p(u_i)=\int_0^1(1-u)q_{k-1}(u)^2\,\mu_{d-1}(\rmd u)
 =\frac{d-1}{d}\bbE_{\mu_d}q_{k-1}^2=\frac{d-1}{k(d+k-1)}.
 \label{eq:odd-zero-polynomial-mean}
\end{equation}
Separating the outcomes with $u_i=0$ and using $p(1)=0$ yields
\begin{equation}
 \frac{d-1}{k(d+k-1)}-\alpha_0(\rho)
 =\sum_{i\colon 0<u_i<1}\pi_i(1-u_i)q_{k-1}(u_i)^2.
 \label{eq:odd-zero-weight-gap}
\end{equation}
Each summand in \eref{eq:odd-zero-weight-gap} is nonnegative, and its coefficient $\pi_i(1-u_i)$ is strictly positive. Hence (ii) holds if and only if every overlap in $(0,1)$ is a root of $q_{k-1}$, which is condition (iii) by \lref{lem:eq-nodes}. Together with (i)$\Leftrightarrow$(ii), this proves the equivalence of all three conditions.

\end{proof}

\begin{proof}[Proof of \lref{lem:even-zero}]
Fix $\rho=\proj\psi$ and put $u=\tr(\rho\phi)$. For every real polynomial $q$ of degree at most $k$ with $q(0)=1$, the pointwise inequality $\bfone_{\{u=0\}}\le q(u)^2$ and the design moment imply
\begin{equation}
\alpha_0(\rho)\le\bbE_\pi q(u)^2=\int_0^1q(u)^2\,\mu_{d-1}(\rmd u).
 \label{eq:zero-poly}
\end{equation}
Write $q_k=q_k^{\mu_{d-1}}$, as defined in \eref{eq:normalized-poly-def}. Equation~\eqref{eq:expl-eq-poly} gives its Jacobi expression, and \eref{eq:end-min} in \lref{lem:end} together with \eref{eq:expl-fi-kernel} in \lref{lem:fi-kernel} gives
\begin{equation}
\min_{\substack{\deg q\le k\\q(0)=1}}\int q(u)^2\,\mu_{d-1}(\rmd u)=\frac{d-1}{(k+1)(d+k-1)}.
 \label{eq:zero-kernel}
\end{equation}
Equations~\eqref{eq:zero-poly} and~\eqref{eq:zero-kernel} prove the bound.

For attainment, use the fixed node $u_0=0$ and the $k$ roots $u_1,\ldots,u_k$ of $q_k$ in $(0,1)$. These roots are simple because $q_k$ is orthogonal for the positive weight $u\mu_{d-1}$. Define weights by integrating the Lagrange cardinal polynomials for these $k+1$ nodes. For a polynomial $f$ of degree at most $2k$, division gives
\begin{equation}
f(u)=u q_k(u)h(u)+r(u),\qquad \deg h\le k-1,\qquad \deg r\le k.
 \label{eq:radau-div}
\end{equation}
The first term has zero integral by orthogonality and vanishes at all nodes; interpolation integrates the second term exactly. Thus the quadrature is exact through degree $2k$.

Its weights are positive. At zero use the nonnegative cardinal polynomial $q_k(u)^2$. At an interior node use
\begin{equation}
\ell_j(u)=\frac{u}{u_j}\left[\frac{q_k(u)}{(u-u_j)q_k'(u_j)}\right]^2,\qquad 1\le j\le k.
 \label{eq:radau-card}
\end{equation}
The quotient is assigned its polynomial value at $u=u_j$. These nonnegative cardinal polynomials have degrees at most $2k$ and take value one at their own node and zero at the others. Their integrals against the positive radial density are strictly positive. The quadrature weight at $u=0$ equals the minimum in \eref{eq:zero-kernel}.

Finally, combine this quadrature with a finite weighted exact projective $2k$-design in $\caH_\psi$ and $2k+1$ equally spaced relative phases. At $u=0$, use only projectors onto kets in $\caH_\psi$. At each positive node use the kets in \eref{eq:fin-lift-states}, with the new radial weights and phase count. By \lref{lem:fin-lift}, this ensemble is an exact projective $2k$-design. The total ensemble weight of projectors orthogonal to $\ket\psi$ is precisely the quadrature weight at $u=0$ in \eref{eq:zero-kernel}, proving attainment.
\end{proof}

\subsection{Proof of \thref{thm:hier}}\label{app:theorem-one-proof}

We prove the lower bounds and attainment at second and odd orders in \eref{eq:hier}, followed by the lower and upper bounds at even orders in \eref{eq:even-bracket}. We then verify the comparisons between the coefficients defined in Eqs.~\eqref{eq:des-coef} and~\eqref{eq:even-coef}.

\begin{proof}[Proof of the $2$-design lower bound in \thref{thm:hier}]\label{app:2-ord-bnd}
Let $\caE$ be a finite weighted exact projective $2$-design, and let $\ket\psi,\ket\nu$ be orthonormal. Write $G=G_{\psi,\caM_\caE}$. For a projector $\phi$ in $\caE$, put $a=\braket\phi\psi$, $b=\braket\phi\nu$, $u=|a|^2$, $s=|b|^2$, and $z=\Rea(\oa b)$. Then $u+s\le1$ because $\ket\psi,\ket\nu$ are orthonormal. Define $y$ by
\begin{equation}
 y=\begin{cases}\frac{z^2}u,&u>0,\\0,&u=0,\end{cases}
 \qquad 0\le y\le s,\qquad
 \inner{\nu}{G\nu}_{\bbR}=d\bbE_\pi y.
 \label{eq:2-des-score-var}
\end{equation}
At $u=0$, both $a$ and $z$ vanish, and setting $y=0$ omits this outcome from the sum in \eref{eq:hor-fi}. Applying the design property \eref{eq:des} with tensor power $2$ and then taking a partial trace gives
\begin{equation}
 \begin{alignedat}{2}
 \bbE_\pi u&=\bbE_\pi s=\frac1d,
 &\qquad \bbE_\pi u^2=\bbE_\pi s^2&=\frac2{d(d+1)},\\
 \bbE_\pi us&=\frac1{d(d+1)},
 &\qquad \bbE_\pi uy=\bbE_\pi z^2&=\frac1{2d(d+1)}.
 \end{alignedat}
 \label{eq:2-des-ov-mom}
\end{equation}
In the last identity, $uy=z^2$ also holds at $u=0$; moreover, $\bbE_\pi(\oa b)^2=0$ and $\bbE_\pi|ab|^2=1/[d(d+1)]$ by $\braket{\psi}{\nu}=0$ and \eref{eq:des} with tensor power $2$.

For $F$ and $\chi$ defined in \lref{lem:2des-cert}, \eref{eq:2-des-ov-mom} gives $\bbE_\pi F(u,s)^2=1/2$. Hence averaging \eref{eq:2des-score-cert} yields
\begin{equation}
 \inner{\nu}{G\nu}_{\bbR}
 \ge\frac{c_{d,2}}{2}\left[\frac{\chi+3}{2}-\frac{\chi-1}{2}\right]=c_{d,2}.
 \label{eq:2-des-fi-prf}
\end{equation}
Since this holds for every $2$-design, pure state, and unit horizontal direction, the definitions in Eqs.~\eqref{eq:eta} and~\eqref{eq:des-cls-eta} give
\begin{equation}
 \eta_{\mathrm{des}}(d,2)\ge c_{d,2}.
 \label{eq:2des-class-lower}
\end{equation}

\end{proof}

\begin{proof}[Proof of attainment for $2$-designs in \thref{thm:hier}]\label{app:2-ext}
For the finite $2$-design and direction in \lref{lem:2des-attain},
\begin{equation}
 \eta_{\mathrm{des}}(d,2)\le\eta(\caM_{\caE_*})
 \le\inner{\nu}{G_{\psi,\caM_{\caE_*}}\nu}_{\bbR}
 =c_{d,2}.
\end{equation}
Together with \eref{eq:2des-class-lower}, this proves attainment.
\end{proof}

\begin{proof}[Proof of the odd-order lower bound in \thref{thm:hier}]\label{app:hi-ord-prf}
Fix $k\ge2$ and a $t$-design $\caE$ with $t\ge2k-1$. For $q_{k-1}=q_{k-1}^{\mu_d}$, we have $q_{k-1}(0)=1$ and
\begin{equation}
 \bbE_{\mu_d}q_{k-1}^2
 =\frac1{K_{k-1}^{\mu_d}(0,0)}
 =\frac d{k(d+k-1)},
 \label{eq:odd-kernel}
\end{equation}
where the first equality follows from \eref{eq:opt-mom} in \lref{lem:end}, and the second from \eref{eq:expl-fi-kernel} in \lref{lem:fi-kernel}. Apply \lref{lem:minorant} with $r=q_{k-1}^2$, whose degree is $2k-2\le t-1$. Equations~\eqref{eq:min-bnd} and~\eqref{eq:odd-kernel} give $G_{\psi,\caM_\caE}\succeq c_{d,2k-1}\IdTan\psi$ at every pure state. Hence
\begin{equation}
 \eta_{\mathrm{des}}(d,t)\ge c_{d,2k-1}
 \qquad(t\ge2k-1).
 \label{eq:odd-class-lower}
\end{equation}

\end{proof}

\begin{proof}[Proof of attainment at odd orders in \thref{thm:hier}]\label{app:ext-sat}
Fix $k\ge2$.
Apply \lref{lem:fin-lift} to the positive quadrature in \lref{lem:rad-quad}. The resulting finite design $\caE_*$ has overlaps in $\{0,u_1,\ldots,u_{k-1},1\}$. Condition~(iii) of \lref{lem:sharp-sat} therefore implies
\begin{equation}
 G_{\psi,\caM_{\caE_*}}=c_{d,2k-1}\IdTan\psi.
 \label{eq:sharp-sat}
\end{equation}
Together with \eref{eq:odd-class-lower} at $t=2k-1$, this proves attainment.

\end{proof}

\begin{proof}[Proof of the even-order lower bound in \thref{thm:hier}]\label{app:even-bnd}
Equation~\eqref{eq:even-fi-bound} in \lref{lem:even-fi} holds for every finite exact $2k$-design and pure state, so
\begin{equation}
 \eta_{\mathrm{des}}(d,2k)\ge c_{d,2k}.
 \label{eq:even-class-lower}
\end{equation}
\end{proof}

\begin{proof}[Proof of the even-order upper bound in \thref{thm:hier}]\label{app:even-ext}
Fix $k\ge2$.
By \lref{lem:even-zero}, there are a finite exact $2k$-design $\caE_*$ and a pure state $\rho=\proj\psi$ such that
\begin{equation}
 \alpha_0(\rho)=\frac{d-1}{(k+1)(d+k-1)}.
\end{equation}
Put $G=G_{\psi,\caM_{\caE_*}}$. The trace identity \eref{eq:trace-defect} in \pref{prop:geometry} and $\dim_{\bbR}\caH_\psi=2d-2$ give
\begin{equation}
 \eta_{\mathrm{des}}(d,2k)\le\lambda_{\min}(G)\le\frac{\tr_{\bbR}G}{2d-2}
 =\frac12\left[1-\frac d{(k+1)(d+k-1)}\right].
 \label{eq:even-witness}
\end{equation}

\end{proof}

\begin{proof}[Verification of the coefficient comparisons]
For $k\ge2$, put $e_{d,k}:=(1/2)\{1-d/[k(d+k)]\}$. Equation~\eqref{eq:even-coef} gives
\begin{equation}
 c_{d,2k}-e_{d,k}=\frac{d+1}{8k^2(d+k)^2}>0,
 \label{eq:even-improvement}
\end{equation}
and
\begin{equation}
e_{d,k}-c_{d,2k-1}=\frac d{2k(d+k-1)(d+k)}>0,
 \label{eq:even-odd-gap}
\end{equation}
which proves $c_{d,2k}>e_{d,k}>c_{d,2k-1}$. The remaining adjacent-order comparison follows from
\begin{equation}
 c_{d,2k+1}-c_{d,2k}
 =\frac{4dk(d+k)-(d+1)(k+1)}{8k^2(k+1)(d+k)^2}>0,
\end{equation}
for $d,k\ge2$. Together with $c_{d,2}<1/(d+1)\le c_{d,3}$, this proves strict increase of the coefficients with design order.

The gap in \eref{eq:even-bracket} satisfies
\begin{equation*}
 \frac12\left[1-\frac d{(k+1)(d+k-1)}\right]-c_{d,2k}
 =\frac{d(d-1)}{2k(k+1)(d+k)(d+k-1)}-\frac{d+1}{8k^2(d+k)^2}
 <\min\left\{\frac1{2k(k+1)},\frac{d(d-1)}{2k^4}\right\}.
\end{equation*}
The two estimates use $(d+k)(d+k-1)>d(d-1)$ and $k(k+1)(d+k)(d+k-1)>k^4$, respectively.

\end{proof}

\subsection{Proof of \coref{cor:zero-wt}: zero-probability weight}\label{app:zero-wt-prf}

\begin{proof}[Proof of \coref{cor:zero-wt}]
At even order, \lref{lem:even-zero} gives the bound and its attaining design. At odd order $t=2k-1$, write $\rho=\proj\psi$. The trace identity and \thref{thm:hier} give
\begin{equation}
 d-1-d\alpha_0(\rho)=\tr_{\bbR}G_{\psi,\caM_\caE}\ge(2d-2)c_{d,2k-1}.
\end{equation}
Rearranging proves the odd-order bound. For the design obtained from \lref{lem:rad-quad}, the total weight of projectors orthogonal to $\ket\psi$ is given by \eref{eq:a0-val} and attains the bound.
\end{proof}

\section{Fisher floors of structured measurements}\label{app:structured-fi}

We prove Propositions~\ref{prop:sic-fi}, \ref{prop:stab-fi}, and~\ref{prop:dod-fi}, using the geometry of SICs, the Pauli structure of stabilizer measurements, and the symmetries of the Platonic solids.

\subsection{Proof of \pref{prop:sic-fi}: SIC Fisher floor}\label{app:sic}

\begin{proof}[Proof of \pref{prop:sic-fi}]
Since every SIC is an exact projective $2$-design, \thref{thm:hier} gives $\eta(\caM_{\SIC})\ge c_{d,2}$. To prove attainment, choose any two distinct SIC projectors $\phi_1,\phi_2$ and unit vectors $\ket{\phi_1},\ket{\phi_2}$ with phases chosen so that $\braket{\phi_1}{\phi_2}=\gamma:=(d+1)^{-1/2}>0$. Define
\begin{equation}
 \ket\psi=\frac{\ket{\phi_1}+\ket{\phi_2}}{\sqrt{2(1+\gamma)}\lsp},\qquad
 \ket \nu=\frac{\ket{\phi_1}-\ket{\phi_2}}{\sqrt{2(1-\gamma)}\lsp}.
 \label{eq:sic-wit-kets}
\end{equation}
These kets are orthonormal. Put $\rho=\proj\psi$. Their tangent operator is
\begin{equation}
 \Delta=\Delta_{\psi,\nu}=\frac{\phi_1-\phi_2}{\sqrt{1-\gamma^2}\lsp},\qquad J_\rho(\Delta,\Delta)=4.
 \label{eq:sic-wit-tan}
\end{equation}
The two selected outcomes satisfy
\begin{equation}
 p_1=p_2=\frac{1+\gamma}{2d},\qquad
 \tr(E_1\Delta)=-\tr(E_2\Delta)=\frac{\sqrt{1-\gamma^2}\lsp}d.
 \label{eq:sic-wit-probs}
\end{equation}
For every $i\ge3$, equiangularity gives $\tr(\phi_i\phi_1)=\tr(\phi_i\phi_2)$, hence $\tr(E_i\Delta)=0$. These outcomes contribute nothing to the Fisher information, whether or not their probabilities vanish. Only outcomes $1$ and $2$ contribute, and both have positive probabilities. Therefore
\begin{equation}
 \inner{\nu}{G_{\psi,\caM_{\SIC}}\nu}_{\bbR}
 =\frac14\sum_{i=1}^2\frac{[\tr(E_i\Delta)]^2}{p_i}
 =\frac{1-\gamma^2}{d(1+\gamma)}
 =\frac{1-(d+1)^{-1/2}}d.
 \label{eq:sic-wit-fi}
\end{equation}
This directional value equals $c_{d,2}$. Together with the universal lower bound, it proves \eref{eq:sic-fi-bnds}.
\end{proof}

\subsection{Proof of \pref{prop:stab-fi}: stabilizer Fisher floor}
\label{app:stab-fi}

The qudit bound in \pref{prop:stab-fi} is proved in \aref{app:shadow-apps}. For the qubit case, by \pref{prop:reg-eq}, it suffices to prove the uniform lower bound at regular states. Attainment at stabilizer states follows from the total ensemble weight of projectors orthogonal to the state and the Fisher trace identity.

\subsubsection{Pauli notation and auxiliary lemmas}\label{app:stab-tools}

We use the standard stabilizer formalism~\cite{Gottesman1997}, with Hermitian Pauli representatives $\sfP_n=\left\{\bbone_{\bbC^2},X,Y,Z\right\}^{\otimes n}$ in dimension $d=2^n$, where $\bbone_{\bbC^2}$ is the single-qubit identity. These representatives satisfy $\tr(PQ)=d\delta_{PQ}$. Commutation and group cardinalities refer to Pauli labels modulo overall phase. Write $P\sim Q$ when $PQ=-QP$. In this subsection, $\sigma\in\{-1,1\}$ denotes a measurement sign, not a state.

Fix $\rho=\proj\psi\in\caP(\caH)$ with unit vector $\ket\psi$, and suppose $|\bra\psi P\ket\psi|<1$ for every $P\ne\bbone$. For any horizontal vector $\nu\in\caH_\psi$, put
\begin{equation}
 r_P=\bra\psi P\ket\psi,\qquad
 x_P(\nu)=\Rea(\bra\psi P\ket \nu),\qquad
 h_\psi(\nu)=\sum_{P\ne\bbone}\frac{x_P(\nu)^2}{1-r_P^2}.
 \label{eq:stab-pauli-scores}
\end{equation}
Every regular state of the complete stabilizer POVM satisfies $|r_P|<1$ for all $P\ne\bbone$. Indeed, if $|r_P|=1$, then $P\ket\psi=\sigma\ket\psi$ for some $\sigma\in\{-1,1\}$. The opposite eigenspace has an orthonormal basis of stabilizer states. For a unit ket $\ket\phi$ in that basis,
\begin{equation}
 P\ket\phi=-\sigma\ket\phi
 \quad\Longrightarrow\quad\braket\phi\psi=0
 \quad\Longrightarrow\quad p_\phi(\rho)=\frac d{M_n}|\braket\phi\psi|^2=0,
 \label{eq:stab-regular-pauli}
\end{equation}
contrary to regularity.

With $\Delta_{\psi,\nu}=\ket\psi\bra\nu+\ket\nu\bra\psi$ as in \eref{eq:tan}, define
\begin{equation}
 \mathcal F^{(n)}_\psi(\nu):=\frac14 I_{\rho,\caM_{\rmstab(2,n)}}(\Delta_{\psi,\nu},\Delta_{\psi,\nu})
 =\inner{\nu}{G_{\psi,\caM_{\rmstab(2,n)}}\nu}_{\bbR}
 =\frac d{M_n}\sum_{\phi\colon\braket\phi\psi\ne0}
 \frac{[\Rea(\braket\psi\phi\braket\phi\nu)]^2}{|\braket\phi\psi|^2}.
 \label{eq:stab-norm-fi}
\end{equation}

\begin{lemma}[Bounds on $h_\psi$]\label{lem:stab-pauli-bnds}
Let $n\ge1$, let $\rho=\proj\psi\in\caP(\caH)$ satisfy $|r_P|<1$ for all $P\ne\bbone$, and let $\nu\in\caH_\psi$. Then
\begin{equation}
 \frac d2\norm{\nu}_2^2\le h_\psi(\nu)\le d\norm{\nu}_2^2.
 \label{eq:stab-score-bnds}
\end{equation}
\end{lemma}

\begin{lemma}[Recursive stabilizer measurement]\label{lem:stab-rec}
For $n\ge2$, choose $P\in\sfP_n\setminus\{\bbone\}$ uniformly and measure $\Pi_{P,\sigma}=(\bbone+\sigma P)/2$, $\sigma=\pm1$. Let $\{\phi_{P,\sigma,j}\}_{j=1}^{M_{n-1}}$ be the complete set of stabilizer projectors in that eigenspace. The subsequent complete $(n-1)$-qubit stabilizer measurement gives POVM elements
\begin{equation}
 E_{P,\sigma,j}=\frac1{d^2-1}\frac d{2M_{n-1}}\phi_{P,\sigma,j},\qquad
 \sum_{(P,\sigma,j)\colon\phi_{P,\sigma,j}=\phi}E_{P,\sigma,j}=\frac d{M_n}\phi.
 \label{eq:stab-rec-elements}
\end{equation}
Thus grouping the outcomes with the same projector gives the complete $n$-qubit stabilizer POVM and preserves Fisher information.
\end{lemma}

For a regular state $\rho=\proj\psi$, a unit horizontal ket $\ket\nu$, and fixed $P\in\sfP_n\setminus\{\bbone\}$, write $z_P=\bra\psi P\ket\nu=x_P+\rmi y_P$. Put
\begin{equation}
 \Pi_\sigma=\frac{\bbone+\sigma P}{2},\qquad p_\sigma=\frac{1+\sigma r_P}{2}>0,\qquad
 \ket{\psi_\sigma}=\frac{\Pi_\sigma\ket\psi}{\sqrt{p_\sigma}\lsp},\qquad
 \ket{\nu_\sigma}=\frac{\Pi_\sigma\ket\nu}{\sqrt{p_\sigma}\lsp}-\ket{\psi_\sigma}\frac{\bra\psi\Pi_\sigma\ket\nu}{p_\sigma}.
 \label{eq:stab-cond-tan}
\end{equation}
Here $\nu_\sigma\in\caH_{\psi_\sigma}$ is the horizontal derivative of the conditional state, after identifying the eigenspace with $n-1$ qubits. Let $\mathcal F_P$ denote one quarter of the classical Fisher information of the two-stage measurement in \lref{lem:stab-rec} with $P$ fixed, evaluated in the direction $\Delta_{\psi,\nu}$. The Fisher chain rule gives
\begin{equation}
 \mathcal F_P=\frac{x_P^2}{1-r_P^2}+\sum_{\sigma=\pm1}p_\sigma \mathcal F^{(n-1)}_{\psi_\sigma}(\nu_\sigma).
 \label{eq:stab-qP-def}
\end{equation}

\begin{lemma}[Conditional Fisher bound]\label{lem:stab-cond-fi}
Let $n\ge2$, and suppose that the complete $(n-1)$-qubit stabilizer measurement satisfies $\mathcal F^{(n-1)}_\varphi(\nu)\ge\norm{\nu}_2^2/3$ at every regular state $\rho'=\proj\varphi$ and every horizontal vector $\nu\in\caH_\varphi$. Let $\rho=\proj\psi$ be regular with respect to the complete $n$-qubit stabilizer measurement, let $\nu\in\caH_\psi$ satisfy $\norm{\nu}_2=1$, and let $P\in\sfP_n\setminus\{\bbone\}$. Using the conditional states and vectors in \eref{eq:stab-cond-tan}, let $\mathcal F_P$ be given by \eref{eq:stab-qP-def}. Then
\begin{equation}
 3\mathcal F_P\ge1+\frac{2x_P^2-y_P^2}{1-r_P^2}.
 \label{eq:stab-reg-bnd}
\end{equation}
\end{lemma}

\subsubsection{Proof of the qubit Fisher floor and attainment}\label{app:stab-m-prf}

\begin{proof}[Proof of the qubit case of \pref{prop:stab-fi}]
\emph{Lower bound at regular states.} We prove $\mathcal F^{(n)}_\psi(\nu)\ge\norm{\nu}_2^2/3$ for every horizontal vector $\nu\in\caH_\psi$ at regular states, by induction on $n$. For $n=1$, the complete stabilizer measurement chooses $X,Y,Z$ uniformly, so
\begin{equation}
 \mathcal F^{(1)}_\psi(\nu)=\frac13h_\psi(\nu)\ge\frac13\norm{\nu}_2^2
 \label{eq:stab-qb-base}
\end{equation}
by \lref{lem:stab-pauli-bnds}.

Suppose the assertion holds for $n-1$, and let $\rho=\proj\psi$ be regular with respect to $\caM_{\rmstab(2,n)}$ in dimension $d=2^n$. For a unit horizontal ket $\ket\nu$, \lref{lem:stab-rec} gives
\begin{equation}
 \mathcal F^{(n)}_\psi(\nu)=\frac1{d^2-1}\sum_{P\ne\bbone}\mathcal F_P.
 \label{eq:stab-fi-rec}
\end{equation}
Since $h_\psi(\rmi\nu)=\sum_{P\ne\bbone}y_P^2/(1-r_P^2)$, summing \eref{eq:stab-reg-bnd} gives the first inequality below. As $\norm{\nu}_2=\norm{\rmi\nu}_2=1$, \lref{lem:stab-pauli-bnds} gives $h_\psi(\nu)\ge d/2$ and $h_\psi(\rmi\nu)\le d$; substituting these bounds gives
\begin{equation}
 3\sum_{P\ne\bbone}\mathcal F_P\ge\left(d^2-1\right)+2h_\psi(\nu)-h_\psi(\rmi\nu)\ge\left(d^2-1\right)+2\frac d2-d=d^2-1.
 \label{eq:stab-ind-sum}
\end{equation}
Thus $\mathcal F^{(n)}_\psi(\nu)\ge1/3$ for every unit horizontal ket, equivalently $G_{\psi,\caM_{\rmstab(2,n)}}\succeq\IdTan\psi/3$ at every regular state. Proposition~\ref{prop:reg-eq} gives $\eta(\caM_{\rmstab(2,n)})\ge1/3$, hence $G_{\psi,\caM_{\rmstab(2,n)}}\succeq\IdTan\psi/3$ at every pure state.

\emph{Equality at stabilizer states.} Let $a_n=\alpha_0(\rho)$ for an $n$-qubit stabilizer state $\rho=\proj\psi$. Clifford covariance makes $a_n$ independent of the chosen stabilizer state. At $n=1$, exactly one of the six stabilizer states is orthogonal to the state, so $a_1=1/6$.

For the recorded measurement in \lref{lem:stab-rec}, define $\pi_{P,\sigma,j}:=\tr(E_{P,\sigma,j})/d$. Equation~\eqref{eq:stab-rec-elements} gives, for each fixed $(P,\sigma)$,
\begin{equation}
 \pi_{P,\sigma,j}=\frac1{2(d^2-1)M_{n-1}},\qquad
 \sum_{j=1}^{M_{n-1}}\pi_{P,\sigma,j}=\frac1{2(d^2-1)}.
 \label{eq:stab-fixed-sign-weight}
\end{equation}

For $d-1$ Pauli labels, the state is a Pauli eigenstate. For the sign $\sigma$ opposite to its eigenvalue, all $\phi_{P,\sigma,j}$ are orthogonal to $\rho$; for the other sign, the fraction is $a_{n-1}$. For each of the remaining $d^2-d$ labels, the subgroup of the state stabilizer commuting with $P$ has $n-1$ independent generators. Together with the measured signed Pauli $\sigma P$, these generators stabilize the nonzero projected ket. Each normalized projection is therefore a stabilizer state in its logical $(n-1)$-qubit eigenspace, so the zero-weight fraction for either sign is $a_{n-1}$.

Grouping identical projectors preserves total ensemble weight. Consequently,
\begin{equation}
 a_n=\frac{(d-1)(1+a_{n-1})/2+(d^2-d)a_{n-1}}{d^2-1}
 =\frac{1+(2d+1)a_{n-1}}{2(d+1)}
 =\frac{d-1}{3d},
 \label{eq:stab-zero-rec}
\end{equation}
where the last equality follows inductively from $a_{n-1}=(d-2)/(3d)$. The trace identity \eref{eq:trace-defect} in \pref{prop:geometry} now gives
\begin{equation}
 \tr_{\bbR}G_{\psi,\caM_{\rmstab(2,n)}}=d-1-da_n=\frac{2(d-1)}3.
 \label{eq:stab-eq-data}
\end{equation}
Thus $G_{\psi,\caM_{\rmstab(2,n)}}-\IdTan\psi/3$ is positive semidefinite and has trace zero, so it vanishes. This proves equality in every tangent direction at stabilizer states and establishes the claimed Fisher floor.
\end{proof}

\subsubsection{Proofs of the auxiliary lemmas}\label{app:stab-aux-prfs}

\begin{proof}[Proof of \lref{lem:stab-pauli-bnds}]
\emph{Lower bound.} Both $h_\psi(\nu)$ and $\norm{\nu}_2^2$ scale quadratically with $\nu$, so assume $\norm{\nu}_2=1$. For $\Delta=\Delta_{\psi,\nu}=\ket\psi\bra\nu+\ket\nu\bra\psi$ from \eref{eq:tan},
\begin{equation}
 \tr(P\Delta)=\bra\nu P\ket\psi+\bra\psi P\ket\nu=2x_P(\nu),\qquad
 \tr\Delta=0,\qquad\tr\left(\Delta^2\right)=2.
 \label{eq:stab-tangent-pauli}
\end{equation}
The Hermitian Pauli matrices form an orthogonal basis with $\tr(PQ)=d\delta_{PQ}$. Expanding $\Delta$ in that basis therefore gives
\begin{equation}
 \Delta=\frac1d\sum_{P\ne\bbone}\tr(P\Delta)P,\qquad
 \sum_{P\ne\bbone}x_P(\nu)^2=\frac14\sum_{P\ne\bbone}[\tr(P\Delta)]^2=\frac d4\tr\left(\Delta^2\right)=\frac d2.
 \label{eq:stab-pauli-pars}
\end{equation}
Since $0<1-r_P^2\le1$ for every $P\ne\bbone$, \eref{eq:stab-pauli-pars} yields $h_\psi(\nu)\ge d/2$, the lower bound in \eref{eq:stab-score-bnds} for unit $\nu$.

\emph{An estimate for anticommuting Pauli matrices.} The Pauli expansion of $\rho$ gives
\begin{equation}
 \rho-P\rho P=\frac2d\sum_{Q\sim P}r_QQ.
 \label{eq:stab-anti-id}
\end{equation}
Taking its squared Hilbert--Schmidt norm gives
\begin{equation}
 \frac4d\sum_{Q\sim P}r_Q^2=\tr\left[(\rho-P\rho P)^2\right]=2\left(1-r_P^2\right).
 \label{eq:stab-anti-norm}
\end{equation}
Taking its Hilbert--Schmidt inner product with $\Delta$ instead, and using $\tr(\rho\Delta)=0$ and \eref{eq:stab-tangent-pauli}, gives
\begin{equation}
 \frac4d\sum_{Q\sim P}r_Qx_Q(\nu)
 =\tr[(\rho-P\rho P)\Delta]
 =-\bra\psi P\Delta P\ket\psi=-2r_Px_P(\nu).
 \label{eq:stab-anti-tangent}
\end{equation}
Thus Cauchy--Schwarz yields
\begin{equation}
 \frac{d^2}{4}r_P^2x_P(\nu)^2
 =\left[\sum_{Q\sim P}r_Qx_Q(\nu)\right]^2
 \le\left[\sum_{Q\sim P}r_Q^2\right]\left[\sum_{Q\sim P}x_Q(\nu)^2\right]
 =\frac d2\left(1-r_P^2\right)\sum_{Q\sim P}x_Q(\nu)^2.
 \label{eq:stab-anti-cs}
\end{equation}
For $r_P\ne0$, rearranging gives
\begin{equation}
 \frac{x_P(\nu)^2}{1-r_P^2}\le\frac2{dr_P^2}\sum_{Q\sim P}x_Q(\nu)^2.
 \label{eq:stab-loc-chg}
\end{equation}

\emph{Upper bound.} Put $S=\left\{P\ne\bbone\colon r_P^2>1/2\right\}$ and $T=\left\{Q\ne\bbone\colon r_Q^2\le1/2\right\}$. If $P\sim Q$, then
\begin{align}
 (r_PP+r_QQ)^2&=\left(r_P^2+r_Q^2\right)\bbone,
 \label{eq:stab-anti-square}\\
 r_P^2+r_Q^2&=\tr[\rho(r_PP+r_QQ)]\le\norm{r_PP+r_QQ}=\sqrt{r_P^2+r_Q^2}.
 \label{eq:stab-anti-expect}
\end{align}
Consequently $r_P^2+r_Q^2\le1$. Two labels in $S$ therefore commute, since otherwise their squared expectations would sum to more than one. Extend $S$ to a maximal commuting group $\mathscr C$ of Pauli labels modulo phase, with $|\mathscr C|=d$.

For fixed $Q$, let $\varepsilon_Q(P)\in\{-1,1\}$ be defined by $PQ=\varepsilon_Q(P)QP$. These signs satisfy
\begin{equation}
 \varepsilon_Q(P_1P_2)=\varepsilon_Q(P_1)\varepsilon_Q(P_2)\qquad(P_1,P_2\in\mathscr C).
 \label{eq:stab-sign-product}
\end{equation}
If every sign is $+1$, no label in $\mathscr C$ anticommutes with $Q$. Otherwise choose $R\in\mathscr C$ with $\varepsilon_Q(R)=-1$. Multiplication by $R$ permutes $\mathscr C$, so
\begin{equation}
 \sum_{P\in\mathscr C}\varepsilon_Q(P)
 =\sum_{P\in\mathscr C}\varepsilon_Q(RP)
 =-\sum_{P\in\mathscr C}\varepsilon_Q(P)=0.
 \label{eq:stab-sign-sum}
\end{equation}
The number of negative signs is then $d/2$. In either case,
\begin{equation}
 N_Q:=|\{P\in S\colon P\sim Q\}|\le|\{P\in\mathscr C\colon P\sim Q\}|\le\frac d2.
 \label{eq:stab-anti-count-simple}
\end{equation}
For $Q\in T$, put $c_Q=\left(1-2r_Q^2\right)/(1-r_Q^2)\ge0$. If $P\in S$ and $P\sim Q$, then $r_Q^2\le1-r_P^2<1/2$, so $Q\in T$, and
\begin{equation}
 2-\frac1{r_P^2}\le2-\frac1{1-r_Q^2}=c_Q.
 \label{eq:stab-score-comparison}
\end{equation}
Using \eref{eq:stab-pauli-pars}, then \eref{eq:stab-loc-chg}, and finally \eref{eq:stab-score-comparison}, we obtain
\begin{equation}
 \begin{aligned}
 h_\psi(\nu)-d
 &=\sum_{P\in S}\frac{(2r_P^2-1)x_P(\nu)^2}{1-r_P^2}-\sum_{Q\in T}c_Qx_Q(\nu)^2\\
 &\le\frac2d\sum_{P\in S}\left(2-\frac1{r_P^2}\right)\sum_{Q\sim P}x_Q(\nu)^2-\sum_{Q\in T}c_Qx_Q(\nu)^2\\
 &\le\sum_{Q\in T}\left(\frac{2N_Q}{d}-1\right)c_Qx_Q(\nu)^2\le0.
 \end{aligned}
 \label{eq:stab-chg-res}
\end{equation}
The last inequality follows from \eref{eq:stab-anti-count-simple}. Thus $h_\psi(\nu)\le d$, proving the upper bound in \eref{eq:stab-score-bnds} for unit $\nu$. Homogeneity gives both bounds for every horizontal $\nu$.
\end{proof}

\begin{proof}[Proof of \lref{lem:stab-rec}]
A Pauli eigenspace is a stabilizer code. A Clifford transformation identifies it with a fixed single-qubit eigenket tensored with $n-1$ logical qubits, under which its physical stabilizer states correspond exactly to the logical stabilizer states. Each $n$-qubit stabilizer state is an eigenstate of exactly $d-1$ nonidentity Pauli labels. Counting stabilizer states with these eigenvalue labels gives
\begin{equation}
 M_n(d-1)=2\left(d^2-1\right)M_{n-1},\qquad M_n=2(d+1)M_{n-1}.
 \label{eq:stab-card-rec}
\end{equation}
For an outcome $(P,\sigma,j)$ associated with a projector $\phi$, the POVM element is $(d/2)\left[\left(d^2-1\right)M_{n-1}\right]^{-1}\phi$. Summing the elements of the $d-1$ outcomes associated with $\phi$ gives
\begin{equation}
 \frac{d-1}{d^2-1}\frac{d}{2M_{n-1}}\phi=\frac d{M_n}\phi.
 \label{eq:stab-merge-eff}
\end{equation}
Their probabilities are fixed multiples of $\tr(\rho\phi)$, so their contributions to Fisher information add to that of the grouped outcome whenever $\tr(\rho\phi)>0$. When $\tr(\rho\phi)=0$, all these probabilities vanish and are omitted from the sum in \eref{eq:score-fi}. Thus grouping preserves the Fisher form at every state.
\end{proof}

\begin{proof}[Proof of \lref{lem:stab-cond-fi}]
Use the conditional states in \eref{eq:stab-cond-tan}. Each stabilizer projector in either eigenspace is also an outcome projector of the complete $n$-qubit measurement. Its probability at the regular state $\rho$ is positive; dividing by $p_\sigma>0$ shows that each conditional state is regular with respect to its stabilizer measurement. Along the unit-ket curve $(\ket\psi+t\ket\nu)/\sqrt{1+t^2}$, the derivative of $p_\sigma$ at $t=0$ and the projected tangent amplitude are
\begin{equation}
 \dot p_\sigma=\tr(\Pi_\sigma\Delta_{\psi,\nu})=\sigma x_P,
 \qquad \bra\psi\Pi_\sigma\ket\nu=\frac{\sigma z_P}{2}.
\end{equation}
The identity for $\dot p_\sigma$ gives the first term in \eref{eq:stab-qP-def}. The subtraction in \eref{eq:stab-cond-tan} gives $\braket{\psi_\sigma}{\nu_\sigma}=0$. Taking the squared norm of this orthogonal projection yields
\begin{equation}
 p_\sigma\norm{\nu_\sigma}_2^2
 =\bra\nu\Pi_\sigma\ket\nu
 -\frac{|\bra\psi\Pi_\sigma\ket\nu|^2}{p_\sigma}.
\end{equation}
Since $\sum_\sigma\Pi_\sigma=\bbone$, $\norm{\nu}_2=1$, and $\sum_\sigma p_\sigma^{-1}=4/(1-r_P^2)$, the conditional tangent norms satisfy
\begin{equation}
 \sum_{\sigma=\pm1}p_\sigma\norm{\nu_\sigma}_2^2
 =1-\frac{|z_P|^2}{4}\sum_{\sigma=\pm1}\frac1{p_\sigma}
 =1-\frac{|z_P|^2}{1-r_P^2}.
 \label{eq:stab-cond-norm-sum}
\end{equation}
Applying the assumed $(n-1)$-qubit bound to each conditional state in \eref{eq:stab-qP-def}, we obtain
\begin{equation}
 3\mathcal F_P\ge\frac{3x_P^2}{1-r_P^2}
       +1-\frac{|z_P|^2}{1-r_P^2}
 =1+\frac{2x_P^2-y_P^2}{1-r_P^2},
\end{equation}
which is \eref{eq:stab-reg-bnd} in \lref{lem:stab-cond-fi}.
\end{proof}

\subsection{Proof of \pref{prop:dod-fi}: Platonic Fisher floors}\label{app:dod-fi}

\begin{proof}[Proof of \pref{prop:dod-fi}]
The tetrahedral value follows from \pref{prop:sic-fi}; the octahedral, cubic, and icosahedral values follow from \thref{thm:hier} and the attaining states described in \sref{sec:examples}. It remains to prove the dodecahedral floor and its equality condition.

The argument uses the reflection-invariant polynomials $R,U,W$ of degrees $2,6,10$ defined below. Their degree-$15$ Jacobian is nonzero away from the $15$ reflection planes, so these invariants are local coordinates there. Concavity in $W$ then reduces the minimum to a reflection plane, where the Fisher eigenvalues depend on one variable.

Put $c=(15-\sqrt{5}\lsp)/30$. Pairing antipodal terms in \eref{eq:qb-fi-mat} gives one tenth of the projector onto each axis's tangent projection, with parallel axes omitted. Thus $\tr F_{\bmr}=1$ at regular states. Singular states correspond to vertex directions because the vertex set is antipodal. At a vertex, one pair contributes zero and threefold symmetry makes the tangent operator scalar, so both eigenvalues are $9/20>c$.

\emph{Clearing denominators.} Write $\chi=(1+\sqrt{5}\lsp)/2$, $h=\chi^2$, and $g=h^{-1}$. Then $h+g=3$ and $1/3<g<2/5$. Choose the ten antipodal representatives $\bmu_i$
\begin{equation}
 (1,\pm1,\pm1),\quad \left(0,\chi^{-1},\pm\chi\right),\quad
 \left(\chi,0,\pm\chi^{-1}\right),\quad \left(\chi^{-1},\pm\chi,0\right),
 \label{eq:dod-vertex-reps}
\end{equation}
with independent signs in the first group. Each $\bmu_i$ has squared norm $3$, and the unit vertices are $\bmv=\pm\bmu_i/\sqrt{3}$. For $\bmr=(x,y,z)\ne0$, put $R=\bmr\cdot\bmr$, $\widehat{\bmr}=\bmr/\sqrt{R}$, $s_i=3R-(\bmu_i\cdot\bmr)^2$, and $\bmu_{i,\perp}=\bmu_i-(\bmu_i\cdot\widehat{\bmr})\widehat{\bmr}$. Their pairwise signed areas are $\widehat{\bmr}\cdot(\bmu_i\times\bmu_j)$, so antipodal pairing and Cauchy--Binet on the tangent plane give, at regular states,
\begin{align}
 F_{\widehat{\bmr}}&=\frac{R}{10}\sum_{i=1}^{10}\frac{\bmu_{i,\perp}\bmu_{i,\perp}^{\sfT}}{s_i},\quad \norm{\bmu_{i,\perp}}_2^2=\frac{s_i}{R},\label{eq:dod-paired-fisher}\\
 \det F_{\widehat{\bmr}}&=\frac{R}{100}\sum_{i<j}\frac{[\bmr\cdot(\bmu_i\times\bmu_j)]^2}{s_i s_j}.\label{eq:dod-det}
\end{align}
Since $\tr F_{\widehat{\bmr}}=1$ and $c(1-c)=11/45$, the desired bound is equivalent to $p(\bmr)\ge0$, where
\begin{equation}
 p(\bmr):=900\left(\prod_i s_i\right)\left(\det F_{\widehat{\bmr}}-\frac{11}{45}\right).
 \label{eq:dod-det-poly}
\end{equation}
By \eref{eq:dod-det}, $p$ extends to a homogeneous invariant polynomial of degree $20$. At a vertex direction, every term after clearing denominators contains either the vanishing $s_i$ of the parallel axis or a vanishing triple product involving that axis, so $p=0$.

\emph{Reduction to a reflection plane.} Rotations carry the symmetry plane $x=0$ to $15$ distinct reflection planes, perpendicular to the edge-midpoint axes. Define
\begin{equation}
 U=\left(x^2-hz^2\right)\left(z^2-hy^2\right)\left(y^2-hx^2\right),\qquad
 W=\prod_{i=1}^{10}(\bmu_i\cdot\bmr).
 \label{eq:dod-invariant-defs}
\end{equation}
These are products of linear forms along the six face-normal axes and ten vertex axes, respectively; representatives for the former are $(1,0,\pm\chi)$, $(0,\pm\chi,1)$, and $(\pm\chi,1,0)$. Each reflection preserves both products up to sign; restricting to its reflection plane fixes the sign as positive, since neither product vanishes identically there. Thus $U$ and $W$ are invariant.

The Jacobian $\mathscr{J}=\det[\partial(R,U,W)/\partial(x,y,z)]$ has degree $15$ and changes sign under each reflection, so the linear forms defining the $15$ reflection planes divide it. It is nonzero: after pairing factors and restricting to $R=1$, the terms linear in $x^2,y^2$ near $x=y=0$ are
\begin{equation}
 U_{\mathrm{lin}}=h^2x^2-hy^2,\quad W_{\mathrm{lin}}=gx^2-hy^2.
 \label{eq:dod-invariants}
\end{equation}
Their coefficient determinant is $1-h^3\ne0$, so the Jacobian in $\left(x^2,y^2,z^2\right)$, and hence $\mathscr{J}$, is not identically zero. Thus $\mathscr{J}$ is a nonzero constant times the product of these linear forms.

Off the reflection planes, $(R,U,W)$ are local coordinates. Differentiation at fixed $R,U$ gives, for any invariant homogeneous polynomial $q$ of degree $m$,
\begin{equation}
 \partial_W q=\frac{\det[\partial(R,U,q)/\partial(x,y,z)]}{\mathscr{J}}.
 \label{eq:dod-invariant-derivative}
\end{equation}
The numerator is likewise divisible by every linear form defining a reflection plane. Hence $\partial_W q$ is an invariant polynomial of degree $m-10$, or zero if $m<10$; in particular, $\partial_W^2q=0$ if $m<20$. Moreover, $p+220W^2$ is divisible by $R$, because $s_i\equiv-(\bmu_i\cdot\bmr)^2\pmod R$. The quotient has degree $18$, so $\partial_W^2p=-440$. This strict concavity excludes a minimum off the reflection planes: locally, $W$ can vary in both directions with $R=1$ and $U$ fixed. Therefore the minimum of $p$ on the sphere lies on a reflection plane.

\emph{Bound on one plane.} On $x=0$, reflection symmetry makes the normal a tangent eigenvector at regular states. Write its eigenvalue as $f(t)$, with $t=y^2/z^2$. The four groups in \eref{eq:dod-vertex-reps} have squared $x$ components $1,0,h,g$. For the first group, $s_i=2(R\pm yz)$, each sign occurring twice; the last two groups have $s_i=3y^2+hz^2$ and $s_i=gy^2+3z^2$. Summing $R(\bmu_i)_x^2/(10s_i)$ therefore gives
\begin{equation}
 5f(t)=\frac{(1+t)^2}{1+t+t^2}+\frac{h(1+t)}{h+3t}+\frac{g(1+t)}{3+gt}.
 \label{eq:dod-mirror-fisher}
\end{equation}
The other eigenvalue is $1-f(t)$.

The endpoints are $f(0)=(2+g/3)/5=c$ and $f(\infty)=(2+h/3)/5=1-c$. For $t>0$, set $A=1+t+t^2$ and $B=(h+3t)(3+gt)$. Subtraction gives
\begin{equation}
\begin{aligned}
 5[f(t)-c]&=t\left[\frac1A-\frac{9g-h-3(1-g^2)t}{3B}\right]>t\left(\frac1A-\frac1{3B}\right)>0,\\
 5[1-c-f(t)]&=\frac{h}{3+gt}-\frac{1}{3(h+3t)}-\frac{t}{A}>\frac{h-2-g}{3+gt}>0.
\end{aligned}
 \label{eq:dod-mirror-bounds}
\end{equation}
Indeed, $9g-h=10g-3<1$ and $B=3h+10t+3gt^2>A$ prove the first line. For the second, $h-2-g=1-2g>0$, while $3t\le A$, $t^2\le A$, and $3(h+3t)>3+gt$ give
\begin{equation}
 \frac{t}{A}=\frac{3t+gt^2}{A(3+gt)}\le\frac{1+g}{3+gt},\quad \frac1{3(h+3t)}<\frac1{3+gt}.
 \label{eq:dod-mirror-estimates}
\end{equation}
Thus $p\ge0$ on the reflection planes, including the vertices, and hence on the sphere. A regular equality direction minimizes $p$ and must lie on a reflection plane, where equality requires $t=0$ or $\infty$. Their rotated images are exactly the edge-midpoint directions; vertices were excluded above.
\end{proof}

\section{Proofs of sampling constructions and outcome constraints}\label{app:samp}

In this appendix, we prove \pref{prop:res}, \thref{thm:samp}, and \pref{prop:basis-samp}, and establish a lower bound on the number of stabilizer bases. The sampling proofs combine design moments, a lower-tail estimate, and a tangent net. Projector sampling also requires a covariance bound for completion, whereas complete-basis sampling is complete by construction. Related moment and frame estimates appear in design-based low-rank recovery~\cite{Gilles2026,HuangLi2026Recovery}.

\subsection{Proof of \pref{prop:res}: Fisher floor at fixed outcome count}\label{app:out-constr}

\begin{proof}[Proof of \pref{prop:res}]
Write $\caM=\{w_i\phi_i\}_{i=1}^M$ and order its weights as $w_{(1)}\ge\cdots\ge w_{(M)}>0$. We prove the stronger bound
\begin{equation}
 \eta(\caM)\le\frac12-\frac{\sum_{j=1}^{d-1}w_{(j)}}{2(d-1)}.
 \label{eq:out-weight-ub}
\end{equation}
A rank-one POVM must have $M\ge d$. Since $\sum_iw_i=d$, its $d-1$ largest weights satisfy
\begin{equation}
 \sum_{j=1}^{d-1}w_{(j)}\ge\frac{d(d-1)}{M}.
\end{equation}
There is a unit ket $\ket\psi$ orthogonal to the corresponding $d-1$
measurement vectors. At the pure state $\rho=\proj\psi$, \eref{eq:trace-defect} in \pref{prop:geometry} gives
\begin{equation}
 \eta(\caM)\le\lambda_{\min}(G_{\psi,\caM})
 \le\frac12-\frac{\sum_{j=1}^{d-1}w_{(j)}}{2(d-1)}.
\end{equation}
This proves \eref{eq:out-weight-ub}; combining it with the weight sum above gives \eref{eq:out-ub}.

\end{proof}

\subsection{Uniform signal and covariance bounds}\label{app:samp-aux}

For the covariance argument, choose an orthonormal basis of $\caH$ and use its complex coordinates. For $z\in\caH$, define $\fkr(z)\in\bbR^{2d}$ by stacking its real and imaginary parts. For $O\in\caL(\caH)$, set
\begin{equation}
 \fkR(O)=\begin{pmatrix}\Rea O&-\Ima O\\
                   \Ima O&\Rea O\end{pmatrix},
 \qquad J_0=\begin{pmatrix}0&-\IdMat{d}\\\IdMat{d}&0\end{pmatrix}.
 \label{eq:real-def}
\end{equation}
The real and imaginary parts of a matrix are taken entrywise. The map $\fkR$ preserves the operator norm, and $J_0$ represents multiplication by $\rmi$ on the full real coordinate space; $J_\rho$ continues to denote the quantum Fisher form.

\subsubsection{Moment and probability bounds}
\label{app:samp-tools}

For Hermitian operators $A,B,C$, the $3$-design property in \eref{eq:des} gives
\begin{align}
 \bbE_\pi\tr(\phi A)\tr(\phi B)
 &=\frac{\tr A\tr B+\tr(AB)}{d(d+1)},\\
 \bbE_\pi\tr(\phi A)\tr(\phi B)\tr(\phi C)
 &=\frac{\tr A\tr B\tr C+\tr(AB)\tr C+\tr(AC)\tr B+\tr(BC)\tr A
       +\tr(ABC)+\tr(ACB)}{d(d+1)(d+2)}.
 \label{eq:des-2-3-mom}
\end{align}
These formulas follow by writing $P_{[s]}=(1/s!)\sum_{\tau\in S_s}U_\tau$ for $s=2,3$, where $S_s$ is the permutation group on $s$ tensor factors and $U_\tau$ implements the permutation $\tau$, and tracing against $A\otimes B$ or $A\otimes B\otimes C$. Each cycle of a permutation produces the trace of the corresponding operator product. We use these identities for every weighted exact design of order at least three.

The following two lemmas give the lower-tail and frame estimates.

\begin{lemma}[Lower tail of a nonnegative empirical mean]\label{lem:nonneg-tail}
Let $m\ge1$ be an integer, and let $z_1,\ldots,z_m$ be independent copies of a nonnegative real random variable $z$ with $\bbE z=\mu$ and $\bbE z^2\le b$, where $0<b<\infty$. For every $a>0$,
\begin{equation}
\bbP\left\{\frac1m\sum_{i=1}^mz_i\le\mu-a\right\}\le\exp\left(-\frac{ma^2}{2b}\right).
 \label{eq:nonneg-tail}
\end{equation}
\end{lemma}

\begin{lemma}[Sample covariance under moment bounds, {\cite{Tikhomirov2018}, Corollary~2}]\label{lem:cov-bnd}
Let $k\ge2$ and let $x\in\bbR^k$ be centered and isotropic, meaning $\bbE x=0$ and $\bbE xx^{\sfT}=\IdMat{k}$. Suppose $q>4$, $B\ge1$, and $\sup_{\norm{h}_2=1}\bbE\left|h^{\sfT}x\right|^q\le B$. For $m\ge2k$, independent copies $x_1,\ldots,x_m$ satisfy
\begin{equation}
 \norm*{\frac1m\sum_{i=1}^m x_ix_i^{\sfT}-\IdMat{k}}
 \le C_q\left(\frac{\max_i\norm{x_i}_2^2}{m}
                  +B^{2/q}\sqrt{\frac km}\lsp\right)
 \label{eq:tikh-cov}
\end{equation}
with probability at least $1-1/k$, where $C_q$ depends only on $q$.
\end{lemma}

For the frame estimate, we apply this covariance bound with $q=6$ and $k=2d$; the third projective moment supplies the required moment bound.

\begin{proof}[Proof of \lref{lem:nonneg-tail}]
For $u\ge0$, $\rme^{-u}\le1-u+u^2/2$. Therefore, for $\lambda\ge0$,
\begin{equation}
\bbE\rme^{-\lambda z}\le1-\lambda\mu+\frac{\lambda^2b}{2}\le\exp\left(-\lambda\mu+\frac{\lambda^2b}{2}\right).
 \label{eq:nonneg-laplace}
\end{equation}
Exponential Markov and independence give an upper bound $\exp\left(-m\lambda a+m\lambda^2b/2\right)$ for the probability in \eref{eq:nonneg-tail}. Choosing $\lambda=a/b$ proves the claim.
\end{proof}

\subsubsection{The set \texorpdfstring{$\Xi_d$}{Xi-d} and the success event}\label{app:joint-evt}

Let $\Xi_d$, $S_m$, $F_m$, and $\Omega_m$ be as defined in \sref{sec:samp}, and set \mbox{$\beta=1/4$}.

Every $\Delta\in\Xi_d$ equals $\Delta_{\psi,\nu}$ for some unit kets $\ket\psi$ and $\ket \nu\in\caH_\psi$ and has
$\norm{\Delta}_2=\sqrt{2}$. Indeed, if $\ket{e_+},\ket{e_-}$ are its unit eigenkets,
take $\ket\psi=(\ket{e_+}+\ket{e_-})/\sqrt{2}\lsp$ and $\ket \nu=(\ket{e_+}-\ket{e_-})/\sqrt{2}$.
All projectors below are sampled independently from a weighted exact $3$-design. The second and third design moments bound the mean absolute signal from below. The variables $d|\tr(\phi\Delta)|$ are nonnegative, and their bounded second moments allow the lower-tail estimate in \lref{lem:nonneg-tail} to be applied. The net argument proves $F_m(\Delta)\ge1/4$ simultaneously for all $\Delta\in\Xi_d$ on the event $\caA_m\cap\{S_m\preceq2\bbone\}$ in \lref{lem:unif-sig}. The frame estimate bounds the probability of $S_m\preceq2\bbone$.
\subsubsection{Uniform absolute-signal bound}\label{app:abs-sig}

\begin{lemma}[Absolute-signal moments]\label{lem:sig-mom}
For $\Delta\in\Xi_d$ and a projector $\phi$ drawn from the design weights, put $Z=d|\tr(\phi \Delta)|$. Then
\begin{equation}
 \bbE Z^2=\frac{2d}{d+1}\le2,\qquad
 \bbE Z^3\le\frac{8d^2}{(d+1)(d+2)},\qquad
 \bbE Z\ge\frac{d+2}{2(d+1)}\ge\frac12.
 \label{eq:sig-moments}
\end{equation}
\end{lemma}

\begin{lemma}[A net for $\Xi_d$]\label{lem:tan-net}
For $0<\varepsilon<1$, $\Xi_d$ has an $\varepsilon$-net in
Hilbert--Schmidt norm, with centers in $\Xi_d$, of cardinality at most
$(C_0/\varepsilon)^{4d}$ for a universal constant $C_0$.
\end{lemma}

\begin{lemma}[Uniform absolute signal]\label{lem:unif-sig}
Let $d\ge2$ and $m\ge1$ be integers, and draw $\phi_1,\ldots,\phi_m$ independently from a finite weighted exact projective $3$-design. Let $\Xi_d$ be the Hermitian operators with spectrum $(1,-1,0,\ldots,0)$, and let $S_m,F_m$ be as in Eqs.~\eqref{eq:samp-povm} and~\eqref{eq:samp-frm}. If $m\ge6154d$, there is an event $\caA_m$ with $\bbP(\caA_m)\ge3/4$ such that
\begin{equation}
\inf_{\Delta\in\Xi_d}F_m(\Delta)\ge\frac14\quad\text{on }\caA_m\cap\{S_m\preceq2\bbone\}.
 \label{eq:unif-sig}
\end{equation}
\end{lemma}

\begin{proof}[Proof of \lref{lem:sig-mom}]
Write $y=\tr(\phi \Delta)$ and $s=\tr(\phi|\Delta|)$. Since $-|\Delta|\preceq \Delta\preceq|\Delta|$, one has $|y|\le s$ and $|y|^3\le y^2s$. Also $\tr \Delta=0$ and $\Delta^2=|\Delta|$ is a rank-two projector. The two identities in \eref{eq:des-2-3-mom} give
\begin{equation}
 \bbE y^2=\frac2{d(d+1)},\qquad
 \bbE\left(y^2s\right)=\frac{\tr(\Delta^2)\tr|\Delta|+2\tr(\Delta^2|\Delta|)}{d(d+1)(d+2)}
 =\frac8{d(d+1)(d+2)}.
\end{equation}
Multiplying by $d^2$ or $d^3$ proves the second- and third-moment bounds. Finally, Cauchy--Schwarz applied to $Z^{1/2}$ and $Z^{3/2}$ yields
\begin{equation}
 \left(\bbE Z^2\right)^2\le(\bbE Z)\left(\bbE Z^3\right),\qquad
 \bbE Z\ge\frac{(\bbE Z^2)^2}{\bbE Z^3}\ge\frac{d+2}{2(d+1)}.
\end{equation}
\end{proof}

\begin{proof}[Proof of \lref{lem:tan-net}]
Equip the compact set of orthonormal ket pairs $(\ket\psi,\ket \nu)$ with the product Euclidean norm. It lies in the sphere of radius $\sqrt{2}\lsp$ in $\bbR^{4d}$. A maximal $\zeta$-separated subset covers this set by balls of radius $\zeta$, with centers that remain orthonormal pairs. Balls of radius $\zeta/2$ around its centers are disjoint and lie in the ball of radius $\sqrt{2}\lsp+\zeta/2$. Comparing their Euclidean volumes gives at most $(1+2\sqrt{2}\lsp/\zeta)^{4d}$ centers.

For two such pairs $(\ket\psi,\ket \nu)$ and $(\ket\varphi,\ket\upsilon)$, the identity $\norm*{xy^\dagger}_2=\norm{x}_2\norm{y}_2$ for vectors $x,y\in\caH$ and the triangle inequality yield
\begin{equation}
 \norm{\Delta_{\psi,\nu}-\Delta_{\varphi,\upsilon}}_2
 \le2\bigl(\norm{\ket\psi-\ket\varphi}_2+\norm{\ket \nu-\ket\upsilon}_2\bigr)
 \le2\sqrt{2}\lsp\norm{(\ket\psi,\ket \nu)-(\ket\varphi,\ket\upsilon)}_2.
 \label{eq:tan-map-lip}
\end{equation}
Take $\zeta=\varepsilon/(2\sqrt{2}\lsp)$. The images of these centers are in $\Xi_d$ and form an $\varepsilon$-net there because every $\Delta\in\Xi_d$ can be represented by orthonormal kets. The volume bound becomes $(1+8/\varepsilon)^{4d}\le(9/\varepsilon)^{4d}$ for $0<\varepsilon<1$, so one may take $C_0=9$.
\end{proof}

\begin{proof}[Proof of \lref{lem:unif-sig}]
For fixed $\Delta\in\Xi_d$, let $Z=d|\tr(\phi \Delta)|$. By \lref{lem:sig-mom}, $\bbE Z\ge1/2$ and $\bbE Z^2\le2$. Apply \lref{lem:nonneg-tail} with $b=2$ and $a=1/8$. Since $F_m(\Delta)$ is the empirical mean of $Z$, and $F_m(\Delta)<3/8$ entails a downward deviation of at least $1/8$, the result is
\begin{equation}
\bbP\left\{F_m(\Delta)<\frac38\right\}\le\exp\left(-\frac m{256}\right).
 \label{eq:fixed-sig-tail}
\end{equation}

Choose a fixed $1/32$-net $\caN$ in $\Xi_d$ for the Hilbert--Schmidt norm. By \lref{lem:tan-net} with $C_0=9$, its size is at most $288^{4d}$. Let $\caA_m$ be the event that $F_m(\Delta)\ge3/8$ for every $\Delta\in\caN$. The union bound gives
\begin{equation}
\bbP\left(\caA_m^c\right)\le288^{4d}\exp\left(-\frac m{256}\right)\le4^{-d}\qquad(m\ge6154d),
 \label{eq:sig-net-prob}
\end{equation}
where $6154>256[4\log288+\log4]$ is sufficient.

On the event $S_m\preceq2\bbone$, each projector $\phi_i\succeq0$ satisfies $|\tr[\phi_i(\Delta-Y)]|\le\tr(\phi_i|\Delta-Y|)$. Summing over $i$ and using $\norm{\Delta-Y}_1\le2\norm{\Delta-Y}_2$, which follows from $\rk(\Delta-Y)\le4$, gives
\begin{equation}
|F_m(\Delta)-F_m(Y)|\le\tr(S_m|\Delta-Y|)\le2\norm{\Delta-Y}_1\le4\norm{\Delta-Y}_2\qquad(\Delta,Y\in\Xi_d).
 \label{eq:frm-lip}
\end{equation}
For any $\Delta\in\Xi_d$, choose $Y\in\caN$ with $\norm{\Delta-Y}_2\le1/32$. Equations~\eqref{eq:sig-net-prob} and~\eqref{eq:frm-lip} give $F_m(\Delta)\ge3/8-4/32=1/4$ on the stated joint event, proving \eqref{eq:unif-sig}.
\end{proof}

\subsubsection{Frame bound from sixth moments}\label{app:frm-bnd}

\begin{lemma}[Frame bound for a projective $3$-design]\label{lem:frm-lin}
There is a universal constant $C_3$ such that, for $m\ge4d$,
\begin{equation}
 \bbP\!\left\{\norm{S_m-\bbone}
       \le C_3\left(\sqrt{\frac dm}\lsp+\frac dm\right)\right\}
 \ge1-\frac1{2d}.
 \label{eq:lin-frm-conc}
\end{equation}
Consequently, for a sufficiently large universal $C_4$ and
$m\ge C_4d$,
\begin{equation}
 \bbP\{S_m\preceq2\bbone\}\ge1-\frac1{2d}\ge\frac34.
 \label{eq:lin-frm-evt}
\end{equation}
\end{lemma}
\begin{proof}[Proof of \lref{lem:frm-lin}]
Choose a unit vector $\ket\phi$ for each sampled projector $\phi$ and add
an independent uniform phase $\theta\in[0,2\pi)$. Using
the real coordinates in \eref{eq:real-def}, define the real random vector
\begin{equation}
 z=\rme^{\rmi\theta}\ket\phi,\qquad x=\sqrt{2d}\lsp\fkr(z)\in\bbR^{2d}.
 \label{eq:ph-real}
\end{equation}
For a fixed $\ket\phi$, phase averaging gives $\bbE_\theta\fkr(z)=0$ and $\bbE_\theta\left[\fkr(z)\fkr(z)^{\sfT}\right]=\fkR(\phi)/2$. Together with $\bbE\phi=\bbone/d$, these identities imply
\begin{equation}
 \bbE x=0,\qquad \bbE xx^{\sfT}=\IdMat{2d},
 \qquad\norm{x}_2^2=2d.
 \label{eq:real-iso}
\end{equation}
Every real unit vector $h$ is $\fkr(\upsilon)$ for a unit ket $\ket\upsilon\in\caH$.
Since $\bbE_\theta\left[\left(\Rea\left(\rme^{\rmi\theta}a\right)\right)^6\right]=5|a|^6/16$ for any $a\in\bbC$,
the third projective moment gives
\begin{equation}
 \bbE\left|h^{\sfT}x\right|^6
 =\frac5{16}(2d)^3\bbE|\braket{\upsilon}{\phi}|^6
 =\frac5{16}\frac6{d(d+1)(d+2)}(2d)^3
 =\frac{15d^2}{(d+1)(d+2)}\le15.
 \label{eq:real-6-mom}
\end{equation}
The hypotheses of \lref{lem:cov-bnd} are now verified: the vectors are independent, centered and isotropic by \eref{eq:real-iso}, and have uniformly bounded sixth moments by \eref{eq:real-6-mom}. Their squared norm is deterministically $2d$. Taking $k=2d$, $q=6$, and $B=15$ in
\eref{eq:tikh-cov} yields, for
$T_m=m^{-1}\sum_i x_ix_i^{\sfT}$,
\begin{equation}
 \norm{T_m-\IdMat{2d}}
 \le C_3\left(\sqrt{\frac dm}\lsp+\frac dm\right)
 \label{eq:real-samp-cov}
\end{equation}
with probability at least $1-1/(2d)$. Here one may take $C_3=15^{1/3}\sqrt{2}\lsp C_6$, where $C_6$ denotes $C_q$ at $q=6$ in \eref{eq:tikh-cov}: the two terms there are $C_6\left[2d/m+15^{1/3}\lsp\sqrt{2d/m}\lsp\right]$, and $2\le15^{1/3}\sqrt{2}$.

To return to the complex frame, take $z\in\caH$ and put $a=\fkr(z)$. Direct block multiplication gives $\fkR\left(zz^\dagger\right)=aa^{\sfT}+J_0aa^{\sfT}J_0^{\sfT}$. Applying this identity to each phase-rotated unit ket and its scaled real vector $x_i$ gives, for every realization of the projectors and auxiliary phases,
\begin{equation}
 \fkR(S_m)=\frac12\left(T_m+J_0T_mJ_0^{\sfT}\right).
 \label{eq:cplx-frm-real}
\end{equation}
Since $J_0$ is orthogonal and $\fkR$ preserves the operator norm,
\begin{equation}
 \norm{S_m-\bbone}
 \le\norm{T_m-\IdMat{2d}}.
\end{equation}
Put $a_m=C_3(\sqrt{d/m}\lsp+d/m)$. The preceding pointwise inequality gives
\begin{equation}
 \bbP_{\phi}\{\norm{S_m-\bbone}>a_m\}
 =\bbP_{\phi,\theta}\{\norm{S_m-\bbone}>a_m\}
 \le\bbP_{\phi,\theta}\{\norm{T_m-\IdMat{2d}}>a_m\}\le\frac1{2d}.
 \label{eq:frame-phase-marginal}
\end{equation}
The equality holds because $S_m$ depends only on the sampled projectors, whose marginal law is unchanged by adding the independent phases. This proves \eref{eq:lin-frm-conc} for the original projector samples. Choosing $C_4=\max\left\{4,4C_3^2\right\}$ gives
$C_3\left(C_4^{-1/2}+C_4^{-1}\right)\le3/4<1$ and proves
\eref{eq:lin-frm-evt}.
\end{proof}

\Needspace{12\baselineskip}
\subsection{Proof of \thref{thm:samp}: linear sampling with spectral completion}\label{app:comp}

\begin{lemma}[Absolute signals, Fisher information, and separation]\label{lem:l1-fi}
For any probability vectors $p,q$,
\begin{equation}
 \norm{p-q}_1\le2\sqrt{\epsilon_{\rmcl}(p,q)}\lsp\le2H(p,q).
 \label{eq:classical-comparison}
\end{equation}
For every finite POVM $\caM$, pure state $\rho$, and tangent $\Delta\in T_\rho$,
\begin{equation}
 I_{\rho,\caM}(\Delta,\Delta)\ge\left[\sum_i|\tr(E_i\Delta)|\right]^2.
 \label{eq:l1-fi}
\end{equation}
If, for some $a\ge0$, $\sum_i|\tr(E_i\Delta)|\ge a$ for every $\Delta\in\Xi_d$, then
\begin{equation}
 \eta(\caM)\ge\frac{a^2}{4},\qquad
 H^2(p_\rho,p_\sigma)\ge\frac{a^2}{4}\epsilon(\rho,\sigma)
 \quad(\rho,\sigma\in\caP(\caH)).
 \label{eq:abs-fi-hel}
\end{equation}
\end{lemma}

\begin{proof}[Proof of \thref{thm:samp}]
By Lemmas~\ref{lem:unif-sig} and \ref{lem:frm-lin}, we may choose
\begin{equation}
 C=\max\{6154,C_4\}=\max\left\{6154,8\,15^{2/3}C_6^2\right\}.
 \label{eq:samp-const-val}
\end{equation}
Here $C_6$ is the universal constant in \lref{lem:cov-bnd} with $q=6$.
The event $\caA_m\cap\{S_m\preceq2\bbone\}$ is contained in $\Omega_m$ by \lref{lem:unif-sig} and \eref{eq:samp-event}. The union bound, \eref{eq:sig-net-prob}, and \lref{lem:frm-lin} therefore give
\begin{equation}
 \bbP(\Omega_m)\ge1-\frac1{2d}-4^{-d}.
 \label{eq:joint-evt}
\end{equation}
On $\Omega_m$, the sampled elements sum to $S_m/2\preceq\bbone$.
The residual $\bbone-S_m/2$ is positive semidefinite and has the spectral decomposition
\begin{equation}
 \bbone-\frac{S_m}{2}=\sum_{j=1}^{r_{\rmres}}\lambda_j\proj{e_j},
 \quad r_{\rmres}\le d,\quad\lambda_j>0,
\end{equation}
where $r_{\rmres}$ is its rank and $\ket{e_j}$ are orthonormal eigenkets, retaining only positive eigenvalues. Adding these residual elements gives the POVM $\caM=\caM_m$ in \eref{eq:spec-comp}, which is determined by the sampled projectors. For every $\Delta\in\Xi_d$,
\begin{equation}
 \sum_j|\tr(E_j\Delta)|\ge\frac12F_m(\Delta)\ge\frac\beta2.
 \label{eq:comp-abs}
\end{equation}
By \eref{eq:samp-event}, $\beta=1/4$ on $\Omega_m$. Applying \lref{lem:l1-fi} with $a=\beta/2$ gives \eref{eq:samp-fi} with $c_*=\beta^2/16=1/256$, including singular states. The lemma also gives the global separation used in \pref{prop:samp-sep}.

Finally, the completion adds at most $d$ outcomes, so $M\le m+d$.
\end{proof}

\begin{proof}[Proof of \lref{lem:l1-fi}]
\emph{Probability comparison.} For arbitrary probability vectors $p,q$, put $B(p,q)=\sum_i\sqrt{p_iq_i}$. Applying Cauchy--Schwarz to $|p_i-q_i|=|\sqrt{p_i}\lsp-\sqrt{q_i}\lsp|(\sqrt{p_i}\lsp+\sqrt{q_i}\lsp)$ gives the classical fidelity--variation inequality~\cite{FuchsVDG1999},
\begin{equation}
 \norm{p-q}_1
 \le H(p,q)\sqrt{2+2B(p,q)}\lsp
 =2\sqrt{\epsilon_{\rmcl}(p,q)}\lsp
 \le2H(p,q).
 \label{eq:l1-hel}
\end{equation}
The last step follows from $0\le B(p,q)\le1$ and $\epsilon_{\rmcl}=H^2\left(1-H^2/4\right)\le H^2$. This proves \eref{eq:classical-comparison}.

\emph{Fisher information.} At $\rho=\proj\psi$, if $p_i(\rho)=0$, the positive semidefiniteness of $E_i$ implies $E_i^{1/2}\ket\psi=0$ and hence $E_i\ket\psi=0$. Equation~\eqref{eq:tan} then gives $\tr(E_i\Delta)=0$. Applying Cauchy--Schwarz to the outcomes with $p_i>0$, and using $\sum_i p_i=1$, yields
\begin{equation}
 \left[\sum_i|\tr(E_i\Delta)|\right]^2
 \le\sum_{i\colon p_i>0}\frac{[\tr(E_i\Delta)]^2}{p_i}=I_{\rho,\caM}(\Delta,\Delta).
 \label{eq:l1-fi-prf}
\end{equation}
For nonzero $\Delta\in T_\rho$, the rank-two form in \eref{eq:tan} gives $\Delta/\sqrt{\tr\left(\Delta^2\right)/2}\lsp\in\Xi_d$. Applying the assumed signal bound to this normalized operator gives $\sum_i|\tr(E_i\Delta)|\ge a\sqrt{\tr\left(\Delta^2\right)/2}$. Substituting this into \eref{eq:l1-fi-prf} gives
\begin{equation}
 I_{\rho,\caM}(\Delta,\Delta)
 \ge\frac{a^2}{2}\tr\left(\Delta^2\right)
 =\frac{a^2}{4}J_\rho(\Delta,\Delta),
\end{equation}
where the last equality is \eref{eq:qfi}. The bound also holds at $\Delta=0$.

\emph{Global separation.} For distinct pure states, put $\epsilon=\epsilon(\rho,\sigma)$. The nonzero eigenvalues of $\rho-\sigma$ are $\pm\sqrt{\epsilon}$, since it is traceless, has rank two, and satisfies $\tr\left[(\rho-\sigma)^2\right]=2\epsilon$. Thus $(\rho-\sigma)/\sqrt{\epsilon}\lsp\in\Xi_d$ and $\norm{p_\rho-p_\sigma}_1\ge a\sqrt{\epsilon}$.

Applying \eref{eq:classical-comparison} to these output distributions gives
\begin{equation}
 H^2(p_\rho,p_\sigma)\ge\frac14\norm{p_\rho-p_\sigma}_1^2
 \ge\frac{a^2}{4}\epsilon(\rho,\sigma),
\end{equation}
as asserted in \eref{eq:abs-fi-hel}. When $\rho=\sigma$, both sides are zero.
\end{proof}

\subsection{Proof of \pref{prop:basis-samp}: orthonormal-basis sampling}\label{app:basis-sampling}

\begin{proof}[Proof of \pref{prop:basis-samp}]
Let $q_b$ be the probabilities of the bases $\caB_b=\{\ket{\phi_{bj}}\}_{j=1}^d$. The ensemble $\{(q_b/d,\phi_{bj})\}_{b,j}$ satisfies \eref{eq:des} with $s=3$. For independent draws $b_1,\ldots,b_L$, the POVM elements are
\begin{equation}
 \caM_L^{\rmbas}=\{E_{\ell j}\}_{\ell,j},\qquad E_{\ell j}=\frac{\phi_{b_\ell j}}L.
 \label{eq:basis-povm-elements}
\end{equation}
\emph{Absolute signal.} For $\Delta\in\Xi_d$, define
\begin{equation}
 A_{\caB}(\Delta)=\sum_{j=1}^d|\tr(\phi_j\Delta)|,\qquad
 F_L^{\rmbas}(\Delta)=\frac1L\sum_{\ell=1}^L A_{\caB_{b_\ell}}(\Delta).
 \label{eq:basis-signal}
\end{equation}
Each basis projector $\phi_j\succeq0$ satisfies $|\tr(\phi_j\Delta)|\le\tr(\phi_j|\Delta|)$. Summing over $j$ and using $\sum_j\phi_j=\bbone$ gives $0\le A_{\caB}(\Delta)\le\tr|\Delta|=\norm{\Delta}_1=2$. The second and third moments of the associated projector ensemble, through \lref{lem:sig-mom}, give
\begin{equation}
 \bbE A_{\caB}(\Delta)=\bbE_\pi\bigl[d|\tr(\phi \Delta)|\bigr]
 \ge\frac{d+2}{2(d+1)}\ge\frac12.
 \label{eq:basis-mean}
\end{equation}
Hoeffding's inequality~\cite{Hoeffding1963} for the independent variables $A_{\caB_{b_\ell}}(\Delta)\in[0,2]$ therefore gives
\begin{equation}
 \bbP\left\{F_L^{\rmbas}(\Delta)<\frac38\right\}
 \le\exp\left(-\frac L{128}\right).
 \label{eq:basis-fixed-tail}
\end{equation}

\emph{Uniform bound.} For every realization and $\Delta,Y\in\Xi_d$,
\begin{equation}
 \left|F_L^{\rmbas}(\Delta)-F_L^{\rmbas}(Y)\right|
 \le\norm{\Delta-Y}_1\le2\norm{\Delta-Y}_2,
 \label{eq:basis-lipschitz}
\end{equation}
since $\sum_j|\tr[\phi_j(\Delta-Y)]|\le\tr|\Delta-Y|$ and $\rk(\Delta-Y)\le4$. By \lref{lem:tan-net}, a fixed $1/16$-net in $\Xi_d$ has at most $144^{4d}$ points. A union bound in \eref{eq:basis-fixed-tail}, followed by \eref{eq:basis-lipschitz}, gives
\begin{equation}
 \bbP\left\{\inf_{\Delta\in\Xi_d}F_L^{\rmbas}(\Delta)<\frac14\right\}
 \le144^{4d}\exp\left(-\frac L{128}\right).
 \label{eq:basis-uniform-tail}
\end{equation}
Indeed, signal at least $3/8$ at all net points implies signal at least $3/8-2/16=1/4$ everywhere. For $L\ge2700d$, the failure probability is at most $\rme^{-d}$ because $2700/128>4\log144+1$. Thus, with probability at least $1-\rme^{-d}$,
\begin{equation}
 \sum_{\ell,j}|\tr(E_{\ell j}\Delta)|=F_L^{\rmbas}(\Delta)\ge\frac14
 \qquad(\Delta\in\Xi_d).
 \label{eq:basis-absolute-signal}
\end{equation}
Lemma~\ref{lem:l1-fi} with $a=1/4$ gives both the Fisher floor in \eref{eq:basis-fisher-guarantee} and $\kappa\left(\caM_L^{\rmbas}\right)\ge1/64$ on this event.
\end{proof}

\subsection{Proof of \pref{prop:stab-basis-obstruction}: a lower bound on the number of stabilizer bases}
\label{app:stab-basis-obstruction}

\begin{proof}[Proof of \pref{prop:stab-basis-obstruction}]
Label Hermitian Pauli operators by $(a,b)\in\bbF_2^n\times\bbF_2^n$, with $P(a,b)\propto\bigotimes_{k=1}^n\left(X^{a_k}Z^{b_k}\right)$. Complex conjugation in the computational basis gives $\overline{P(a,b)}=(-1)^{a\cdot b}P(a,b)$; all dot products here are over $\bbF_2$. Let $W_\ell$ be the commuting label space of the $\ell$th stabilizer basis and put $q(a,b)=a\cdot b$. For $(a,b),(a',b')\in W_\ell$, Pauli commutativity gives $a\cdot b'+a'\cdot b=0$, so
\begin{equation}
 q(a+a',b+b')=q(a,b)+q(a',b')+a\cdot b'+a'\cdot b=q(a,b)+q(a',b').
\end{equation}
Thus $q$ is linear on $W_\ell$. Since $q(0,b)=0$, its value depends only on the first component $a$. Extending the resulting linear functional to $\bbF_2^n$, choose $s_\ell\in\bbF_2^n$ such that $q(a,b)=s_\ell\cdot a$ on $W_\ell$.

Each basis projector $\phi_{\ell j}$ is a real linear combination of the Paulis labeled by $W_\ell$. Only labels with first component $a$ contribute to $\bra0\phi_{\ell j}\ket a$, where computational kets are indexed by $\bbF_2^n$. Hence
\begin{equation}
 \overline{\bra0\phi_{\ell j}\ket a}
 =(-1)^{s_\ell\cdot a}\bra0\phi_{\ell j}\ket a.
\end{equation}
There are $L-1<n$ homogeneous constraints $(s_\ell-s_1)\cdot a=0$, so they admit a nonzero solution $a$. Put $r=s_1\cdot a\in\{0,1\}$ and $z=\rmi^{1-r}$. Then $\Rea[z\bra0\phi_{\ell j}\ket a]=0$ for every $\ell,j$. The orthogonal states
\begin{equation}
 \ket{\psi_\pm}=\frac{\ket0\pm z\ket a}{\sqrt{2}\lsp},\qquad
 \rho_\pm=\proj{\psi_\pm}
\end{equation}
satisfy
\begin{equation}
 \tr[\phi_{\ell j}(\rho_+-\rho_-)]
 =2\Rea[z\bra0\phi_{\ell j}\ket a]=0
 \qquad\text{for every }\ell,j.
\end{equation}
They therefore have identical outcome probabilities in every selected basis and in every mixture of these bases. Proposition~\ref{prop:pure-ic} gives $\eta=0$.
\end{proof}

\section{Proofs of statistical guarantees}\label{app:est}

In this appendix, we prove \pref{prop:sym-fi}, \thref{thm:fi-glob}, and Propositions~\ref{prop:glob-3}--\ref{prop:mle-risk}.

\subsection{Proof of \pref{prop:sym-fi}: symmetric amplitudes and Fisher information}\label{app:sym-score}

\begin{proof}[Proof of \pref{prop:sym-fi}]
Put $\rho=\proj\psi$ and $b_i=\braket{\phi_i}{\nu}$, with $a_i=\braket{\phi_i}{\psi}$ as in \eref{eq:hor-fi}. Along the two curves, the square-root probabilities are
\begin{equation}
 f_{i,\pm}(\theta):=\sqrt{p_i(\rho_\pm(\theta))}\lsp
 =\sqrt{w_i}\lsp|a_i\cos\theta\pm b_i\sin\theta|.
 \label{eq:sym-born-amps}
\end{equation}
For $a_i\ne0$, Taylor expansion gives
\begin{equation}
 f_{i,\pm}(\theta)
 =\sqrt{w_i}\lsp\left[|a_i|\pm\theta
       \frac{\Rea(\ba_ib_i)}{|a_i|}\right]+o(\theta).
 \label{eq:sym-amp-deriv}
\end{equation}
If $a_i=0$, the individual amplitudes need not be differentiable at zero, but they coincide:
\begin{equation}
 f_{i,+}(\theta)=f_{i,-}(\theta)
 =\sqrt{w_i}\lsp|b_i|\,|\sin\theta|.
 \label{eq:sym-zero-amps}
\end{equation}
These outcomes contribute zero to $H^2(p_{\rho_+(\theta)},p_{\rho_-(\theta)})$. Since the number of outcomes is finite, the limit passes through the sum, giving
\begin{equation}
 \lim_{\theta\to0}
 \frac{H^2(p_{\rho_+(\theta)},p_{\rho_-(\theta)})}{4\theta^2}
 =\sum_{i\colon a_i\ne0}w_i
       \frac{[\Rea(\ba_ib_i)]^2}{|a_i|^2}
 =\inner{\nu}{G_{\psi,\caM}\nu}_{\bbR}.
 \label{eq:sym-amp-sum}
\end{equation}
The last equality follows from \eref{eq:hor-fi}. Finally, orthonormality gives $\epsilon(\rho_+(\theta),\rho_-(\theta))=\sin^2(2\theta)$, so its ratio to $4\theta^2$ tends to one. Equation~\eqref{eq:sym-amp-sum} therefore proves \eref{eq:sym-fi}.
\end{proof}

For comparison, put $\rho=\proj\psi$ and $\Delta=\Delta_{\psi,\nu}$. The same Taylor expansion with only the curve $\rho_+(\theta)$ and the fixed state $\rho$ gives
\begin{equation}
 \lim_{\theta\to0}\frac{H^2(p_{\rho_+(\theta)},p_\rho)}{\theta^2}
 =\frac14 I_{\rho,\caM}(\Delta,\Delta)
    +\sum_{i\colon p_i(\rho)=0}w_i|b_i|^2.
 \label{eq:anchor-amp-limit}
\end{equation}
The centered pairs in \pref{prop:sym-fi} cancel the additional sum over zero-probability outcomes in \eref{eq:anchor-amp-limit}.

\subsection{Proof of \thref{thm:fi-glob}: Fisher floors and global separation}\label{app:glob-sep}

We use two auxiliary lemmas: a phase-average identity for Fisher information and a Hellinger bound along a phase path.

\Needspace{14\baselineskip}
\begin{lemma}[Phase-average Fisher identity]\label{lem:fi-ph-avg}
Let $\caM$ be a finite rank-one POVM, let $\ket{e_+},\ket{e_-}$ be orthonormal, and put $P_\pm=\proj{e_\pm}$ and $\Delta=P_+-P_-$. Define
\begin{equation}
 \ket{\psi_\theta}=\frac{\ket{e_+}+\rme^{\rmi\theta}\ket{e_-}}{\sqrt{2}\lsp}.
 \label{eq:glob-ph-states}
\end{equation}
Put $\rho_\theta=\proj{\psi_\theta}$. Then $\Delta\in T_{\rho_\theta}$, $J_{\rho_\theta}(\Delta,\Delta)=4$, and
\begin{equation}
 \frac1{2\pi}\int_0^{2\pi}I_{\rho_\theta,\caM}(\Delta,\Delta)\,\rmd\theta
 =2\sum_i|\tr(E_i\Delta)|.
 \label{eq:fi-ph-average}
\end{equation}
\end{lemma}

\begin{lemma}[Hellinger bound along a phase path]\label{lem:ph-hel}
For a finite POVM, orthonormal kets $\ket\upsilon,\ket \nu$, and $c=\cos t$, $s=\sin t$ with $0<t\le\pi/4$, set $\ket{z_\alpha}=c\ket\upsilon+\rme^{\rmi\alpha}s\ket \nu$ and $\rho_\alpha=\proj{z_\alpha}$. Let $I_\alpha$ denote its classical Fisher information for the parameter $\alpha$. Then
\begin{equation}
 H^2(p_{\rho_0},p_{\rho_\pi})\le\frac1\pi\int_0^{2\pi}I_\alpha\,\rmd\alpha.
 \label{eq:ph-hel-int}
\end{equation}
\end{lemma}

\Needspace{5\baselineskip}
\begin{proof}[Proof of \thref{thm:fi-glob}]
Write $\eta=\eta(\caM)$. The case $\rho=\sigma$ is immediate. For distinct pure states, put $\epsilon=\epsilon(\rho,\sigma)$ and
\begin{equation}
 \Delta=\frac{\rho-\sigma}{\sqrt{\epsilon}\lsp}=P_+-P_-,\qquad
 \norm{\rho-\sigma}_1=2\sqrt{\epsilon},
 \label{eq:glob-rank-2}
\end{equation}
where $P_\pm$ are its rank-one eigenprojectors. In particular, $\Delta\in\Xi_d$.

\emph{Lower bounds.} Along the circle in \lref{lem:fi-ph-avg}, the uniform Fisher bound gives $I_{\rho_\theta,\caM}(\Delta,\Delta)\ge4\eta$. Averaging yields $\sum_i|\tr(E_i\Delta)|\ge2\eta$, and therefore
\begin{equation}
 \norm{p_\rho-p_\sigma}_1
 =\sqrt{\epsilon}\lsp\sum_i|\tr(E_i\Delta)|
 \ge2\eta\sqrt{\epsilon}\lsp=\eta\norm{\rho-\sigma}_1.
 \label{eq:glob-ph-bnd}
\end{equation}
This proves \eref{eq:fi-glob-l1}. The classical comparison \eref{eq:classical-comparison} in \lref{lem:l1-fi} then gives both $H^2(p_\rho,p_\sigma)\ge\eta^2\epsilon$ and $\epsilon_{\rmcl}(p_\rho,p_\sigma)\ge\eta^2\epsilon$.

\emph{Upper bounds.} POVM completeness and the triangle inequality give
\begin{equation}
 B(p_\rho,p_\sigma)
 =\sum_iw_i|\braket{\phi_i}{\psi}|\,|\braket{\phi_i}{\varphi}|
 \ge\left|\sum_iw_i\braket{\psi}{\phi_i}\braket{\phi_i}{\varphi}\right|
 =|\braket{\psi}{\varphi}|,
 \label{eq:cl-fid-mono}
\end{equation}
where $\rho=\proj\psi$ and $\sigma=\proj\varphi$. Squaring proves $\epsilon_{\rmcl}(p_\rho,p_\sigma)\le\epsilon$.

Write $G=G_{\psi,\caM}$. Equation~\eqref{eq:pairing} in \pref{prop:geometry}, together with $Z_\psi\succeq0$ and the uniform bound $G\succeq\eta\IdTan\psi$, gives
\begin{equation}
 G\preceq\IdTan\psi-\caJ G\caJ^{-1}\preceq(1-\eta)\IdTan\psi.
 \label{eq:global-operator-upper}
\end{equation}
Here conjugation by the real orthogonal operator $\caJ$ preserves the lower bound. Equation~\eqref{eq:normalized-fi-form} therefore gives $I_{\rho,\caM}\preceq(1-\eta)J_\rho$ at every state.

Rephase the unit vectors so that $a=\braket\psi\varphi=|\braket\psi\varphi|\in[0,1)$, and put
\begin{equation}
 c=\sqrt{\frac{1+a}{2}},\qquad s=\sqrt{\frac{1-a}{2}},\qquad
 \ket\upsilon=\frac{\ket\psi+\ket\varphi}{2c},\qquad
 \ket\nu=\frac{\ket\psi-\ket\varphi}{2s}.
 \label{eq:global-pair-coordinates}
\end{equation}
The last two kets are orthonormal, and $c=\cos t$, $s=\sin t$ for some $0<t\le\pi/4$. Thus $\ket\psi=c\ket\upsilon+s\ket\nu$ and $\ket\varphi=c\ket\upsilon-s\ket\nu$ are the endpoints of the phase path in \lref{lem:ph-hel}. Along that path, $\ket{\dot z_\alpha}=\rmi\rme^{\rmi\alpha}s\ket\nu$, so
\begin{equation}
 J_{\rho_\alpha}(\partial_\alpha\rho_\alpha,\partial_\alpha\rho_\alpha)
 =4\left(\norm{\dot z_\alpha}_2^2-|\braket{z_\alpha}{\dot z_\alpha}|^2\right)
 =4\left(s^2-s^4\right)=4c^2s^2=1-a^2=\epsilon.
 \label{eq:global-phase-qfi}
\end{equation}
Consequently $I_\alpha\le(1-\eta)\epsilon$. Applying \lref{lem:ph-hel} gives
\begin{equation}
 H^2(p_\rho,p_\sigma)\le\frac1\pi\int_0^{2\pi}I_\alpha\,\rmd\alpha\le2(1-\eta)\epsilon.
\end{equation}
This completes \eref{eq:fi-glob-hel}.

\emph{Separation coefficient.} Taking the infimum in the lower bound gives $\kappa(\caM)\ge\eta^2$. By \pref{prop:reg-eq}, there is an orthonormal pair $\ket{\psi_*},\ket{\nu_*}$ attaining the Fisher floor. For its symmetric pairs in \pref{prop:sym-fi},
\begin{equation}
 \kappa(\caM)\le\lim_{\theta\to0}\frac{H^2(p_{\rho_+(\theta)},p_{\rho_-(\theta)})}{\epsilon(\rho_+(\theta),\rho_-(\theta))}
 =\inner{\nu_*}{G_{\psi_*,\caM}\nu_*}_{\bbR}=\eta.
 \label{eq:kappa-symmetric-limit}
\end{equation}
This proves \eref{eq:fi-glob-sandwich}.
\end{proof}

\begin{proof}[Proof of \lref{lem:fi-ph-avg}]
Direct calculation gives $\rho_\theta\Delta+\Delta\rho_\theta=\Delta$. Thus $\Delta$ has only off-diagonal blocks with respect to $\rho_\theta$ and $\bbone-\rho_\theta$, so $\Delta\in T_{\rho_\theta}$ by \eref{eq:tan}. Since $\Delta^2=P_++P_-$, \eref{eq:qfi} gives $J_{\rho_\theta}(\Delta,\Delta)=4$.
For each POVM element $E_i$, define
\begin{equation}
 \alpha_i=\tr(E_iP_+),\qquad \beta_i=\tr(E_iP_-),\qquad
 z_i=\bra{e_+}E_i\ket{e_-}.
 \label{eq:glob-eff-compr}
\end{equation}
Since $E_i$ has rank one, $|z_i|^2=\alpha_i\beta_i$. Moreover,
\begin{equation}
 p_i(\rho_\theta)=\frac{\alpha_i+\beta_i+2\Rea(\rme^{\rmi\theta}z_i)}2,
 \qquad \tr(E_i\Delta)=\alpha_i-\beta_i.
 \label{eq:glob-ph-prob}
\end{equation}
If $\alpha_i\ne\beta_i$, then
$(\alpha_i+\beta_i)^2-4|z_i|^2=(\alpha_i-\beta_i)^2>0$, so this probability is positive for every $\theta$. We use the elementary integral, valid for $a>b\ge0$,
\begin{equation}
 \frac1{2\pi}\int_0^{2\pi}\frac{\rmd\theta}{a+b\cos\theta}
 =\frac1\pi\int_{-\infty}^{\infty}\frac{\rmd u}{(a+b)+(a-b)u^2}
 =\frac1{\sqrt{a^2-b^2}\lsp}.
 \label{eq:ph-recip-int}
\end{equation}
The middle expression follows by writing the angular integral on $(-\pi,\pi)$ and substituting $u=\tan(\theta/2)$. Shifting the phase of $z_i$ in \eref{eq:glob-ph-prob} therefore gives
\begin{equation}
 \frac1{2\pi}\int_0^{2\pi}
 \frac{(\alpha_i-\beta_i)^2}{p_i(\rho_\theta)}\,\rmd\theta
 =\frac{2(\alpha_i-\beta_i)^2}{|\alpha_i-\beta_i|}
 =2|\tr(E_i\Delta)|.
 \label{eq:glob-ph-out}
\end{equation}
If $\alpha_i=\beta_i$, the numerator vanishes. The corresponding summand in \eref{eq:score-fi} is zero when $p_i(\rho_\theta)>0$ and is omitted otherwise, so its average is also zero. This includes elements whose probability vanishes identically on the two-dimensional subspace. Summing over the finite set of outcomes proves \eref{eq:fi-ph-average}, including singular states.
\end{proof}

\begin{proof}[Proof of \lref{lem:ph-hel}]
Put $A_i=c^2\bra\upsilon E_i\ket\upsilon+s^2\bra\nu E_i\ket\nu$ and choose $R_i\ge0$ and $\delta_i\in\bbR$ so that $2cs\bra\upsilon E_i\ket\nu=R_i\rme^{\rmi\delta_i}$. Then $p_i(\rho_\alpha)=A_i+R_i\cos(\alpha+\delta_i)$. These Born probabilities are nonnegative for every $\alpha$, so their minimum $A_i-R_i$ is nonnegative and $A_i\ge R_i\ge0$. Differentiating the probabilities and substituting into \eref{eq:score-fi} gives
\begin{equation}
 I_\alpha=\sum_{i\colon p_i(\rho_\alpha)>0}
 \frac{R_i^2\sin^2(\alpha+\delta_i)}{A_i+R_i\cos(\alpha+\delta_i)}.
\end{equation}
For $A_i>R_i$, the substitution $u=\tan(\alpha/2)$ gives
\begin{equation}
 \int_0^{2\pi}\frac{\rmd\alpha}{A_i+R_i\cos\alpha}
 =2\int_{-\infty}^{\infty}\frac{\rmd u}{(A_i+R_i)+(A_i-R_i)u^2}
 =\frac{2\pi}{\sqrt{A_i^2-R_i^2}\lsp}.
\end{equation}
After shifting the phase in the outcome probability, we therefore obtain
\begin{equation}
 \int_0^{2\pi}\frac{R_i^2\sin^2\alpha}{A_i+R_i\cos\alpha}\,\rmd\alpha
 =\int_0^{2\pi}\left[A_i-R_i\cos\alpha-\frac{A_i^2-R_i^2}{A_i+R_i\cos\alpha}\right]\rmd\alpha
 =2\pi\left[A_i-\sqrt{A_i^2-R_i^2}\lsp\right].
\end{equation}
When $A_i=R_i>0$, the integrand equals $R_i(1-\cos\alpha)$ except at the isolated zero probability, so its integral is $2\pi R_i$, as in the same formula. If $A_i=0$, then $R_i=0$ and the outcome contributes zero. The endpoint probabilities are $A_i\pm R_i\cos\delta_i$. Consequently,
\begin{equation}
 H^2(p_{\rho_0},p_{\rho_\pi})=2\sum_i\left[A_i-\sqrt{A_i^2-R_i^2\cos^2\delta_i}\lsp\right]
 \le2\sum_i\left[A_i-\sqrt{A_i^2-R_i^2}\lsp\right]=\frac1\pi\int_0^{2\pi}I_\alpha\,\rmd\alpha.
\end{equation}
This proves \eref{eq:ph-hel-int} in \lref{lem:ph-hel}.
\end{proof}

\subsection{Proofs of Propositions~\ref{prop:glob-3} and~\ref{prop:hel}: Hellinger separation for \texorpdfstring{$3$}{3}-designs and odd orders}\label{app:glob-3-prf}\label{app:odd-hel-prf}

We prove the odd-order bound in \pref{prop:hel} and then establish sharpness at order three for \pref{prop:glob-3}. The following kernel identity evaluates the polynomial bound.

\begin{lemma}[Endpoint kernel for Hellinger separation]\label{lem:hel-end-kernel}
Let $d\ge3$ and $k\ge1$ be integers, and let $\mu=\mu_{2,d-3}$ be the $\Beta(3,d-2)$ probability measure defined in \eref{eq:jacobi-prob-wt}. The reproducing kernel of $\scrP_k$ in $L^2(\mu)$ obeys
\begin{equation}
 K_k^\mu(0,0)=\binom{k+3}{3}\frac{(d+1)_k}{(d-2)_k}.
 \label{eq:hel-kernel}
\end{equation}
\end{lemma}

\begin{proof}[Proof of Propositions~\ref{prop:glob-3} and~\ref{prop:hel}]
\emph{Odd-order lower bound.} Fix $k\ge2$ and $\caE\in\fkE_t(d)$ with $t\ge2k-1$, as in \pref{prop:hel}. The case $\rho=\sigma$ is immediate. Otherwise put
$\Delta=\rho-\sigma$, $\Sigma=\rho+\sigma$, and
$\epsilon=1-\tr(\rho\sigma)>0$. Let $P$ project onto the
two-dimensional span of the corresponding unit vectors. For a rank-one projector $\phi$ drawn from the design ensemble, set
\begin{equation}
 s=\tr(P\phi),\qquad x=\tr(\Delta\phi),\qquad y=\tr(\Sigma\phi).
 \label{eq:hel-rad}
\end{equation}
For $a,b\ge0$,
\begin{equation}
 \bigl(\sqrt{a}\lsp-\sqrt{b}\lsp\bigr)^2\ge\frac{(a-b)^2}{2(a+b)},
\end{equation}
with zero assigned when $a=b=0$. Since $p_i=d\pi_i\tr(\phi_i\rho)$,
Cauchy--Schwarz gives, for any real polynomial $h$ of degree at most
$k-2$ with positive denominator,
\begin{equation}
 H^2(p_\rho,p_\sigma)\ge\frac d2\bbE_\pi\frac{x^2}{y}
 \ge\frac d2\frac{[\bbE_\pi x^2h(s)]^2}
                       {\bbE_\pi yx^2h(s)^2}.
 \label{eq:hel-cert}
\end{equation}
Terms with $y=0$ have $x=0$ and do not affect either moment. The second inequality in \eref{eq:hel-cert} applies Cauchy--Schwarz to $x/\sqrt{y}\lsp$ and $x\sqrt{y}\lsp h(s)$, both assigned zero when $y=0$.
The polynomial degrees are at most $k$ and $2k-1$, respectively,
so the $t$-design property in \eref{eq:des} replaces both moments by their Haar values.

First suppose $d\ge3$. In coordinates adapted to the range of $P$ and its orthogonal complement, a Haar ket is obtained by normalizing independent standard complex Gaussian vectors $\xi\in\bbC^2$ and $\zeta\in\bbC^{d-2}$. Thus
\begin{equation}
 s=\frac{\norm{\xi}_2^2}{\norm{\xi}_2^2+\norm{\zeta}_2^2},\qquad
 \lambda(\rmd s)=(d-1)(d-2)s(1-s)^{d-3}\rmd s.
 \label{eq:hel-radial-law}
\end{equation}
Indeed, the two squared norms are independent gamma variables with shape parameters $2$ and $d-2$ and unit scale. Rotational invariance makes $\xi/\norm{\xi}_2$ a Haar unit vector in the range of $P$, independent of these norms and hence of $s$. Write $\tau$ for its pure state. The identities
$\tr\Delta=0$, $\tr\Sigma=2$, $\tr\Delta^2=2\epsilon$, and
$\tr\left(\Delta^2\Sigma\right)=2\epsilon$ give, by the second and third Haar
moments in dimension two,
\begin{equation}
 \bbE_\tau\left[\tr(\tau\Delta)^2\right]=\frac\epsilon3,
 \qquad
 \bbE_\tau\left[\tr(\tau\Delta)^2\tr(\tau\Sigma)\right]=\frac\epsilon3.
 \label{eq:hel-ang-mom}
\end{equation}
Consequently,
\begin{equation}
 \bbE_\Ha\left[x^2\mid s\right]=\frac\epsilon3s^2,
 \qquad
 \bbE_\Ha\left[yx^2\mid s\right]=\frac\epsilon3s^3.
 \label{eq:hel-cond-mom}
\end{equation}
Weight the law $\lambda$ in \eref{eq:hel-radial-law} by $s$ and normalize:
\begin{equation}
 \mu(\rmd s)=\frac{s\lambda(\rmd s)}{\bbE_\lambda s}
 =\frac{d(d-1)(d-2)}2s^2(1-s)^{d-3}\rmd s,
 \label{eq:hel-beta}
\end{equation}
where $\bbE_\lambda s=2/d$. Thus $\mu$ is the
$\Beta(3,d-2)$ probability measure. Substituting
$g(s)=sh(s)$ in \eref{eq:hel-cert} yields
\begin{equation}
 H^2(p_\rho,p_\sigma)\ge
 \frac\epsilon3\frac{(\bbE_\mu g)^2}{\bbE_\mu g^2}.
 \label{eq:hel-poly-quot}
\end{equation}
Set $g=1-q_{k-1}^\mu$, using the definition in \eref{eq:normalized-poly-def}. Since $g(0)=0$ and $\deg g\le k-1$, the quotient $h(s)=g(s)/s$ is a polynomial of degree at most $k-2$. Equation~\eqref{eq:opt-mom} gives
\begin{equation}
 \bbE_\mu g=\bbE_\mu g^2
 =1-\frac1{K_{k-1}^\mu(0,0)}>0.
\end{equation}
Substituting this choice into \eref{eq:hel-poly-quot} yields
\begin{equation}
 H^2(p_\rho,p_\sigma)\ge\frac\epsilon3
 \left[1-\frac1{K_{k-1}^\mu(0,0)}\right].
\end{equation}
Using the kernel value in \eref{eq:hel-kernel} proves \eref{eq:des-hel} for $d\ge3$.

\emph{Dimension two.} If $d=2$, then $P=\bbone$ and $s=1$. The angular moment identities in \eref{eq:hel-ang-mom} hold for the full Haar measure. Taking $h=1$ in \eref{eq:hel-cert} gives $H^2\ge\epsilon/3$. This is precisely $b_{2,k}\epsilon$ for every $k\ge2$, since $(0)_{k-1}=0$.

\emph{Third-order bound and sharpness.} The case $k=2$ of the bound just proved gives \eref{eq:3-des-glob-sep} in \pref{prop:glob-3}, since $b_{d,2}=(d+2)/[4(d+1)]=c_{d,3}$. For sharpness, let $\caE_*$ be the
finite design in \thref{thm:hier} with $k=2$. Its measurement
has $\eta(\caM_{\caE_*})=c_{d,3}$. The preceding lower bound and
\pref{prop:sym-fi} give
\begin{equation}
 c_{d,3}\le\kappa(\caM_{\caE_*})
           \le\eta(\caM_{\caE_*})=c_{d,3}.
\end{equation}
Equivalently, the centered pairs in \pref{prop:sym-fi}, chosen at a state and tangent direction attaining the Fisher floor, satisfy $H^2(p_{\rho_+},p_{\rho_-})/\epsilon(\rho_+,\rho_-)\to c_{d,3}$. Thus \eref{eq:des-hel} proves \pref{prop:hel}, while \eref{eq:3-des-glob-sep} and \eref{eq:sharp-glob-3} prove \pref{prop:glob-3}.
\end{proof}

\begin{proof}[Proof of \lref{lem:hel-end-kernel}]
Use the orthogonal basis $Q_\ell(s)=P_\ell^{(2,d-3)}(1-2s)$. The endpoint and norm formulas in Eqs.~\eqref{eq:jacobi-end-val} and \eqref{eq:jacobi-end-norm} give
\begin{equation}
 A_\ell:=\frac{Q_\ell(0)^2}{\bbE_\mu Q_\ell^2}
 =\frac{(\ell+1)(\ell+2)(2\ell+d)(\ell+d-1)(\ell+d-2)}{2d(d-1)(d-2)}.
 \label{eq:hel-jacobi-norm}
\end{equation}
Put $R_\ell=\binom{\ell+3}{3}(d+1)_\ell/(d-2)_\ell$. Direct subtraction gives $R_0=A_0=1$ and $R_\ell-R_{\ell-1}=A_\ell$ for $\ell\ge1$. Hence $K_k^\mu(0,0)=\sum_{\ell=0}^kA_\ell=R_k$, proving \eref{eq:hel-kernel}.
\end{proof}

\subsection{Proof of \pref{prop:samp-sep}: global separation for sampled POVMs}\label{app:samp-stat}

\begin{proof}[Proof of \pref{prop:samp-sep}]
On $\Omega_m$, the sampled elements in \eref{eq:samp-povm}, their completion in \eref{eq:spec-comp}, and the definition of $F_m$ in \eref{eq:samp-frm} give
\begin{equation}
 \sum_j|\tr(E_j\Delta)|\ge\frac12F_m(\Delta)\ge\frac18
 \qquad(\Delta\in\Xi_d).
\end{equation}
Thus \lref{lem:l1-fi} applies with $a=1/8$, giving $H^2(p_\rho,p_\sigma)\ge\epsilon(\rho,\sigma)/256=c_*\epsilon(\rho,\sigma)$. Taking the infimum over distinct pure-state pairs proves \eref{eq:samp-kappa}. For complete-basis sampling, apply \lref{lem:l1-fi} with $a=1/4$ to \eref{eq:basis-absolute-signal} in \aref{app:basis-sampling}; it gives $\kappa\left(\caM_L^{\rmbas}\right)\ge1/64$ with probability at least $1-\rme^{-d}$.
\end{proof}

\subsection{Proof of \pref{prop:loc-risk}: asymptotic infidelity of efficient estimators}\label{app:loc-risk}

\begin{proof}[Derivation of \eref{eq:loc-risk-coef}]
Write $\rho=\proj\psi$ and choose unit kets $\ket{e_1},\ldots,\ket{e_{2d-2}}$ forming a real orthonormal basis of $\caH_\psi$. For $z\in\bbR^{2d-2}$, use the local chart
\begin{equation}
 \ket{\psi(z)}=\frac{\ket\psi+\sum_{a=1}^{2d-2}z_a\ket{e_a}}{\sqrt{1+\norm{z}_2^2}\lsp},
 \qquad\rho(z)=\proj{\psi(z)}.
\end{equation}
At $z=0$, the parameter QFI and classical Fisher matrices are $4\IdMat{2d-2}$ and $4G$, respectively, where $G$ is the matrix of $G_{\psi,\caM}$ in this basis. Moreover,
\begin{equation}
 1-\tr[\rho\rho(z)]=\frac{\norm{z}_2^2}{1+\norm{z}_2^2}.
\end{equation}
Let $z_N=z(\hrho_N)$ for a pure-state estimator on this chart. We assume that $\rho$ is regular with respect to $\caM$, that $G\succ0$, and that the estimator has the efficient Gaussian limit $\sqrt{N}\lsp z_N\Rightarrow\caN\left(0,(4G)^{-1}\right)$ with convergence of second moments:
\begin{equation}
 N\bbE_\rho\left[z_Nz_N^{\sfT}\right]\longrightarrow(4G)^{-1}.
 \label{eq:mean-sq-eta}
\end{equation} If the estimator can leave a fixed neighborhood contained in the chart, we additionally require the probability of this event to be $o\left(N^{-1}\right)$ and define $z_N=0$ there. Since infidelity is bounded by one, this event contributes $o\left(N^{-1}\right)$ to the risk.

Convergence in distribution and of the second moments implies uniform integrability of $T_N:=N\norm{z_N}_2^2$. Also $z_N\to0$ in probability. On the chart, the difference between the scaled quadratic loss and the scaled infidelity is
\begin{equation}
 N\left(\norm{z_N}_2^2-\frac{\norm{z_N}_2^2}{1+\norm{z_N}_2^2}\right)
 =T_N\frac{\norm{z_N}_2^2}{1+\norm{z_N}_2^2}.
\end{equation}
Put $b_N=\norm{z_N}_2^2/(1+\norm{z_N}_2^2)$. It is bounded by one and tends to zero in probability, so $\bbE b_N\to0$. For every $K>0$,
\begin{equation}
 0\le\bbE(T_Nb_N)\le K\bbE b_N+\bbE[T_N\bfone_{\{T_N>K\}}].
 \label{eq:risk-ui-truncation}
\end{equation}
First let $N\to\infty$, then let $K\to\infty$; uniform integrability makes the last term uniformly small. Hence the expectation of the scaled loss difference tends to zero. Taking the trace in \eref{eq:mean-sq-eta} proves the risk expansion in \eref{eq:loc-risk-coef}.

\end{proof}

\begin{proof}[Proof of \pref{prop:loc-risk}]
Choose a unit vector $\ket\psi$ with $\rho=\proj\psi$ and put $G=G_{\psi,\caM_\caE}$. By \pref{prop:geometry} and \thref{thm:hier}, its eigenvalues are pairs $\lambda_j,1-\lambda_j$, with $c\le\lambda_j\le1-c$, for $1\le j\le d-1$. Hence
\begin{equation}
 d-1\le R_{\rho,\caM_\caE}
 =\frac14\sum_{j=1}^{d-1}\frac{1}{\lambda_j(1-\lambda_j)}
 \le\frac{d-1}{4c(1-c)}.
\end{equation}
This proves \eref{eq:loc-risk-bnd}. For odd order $t=2k-1$, the coefficient in \thref{thm:hier} satisfies
\begin{equation}
 c_{d,2k-1}=\frac12\left[1-\frac{d}{k(d+k-1)}\right]
 \ge\frac{k-1}{2k}.
\end{equation}
Since $c(1-c)$ increases for $0<c\le1/2$, \eref{eq:loc-risk-bnd} implies
\begin{equation}
 \frac{R_{\rho,\caM_\caE}}{d-1}
 \le\frac1{4c_{d,2k-1}(1-c_{d,2k-1})}
 \le\frac{k^2}{k^2-1}.
\end{equation}
Setting $k=2,3,4$ gives the three ratios in \eref{eq:loc-risk-rat}. For any positive definite normalized Fisher operator $G$ of a single-copy POVM, write $\gamma_1,\ldots,\gamma_{2d-2}>0$ for its eigenvalues. The rank-one refinement used in the proof of \pref{prop:reg-eq} gives $G\preceq\widetilde G$ and $\tr_{\bbR}\widetilde G\le d-1$ by \pref{prop:geometry}. Thus $\sum_j\gamma_j\le d-1$, and Cauchy--Schwarz gives
\begin{equation}
 \tr_{\bbR}G^{-1}=\sum_j\gamma_j^{-1}\ge\frac{(2d-2)^2}{\sum_j\gamma_j}\ge4(d-1).
 \label{eq:risk-trace-cs}
\end{equation}
Equality requires $\gamma_j=1/2$ for every $j$, equivalently $G=\IdTan\psi/2$. Fisher-symmetric measurements with this value at a prescribed state attain the local benchmark $d-1$~\cite{GillMassar2000,LiEtAl2016}.
\end{proof}

\subsection{Proof of \pref{prop:mle-risk}: finite-sample maximum-likelihood reconstruction}\label{app:fin-samp-risk}

For the bound independent of $M$, we use finite pointwise envelopes of the square-root Born probabilities.

\begin{lemma}[Finite amplitude envelopes]\label{lem:mle-env}
Let $\caM=\{w_i\phi_i\}_{i=1}^M$ be a finite rank-one POVM, and put $f_\psi(i)=\sqrt{w_i}\lsp|\braket{\phi_i}{\psi}|$ for each unit ket $\ket\psi$. For every $\varepsilon>0$, there are unit kets $\ket{\psi_1},\ldots,\ket{\psi_\ell}$ with $\ell\le(1+2/\varepsilon)^{2d}$ such that the nonnegative envelopes
\begin{equation}
 g_k(i)=f_{\psi_k}(i)+\varepsilon\sqrt{w_i},\qquad r_k(i)=g_k(i)^2
 \label{eq:mle-envelopes}
\end{equation}
satisfy, for every unit ket $\ket\psi$ with $\sigma=\proj\psi$ and a suitable index $k$,
\begin{equation}
 p_\sigma(i)\le r_k(i)\quad\text{for every }i,\qquad
 \norm{g_k-f_\psi}_2\le2\varepsilon\sqrt{d}.
 \label{eq:mle-env-control}
\end{equation}
The same index can be chosen with $\norm{\ket\psi-\ket{\psi_k}}_2\le\varepsilon$.
\end{lemma}

\begin{proof}[Proof of \pref{prop:mle-risk}]
Fix the true distribution $p=p_\rho$, put $\hp=p_{\hrho_N}$, and for a nonnegative vector $r$ write $L_N(r)=\prod_{j=1}^Nr(X_j)$. The likelihood is continuous on the compact pure-state space, so it attains a maximum; choosing one maximizer for each possible record gives a measurable estimator because the record space is finite. Almost surely, $L_N(p)>0$, since zero-probability outcomes are never observed.

\emph{Outcome-count bound.} Let $q_i=N^{-1}\sum_{j=1}^N\bfone_{\{X_j=i\}}$ be the empirical frequencies. For probability vectors $q,r$, put $D(q\|r)=\sum_{i\colon q_i>0}q_i\log(q_i/r_i)$, with value $+\infty$ if some $q_i>0$ has $r_i=0$. Almost surely, $q_i>0$ implies $p_i>0$ and, by $L_N(\hp)\ge L_N(p)>0$, also $\hp_i>0$. Likelihood optimality therefore gives
\begin{equation}
 D(q\|\hp)\le D(q\|p).
 \label{eq:mle-empirical-kl}
\end{equation}
Jensen's inequality for the logarithm, with weights $q_i$ on their positive support, yields
\begin{equation}
 B(p,\hp)\ge\sum_{i\colon q_i>0}q_i\frac{\sqrt{p_i\hp_i}\lsp}{q_i}
 \ge\exp\!\left[-\frac{D(q\|p)+D(q\|\hp)}2\right]
 \ge\rme^{-D(q\|p)}.
 \label{eq:mle-affinity-kl}
\end{equation}
Thus $H^2(p,\hp)\le2D(q\|p)$ by \eref{eq:hel-def} and $1-\rme^{-x}\le x$. The inequality $\log x\le x-1$ and the multinomial variances give
\begin{equation}
 D(q\|p)\le\sum_{i\colon p_i>0}\frac{(q_i-p_i)^2}{p_i},\qquad
 \bbE_pD(q\|p)\le\sum_{i\colon p_i>0}\frac{p_i(1-p_i)}{Np_i}
 =\frac{|\supp p|-1}{N}.
 \label{eq:mle-kl-mean}
\end{equation}
Consequently,
\begin{equation}
 \bbE_pH^2(p,\hp)\le\frac{2(|\supp p|-1)}N\le\frac{2(M-1)}N.
 \label{eq:mle-outcome-risk}
\end{equation}

\emph{Amplitude envelopes.} Apply \lref{lem:mle-env} with $\varepsilon_N=1/(4N\sqrt{d}\lsp)$. The number of envelopes satisfies
\begin{equation}
 \ell\le(1+8N\sqrt{d}\lsp)^{2d},\qquad
 \norm{g_k-f_\psi}_2\le\frac1{2N}
 \label{eq:mle-net-size}
\end{equation}
for a covering center of each unit ket $\ket\psi$.

\emph{Likelihood bound.} Set $b_k=\sum_i\sqrt{p_i}\lsp g_k(i)$. Markov's inequality and independence give
\begin{equation}
\bbP_p\{L_N(r_k)\ge L_N(p)\}\le\bbE_p\sqrt{\frac{L_N(r_k)}{L_N(p)}}\lsp=b_k^N.
 \label{eq:mle-markov}
\end{equation}
For every unit ket $\ket\psi$ covered by center $k$, put $\sigma=\proj\psi$. Cauchy--Schwarz, \eref{eq:mle-env-control}, and the convention \eref{eq:hel-def} imply
\begin{equation}
b_k\le\sum_i\sqrt{p_i}\lsp f_\psi(i)+\frac1{2N}=1-\frac12H^2(p,p_\sigma)+\frac1{2N}.
 \label{eq:mle-affinity}
\end{equation}

\emph{Tail and mean error.} Fix $0\le s\le2$. On the event $H^2(p,\hp)\ge s$, choose a net center covering a unit vector representing the MLE. Equations~\eqref{eq:mle-env-control} and~\eqref{eq:mle-affinity} give
\begin{equation}
L_N(r_k)\ge L_N(\hp)\ge L_N(p),\qquad b_k\le1-\frac s2+\frac1{2N}.
 \label{eq:mle-distant-env}
\end{equation}
For fixed $p,s$, the indices satisfying the latter inequality form a deterministic set. For each such $k$, \eref{eq:mle-markov} and $1+u\le\rme^u$ give $\bbP_p\{L_N(r_k)\ge L_N(p)\}\le\exp(1/2-Ns/2)$. A union bound over at most $\ell$ indices therefore yields
\begin{equation}
\bbP_p\left\{H^2(p,\hp)\ge s\right\}\le\min\left\{1,\ell\exp\left(\frac12-\frac{Ns}{2}\right)\right\}\qquad(0\le s\le2).
 \label{eq:mle-tail}
\end{equation}
Since $H^2\le2$, integrating this bound and extending the nonnegative integrand to $s\ge0$ gives
\begin{equation}
\bbE_pH^2(p,\hp)\le\int_0^\infty\min\left\{1,\ell\exp\left(\frac12-\frac{Ns}{2}\right)\right\}\rmd s=\frac{2\log\ell+3}{N}\le\frac{4d\log(1+8N\sqrt{d}\lsp)+3}{N}.
 \label{eq:mle-hel-proof}
\end{equation}
The integral is evaluated by splitting at $s=(2\log\ell+1)/N$. Combining this estimate with \eref{eq:mle-outcome-risk} and $\epsilon(\rho,\hrho_N)\le H^2(p,\hp)/\kappa_0$ proves \eref{eq:mle-risk}, uniformly over all pure states.
\end{proof}

\begin{proof}[Proof of \lref{lem:mle-env}]
A maximal $\varepsilon$-separated subset of the unit sphere in $\bbR^{2d}$ is an $\varepsilon$-net. The disjoint balls of radius $\varepsilon/2$ about its centers lie in the ball of radius $1+\varepsilon/2$, so comparison of volumes gives the cardinality bound.

The modulus inequality gives $|f_\psi(i)-f_{\psi_k}(i)|\le\sqrt{w_i}\lsp\norm{\ket\psi-\ket{\psi_k}}_2$. For a covering center, $f_\psi\le g_k$ pointwise and $|g_k(i)-f_\psi(i)|\le2\varepsilon\sqrt{w_i}$. Since $\sum_iw_i=d$, summing squares proves \eref{eq:mle-env-control}. POVM completeness $\sum_iw_i\phi_i=\bbone$ gives $\norm{f_\psi}_2^2=\sum_iw_i|\braket{\phi_i}{\psi}|^2=1$. The envelopes need not be normalized.
\end{proof}

\section{Numerical methods}\label{app:numerics}

\begin{figure}[!ht]
\centering
\raisebox{-0.00000pt}[161.04074pt][0.00000pt]{\includegraphics[width=251.69220pt,height=161.04074pt]{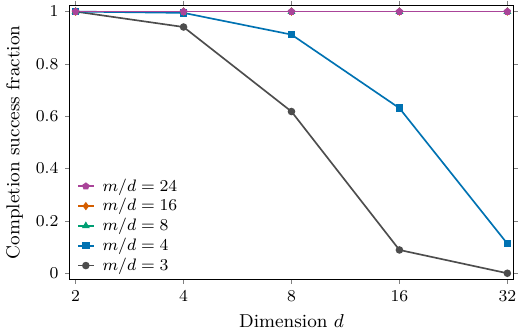}}
\caption{Fraction of the $1000$ measurement realizations at each parameter point satisfying $\lambda_{\max}(S_m)\le2$. Colors and markers indicate the same five sampling ratios as in Fig.~\ref{fig:samp-fi}. The curves for $m/d=8,16,24$ coincide at success fraction $1$.}
\label{fig:comp-ok}
\end{figure}

This appendix describes the measurements, state searches, and numerical checks underlying Figs.~\ref{fig:samp-fi} and~\ref{fig:glob-sep}, together with the completion success fractions in Fig.~\ref{fig:comp-ok}. Figure~\ref{fig:des-hier} plots the analytic bounds in \thref{thm:hier}.

\subsection{Sampling and spectral completion}\label{app:num-samp}

We use the uniform multiqubit stabilizer ensemble, an exact complex projective $3$-design in power-of-two dimensions~\cite{KuengGross2015,Webb2016,Zhu2017}. The dimensions are $d=2,4,8,16,32$, and the sampling ratios are $m/d=3,4,5,6,8,12,16,24$. Each parameter point has $1000$ independent measurement realizations. Figure~\ref{fig:samp-fi} displays the representative ratios $m/d=3,4,8,16,24$. For a fixed dimension and realization, the sample sizes use prefixes of the same independent sequence of $24d$ projectors drawn with replacement; results at different ratios are therefore paired. Stabilizer states are sampled uniformly using symplectic linear algebra over the binary field $\bbF_2$ and independent stabilizer signs~\cite{KoenigSmolin2014}.

The sampled elements retain the weights in \eref{eq:samp-povm}. With $S_m=(d/m)\sum_i\phi_i$, we test $\lambda_{\max}(S_m)\le2$ using a tolerance of $10^{-10}$. When this test passes, the residual $R_m=\bbone-S_m/2$ is spectrally decomposed into positive rank-one elements. Residual eigenvalues below $10^{-12}$ are treated as numerical zeros. Infeasible realizations are discarded without rescaling the sampled elements. Repeated sampled projectors keep their outcome labels, giving at most $m+d$ labeled outcomes; identical elements may be combined without changing Fisher information. Each completed measurement is fixed before choosing the probe state.

\subsection{Fisher evaluation and state search}\label{app:num-search}

Let the columns of $B$ form a complex orthonormal basis of $\caH_\psi$. Write $E_i=w_i\phi_i$ and $a_i=\braket{\phi_i}{\psi}$. In the real basis formed by the columns of $B$ and $\rmi B$, the Fisher matrix is given by the following formulas, with $\boldsymbol{\upsilon}_i$ and $\bmr_i$ defined for $a_i\ne0$:
\begin{equation}
 \boldsymbol{\upsilon}_i=\frac{a_i}{|a_i|}B^\dagger\ket{\phi_i},\qquad
 \bmr_i=\begin{pmatrix}\Rea\boldsymbol{\upsilon}_i\\\Ima\boldsymbol{\upsilon}_i\end{pmatrix},\qquad
 G_{\psi,\caM}=\sum_{i\colon a_i\ne0}w_i\bmr_i\bmr_i^{\sfT}.
 \label{eq:num-fi-mat}
\end{equation}

Minimizing the normalized Fisher quadratic form over unit horizontal tangent kets gives the smallest eigenvalue of $G_{\psi,\caM}$. Amplitudes $|a_i|\le10^{-12}$ are treated as numerical zeros, and their terms are omitted.

Each feasible POVM is evaluated at $800$ Haar-random states, sampled measurement states, completion eigenstates, computational-basis states, and states satisfying selected zero-probability constraints. We evaluate all stabilizer states for $d=2,4,8$ and $2000$ uniformly sampled stabilizer states for $d=16,32$. Local optimizations jointly vary the state and a unit horizontal tangent vector, starting from the five Haar-random states with the smallest $\lambda_{\min}(G_{\psi,\caM})$, with at most $100$ iterations per start. Returned states are reevaluated using $\lambda_{\min}(G)$. At $d=16,32$, we also search over stabilizer states by testing the states $\exp(\pm\rmi\pi P/4)\ket s$, where $\ket s$ is the current stabilizer ket and $P$ is a Hermitian Pauli operator satisfying $\bra sP\ket s=0$. We accept moves that decrease the smallest Fisher eigenvalue, using two starting states and at most eight steps per start.

Writing $\caS_{\caM}$ for all evaluated states, the search estimate is
\begin{equation}
 \heta(\caM)=\min_{\ket\psi\colon\proj\psi\in\caS_{\caM}}\lambda_{\min}(G_{\psi,\caM}),\qquad
 \eta(\caM)\le\heta(\caM).
 \label{eq:num-search-eta}
\end{equation}
The reported values are numerical approximations to $\heta(\caM)$.

\subsection{Statistics}\label{app:num-checks}

Figure~\ref{fig:comp-ok} reports completion success fractions. For Fig.~\ref{fig:samp-fi}, we report the median and $10$th--$90$th percentiles of $\heta$ over feasible POVMs.

The completed POVMs had completeness residuals $\norm{\sum_iE_i-\bbone}<10^{-14}$.

\subsection{Global-separation searches}\label{app:num-sep}

Figure~\ref{fig:glob-sep} uses $d=2,4,8$. An angular $3$-design in $\bbC^{d-1}$ is obtained by projecting the complete stabilizer ensemble $\{(1/M_n,\phi_i)\}$ in dimension $d=2^n$ onto a fixed $(d-1)$-dimensional subspace with projector $P$. For $h_i=\bra{\phi_i}P\ket{\phi_i}>0$, set
\begin{equation}
 \ket{\upsilon_i}=\frac{P\ket{\phi_i}}{\sqrt{h_i}\lsp},\qquad
 \gamma_i=\frac{(d+2)h_i^3}{(d-1)M_n}.
 \label{eq:num-proj-des}
\end{equation}
Zero projections are discarded. The third tensor moment gives $M_n^{-1}\sum_i h_i^3=(d-1)/(d+2)$, so the weights $\gamma_i$ sum to one, and the projected ensemble is an exact $3$-design in the subspace.

Fix the phase of each $\ket{\upsilon_i}$ by making its first nonzero computational amplitude positive and real, merge coincident projectors by adding their weights, and apply \lref{lem:fin-lift} with relative phases $1,\rmi,-1,-\rmi$. The radial nodes and weights are
\begin{equation}
 (u_0,u_1,u_2)=\left(0,\frac2{d+2},1\right),\qquad
 (\alpha_0,\alpha_1,\alpha_2)=
 \left(\frac{d-1}{2(d+1)},\frac{(d-1)(d+2)^2}{2d^2(d+1)},\frac2{d^2(d+1)}\right).
 \label{eq:num-ext-rad-rule}
\end{equation}
The angular designs have $1$, $31$, and $939$ projectors, yielding $6$, $156$, and $4696$ POVM elements, respectively. For the resulting POVM ensembles, $\left\|D_{[s]}\sum_i\pi_i\phi_i^{\otimes s}-P_{[s]}\right\|<10^{-14}$ for $s=1,2,3$. At $\rho=\proj0$, each eigenvalue of $G_{0,\caM}$ differed from $c_{d,3}$ in \eref{eq:sharp-iso} by less than $10^{-14}$.

For $27$ infidelities between $10^{-4}$ and $1$, pairs are parameterized by orthonormal kets $(\ket\psi,\ket \nu)$ as in \pref{prop:sym-fi}, with $\theta=(1/2)\arcsin\sqrt{\epsilon}$. Each dimension uses $20000$ Haar-random orthonormal ket pairs and multistart L-BFGS-B optimization~\cite{ByrdEtAl1995} with at most $300$ iterations per start. For the extremal POVMs, we also use pairs centered at $\ket0$, with unit tangent kets orthogonal to $\ket0$, and add starting pairs obtained by Gaussian perturbations of the best such pair followed by orthonormalization. Minimizers from the preceding infidelity provide additional starts. 

For the stabilizer POVMs, we also draw $128$ Haar-random orthonormal ket pairs in the span of the first two computational-basis vectors and generate candidate state pairs using the parameterization above. The four pairs with the smallest separation ratios provide additional optimization starts. At $d=2$, the measurements coincide.

\end{document}